%% file: ms.tex
\documentclass{emulateapj}
\usepackage{epsfig}

\input defs.tex

\input epsf
\bibpunct{(}{)}{;}{a}{}{,}

\begin{document}

\lefthead{QUASARS PROBING QUASARS}
\righthead{HENNAWI \etal}

\title{Quasars Probing Quasars I: Optically Thick Absorbers Near Luminous Quasars}
\author{Joseph F. Hennawi\altaffilmark{1,2,3}, 
  Jason X. Prochaska\altaffilmark{4} 
  Scott Burles,\altaffilmark{5}
  Michael A. Strauss,\altaffilmark{2}
  Gordon T. Richards,\altaffilmark{6} 
  David J. Schlegel,\altaffilmark{7}
  Xiaohui Fan,\altaffilmark{8}
  Donald P. Schneider,\altaffilmark{9} 
  Nadia L. Zakamska,\altaffilmark{10,11}
  Masamune Oguri,\altaffilmark{2}
  James E. Gunn,\altaffilmark{2}
  Robert H. Lupton,\altaffilmark{2}
  Jon Brinkmann\altaffilmark{12}
}

\altaffiltext{1}{Department of Astronomy, University of California
  Berkeley, Berkeley, CA 94720; joeh@berkeley.edu}
\altaffiltext{2}{Princeton University Observatory, Princeton, NJ 08544}
\altaffiltext{3}{Hubble Fellow}
\altaffiltext{4}{Department of Astronomy and Astrophysics, 
  UCO/Lick Observatory; University of California, 1156 High Street, Santa Cruz, 
  CA 95064; xavier@ucolick.org}
\altaffiltext{5}{Physics Department, Massachusetts Institute of Technology, 77 Massachusetts Avenue, Cambridge, MA 02139.} 
\altaffiltext{6}{Department of Physics and Astronomy, Johns Hopkins University, 3400 N. Charles Street, Baltimore, MD 21218-2686}
\altaffiltext{7}{Lawrence Berkeley National Laboratory, One Cyclotron Road, Mailstop 50R232, Berkeley, CA, 94720, USA.}
\altaffiltext{8}{Steward Observatory, University of Arizona, 933 North Cherry Avenue, Tucson, AZ 85721}
\altaffiltext{9}{Department of Astronomy and Astrophysics, Pennsylvania State
University, 525 Davey Laboratory, University Park, PA 16802, USA}
\altaffiltext{10}{Institute for Advanced Study, Einstein Drive, Princeton, NJ 08540 }
\altaffiltext{11}{Spitzer Fellow}
\altaffiltext{12}{Apache Point Observatory, P. O. Box 59, Sunspot, NM88349-0059.}

\begin{abstract}
  With close pairs of quasars at different redshifts, a background
  quasar sightline can be used to study a foreground quasar's
  environment in \emph{absorption}. We search 149 moderate resolution
  background quasar spectra, from Gemini, Keck, the MMT, and the SDSS
  to survey Lyman Limit Systems (LLSs) and Damped Ly$\alpha$ systems
  (DLAs) in the vicinity of $1.8 < z < 4.0$ luminous foreground
  quasars.  A sample of 27 new quasar-absorber pairs is uncovered with
  column densities, $10^{17.2}~{\rm cm}^{-2} < \mnhi < 10^{20.9}{\rm
    cm}^{-2}$, and transverse (proper) distances of $22~\hkpc < R <
  1.7~\hMpc$, from the foreground quasars. If they emit isotropically,
  the implied ionizing photon fluxes are a factor of $\sim 5-8000$
  times larger than the ambient extragalactic UV background over this
  range of distances. The observed probability of intercepting an
  absorber is very high for small separations: six out of eight
  projected sightlines with transverse separations $R < 150~\hkpc$ have an absorber coincident with the foreground quasar,
  of which four have $\mnhi > 10^{19}{\rm cm}^{-2}$. The covering
  factor of $\mnhi > 10^{19}{\rm cm}^{-2}$ absorbers is thus $\sim
  50\%$ (4/8) on these small scales, whereas $\lesssim 2\%$ would have
  been expected at random. There are many cosmological applications of
  these new sightlines: they provide laboratories for studying
  fluorescent Ly$\alpha$ recombination radiation from LLSs, constrain
  the environments, emission geometry, and radiative histories of
  quasars, and shed light on the physical nature of LLSs and DLAs.
\end{abstract}


\section{Introduction}
\label{sec:intro}

Although optically thick absorption line systems, that is the Lyman
Limit Systems (LLSs) and damped Lyman-$\alpha$ systems (DLAs), are
detected as the strongest absorption lines in quasar spectra, the two
types of objects, quasars and absorbers, play rather different roles
in the evolution of structure in the Universe.  The hard ultraviolet
radiation emitted by luminous quasars gives rise to the ambient
extragalactic ultraviolet (UV) background \citep[see
e.g.][]{HM96,Meiksin05} responsible for maintaining the low neutral
fraction of hydrogen ($\sim 10^{-6}$) in the intergalactic medium (IGM),
established during reionization. However, high column density
absorbers represent the rare locations where the neutral fractions are
much larger.  Gas clouds with column densities $\log N_{\rm HI}>17.2$
are optically thick to Lyman continuum ($\tau_{\rm LL} \gtrsim 1$)
photons, giving rise to a neutral interior self-shielded from the
extragalactic ionizing background.  In particular, the damped
Ly$\alpha$ systems dominate the neutral gas content of the Universe
\citep{phw05}, which provides the primary reservoir for the star
formation which occurred to form the stellar masses of galaxies in the
local Universe.

One might expect optically thick absorbers to keep a safe distance from 
luminous quasars. For a quasar at $z=2.5$ with an $r$-band magnitude
of $r=19$, the flux of ionizing photons is 130 times higher than that of
the extragalactic UV background at an angular separation of
$60\arcsec$, corresponding to a proper distance of 340~$\hkpc$ and
increasing as $r^{-2}$ toward the quasar. Indeed, the decrease in the
number of \emph{optically thin} absorption lines ($\log N_{\rm
  HI}<17.2$ hence $\tau_{\rm LL}\lesssim 1$), in the vicinity of
quasars, known as the \emph{proximity effect} \citep{BDO88}, has been
detected and its strength provides a measurement of the UV background
\citep{Scott00}. If Nature provides a nearby background quasar
sightline, one can also study the \emph{transverse proximity effect},
which is the expected decrease in absorption in a \emph{background}
quasar's Ly$\alpha$ forest, caused by the transverse ionizing flux of
a \emph{foreground} quasar. It is interesting that the transverse
effect has yet to be detected, in spite of many attempts \citep[][but
see Jakobsen \etal 2003]{Crotts89,DB91,FS95,LS01,Schirber04,Croft04}.

On the other hand, it has long been known that quasars are associated
with enhancements in the distribution of galaxies
\citep{BSG69,YG84,YG87,BC91,SBM00,BBW01,Serber06,Coil06}, although these
measurements of quasar galaxy clustering are limited to low redshifts
$\lesssim 1.0$. Recently, \citet{AS05}, measured the clustering of
Lyman Break Galaxies (LBGs) around luminous quasars in the redshift
range ($2 \lesssim z \lesssim 3.5$), and found a best fit
correlation length of $r_0=4.7~\hMpc$ ($\gamma=1.6$), very
similar to the auto-correlation length of $z\sim 2-3$ LBGs
\citep{Adel03}. \citet{Cooke06} recently measured the clustering
of LBGs around DLAs and measured a best fit $r_0=2.9~\hMpc$ with
$\gamma=1.6$, but with large uncertainties \citep[see
also][]{Gawiser01,Bouchet04}.  If LBGs are clustered around quasars,
and LBGs are clustered around DLAs, might we expect optically thick
absorbers to be clustered around quasars?  This is especially
plausible in light of recent evidence that DLAs arise from a high
redshift galaxy population which are not unlike LBGs \citep{Moller02}.

Clues to the clustering of optically thick absorbers around
quasars come from a subset of DLAs with $z_{\rm abs}\sim z_{\rm em}$
known as \emph{proximate DLAs}, which have absorber redshifts within
$3000\kms$ of the emission redshift of the quasars \citep[see
e.g.][]{Moller98}. Recently, \citet{REB05} \citep[see
also][]{Ellison02}, compared the number density of proximate
DLAs per unit redshift to the average number density of DLAs in the
the Universe \citep{phw05}. They found that the abundance of
DLAs is enhanced by a factor of $\sim 2$ near quasars, which they
attributed to the clustering of DLA-galaxies around quasars.

%In this and a series of upcoming papers \citep{LLS2,LLS3,LLS4}, 
Here, we present a new technique for studying absorbers near luminous
quasars, which can be thought of as the \emph{optically thick} analog
of the transverse proximity effect.  Namely, we use background quasar
sightlines to search for optically thick absorption in the vicinity of
foreground quasars.  Although such projected quasar pair sightlines
are extremely rare, \citet{binary} showed that it is straightforward
to select $z\gtrsim2$ projected quasar pairs from the imaging and
spectroscopy provided by the Sloan Digital Sky Survey
\citep[SDSS;][]{York00}.  In this work, we combine high
signal-to-noise ratio (SNR) moderate resolution spectra of the closest
\citet{binary} projected pairs, obtained from Gemini, Keck, and the
Multiple Mirror Telescope (MMT), with a large sample of wider
separation pairs, from the SDSS spectroscopic survey, arriving at a
total of 149 projected pair sightlines in the redshift range $1.8 < z
< 4.0$.  A systematic search for optically thick absorbers in the
vicinity of the foreground quasars is conducted, uncovering 27 new
quasar absorber pairs with column densities $17.2 < \log \mnhi < 20.9$
and transverse (proper) distances $22~\hkpc < R < 1.7~\hMpc$ from the
foreground quasars.

A handful of quasar-absorber pairs exist in the literature, all of
which were discovered serendipitously. In a study of the statistics of
coincidences of optically thick absorbers across close quasar pair
sightlines, \citet{Dodo02} discovered one LLS ($z_{\rm abs} = 2.12$)
and one DLA ($z_{\rm abs} = 2.54$) in background quasar spectra within
$\Delta v \lesssim 1000\kms$ of the foreground quasar redshifts,
corresponding to transverse proper distances of $320~\hkpc$ and
$1.75~\hMpc$, respectively. More recently, \citet{Adel05}
serendipitously discovered a faint background quasar ($z=2.92$)
$49\arcsec$ from a luminous ($r\sim 16$) foreground quasar at
$z=2.84$, corresponding to transverse separation $R =280~\hkpc$.  A
DLA was detected in the background spectrum at the same redshift as
the foreground quasar.

This is the first in a series of four papers on optically thick
absorbers near quasars. In this work, we describe the observations and
sample selection and present 27 new quasar-absorber pairs. Paper II
\citep{LLS2} focuses on the clustering of absorbers around foreground
quasars and a measurement of the transverse quasar-absorber
correlation function is presented. We investigate fluorescent
Ly$\alpha$ emission from our quasar-absorber pairs in Paper III
\citep{LLS3}.  Echelle spectra of several of the quasar-LLS systems
published here are analyzed in Paper IV \citep{LLS4}.

Quasar pair selection and details of the observations are described in
\S~\ref{sec:observe}.  The selection techniques and the sample are
presented in \S~\ref{sec:sample}.  A detailed discussion of how the
systemic redshifts of the foreground quasars were estimated is given
in \S~\ref{sec:systemic}.  The individual members of the sample are
discussed in \S~\ref{sec:notes}.  Cosmological applications of
quasar-absorber pairs are mentioned in \S~\ref{sec:apps} and we
summarize in \S~\ref{sec:conc}.

Throughout this paper we use the best fit WMAP (only) cosmological
model of \citet{Spergel03}, with $\Omega_m = 0.270$, $\Omega_\Lambda
=0.73$, $h=0.72$.  Unless otherwise specified, all distances are
proper. It is helpful to remember that in the
chosen cosmology, at a redshift of $z=2.5$, an angular separation of
$\Delta\theta=1\arcsec$ corresponds to a proper transverse 
separation of $R=6~\hkpc$, and a velocity difference of $1500\kms$ 
corresponds to a radial redshift space distance of $s=4.3~\hMpc$.  For
a quasar at $z=2.5$, with an SDSS magnitude of $r=19$, the
flux of ionizing photons is 130 times higher than the ambient
extragalactic UV background at an angular separation of $60\arcsec$
($R=340~\hkpc$). Finally, we use term optically thick absorbers and
LLSs interchangeably, both referring to quasar absorption line systems
with $\log N_{\rm HI}>17.2$, making them optically thick at the Lyman
limit ($\tau_{\rm LL}\gtrsim 1$).

\section{Quasar Pair Observations}
\label{sec:observe}

Finding optically thick absorbers near quasars requires spectra of
projected pairs of quasars at different redshifts, both with $z
\gtrsim 2$, so that Ly$\alpha$ is above the atmospheric cutoff.  In
this section we describe the spectra of projected quasar pairs from
the SDSS and 2QZ spectroscopic surveys as well our subsequent quasar pair
observations from Keck, Gemini, and the MMT.

\subsection{The SDSS Spectroscopic Quasar Sample}

The Sloan Digital Sky Survey uses a dedicated 2.5m telescope and a
large format CCD camera \citep{Gunn98,Gunn06} at the Apache Point Observatory
in New Mexico to obtain images in five broad bands \citep[$u$, $g$,
$r$, $i$ and $z$, centered at 3551, 4686, 6166, 7480 and 8932 \AA,
respectively;][]{Fuku96,Stoughton02} of high Galactic latitude sky in
the Northern Galactic Cap.  The imaging data are processed by the
astrometric pipeline \citep{Astrom} and photometric pipeline
\citep{Photo}, and are photometrically calibrated to a standard star
network \citep{Smith02,Hogg01}. Additional details on the SDSS data
products can be found in \citet{DR1,DR2,DR3}.

Based on this imaging data, spectroscopic targets chosen by various
selection algorithms (i.e. quasars, galaxies, stars, serendipity) are
observed with two double spectrographs producing spectra covering
\hbox{3800--9200 \AA} with a spectral resolution ranging from 1800 to
2100 (FWHM $\simeq150-170\kms$).  Details of the spectroscopic
observations can be found in \citet{Castander01} and
\citet{Stoughton02}.  A discussion of quasar target selection is
presented in \citet{qsoselect}.  The blue cutoff of the SDSS
spectrograph imposes a lower redshift cutoff of $z \approx 2.2$ for
detecting the Ly$\alpha$ transition.  The Third Data Release Quasar
Catalog contains 46,420 quasars \citep{Schneider05}, of which 6,635
have $z > 2.2$.  We use a larger sample of quasars which also includes
non-public data: our parent quasar sample comprises 11,742 quasars
with $z > 2.2$.  Note also that we have used the Princeton/MIT
spectroscopic reductions\footnote{Available at
  http://spectro.princeton.edu} which differ slightly from the
official SDSS data release.

The SDSS spectroscopic survey selects \emph{against} close pairs of
quasars because of fiber collisions.  The finite size of optical
fibers implies only one quasar in a pair with separation $<55\arcsec$
can be observed spectroscopically on a given plate\footnote{An
  exception to this rule exists for a fraction ($\sim 30\%$) of the
  area of the SDSS spectroscopic survey covered by overlapping plates.
  Because the same area of sky was observed spectroscopically on more
  than one occasion, there is no fiber collision limitation.}. Thus
for sub-arcminute separations, additional spectroscopy is required
both to discover companions around quasars and to obtain spectra of
sufficient quality to search for absorption line systems.  For wider
separations, projected quasar pairs can be found directly in the
spectroscopic quasar catalog.

\subsection{The 2QZ Quasar Sample}

The 2dF Quasar Redshift Survey (2QZ) is a homogeneous spectroscopic
catalog of 44,576 stellar objects with 18.25 $\leq b_J\leq$ 20.85
\citep{Croom04}. Selection of quasar candidates is based on broad band
colors $(u b_J r)$ from automated plate measurements of the United
Kingdom Schmidt Telescope photographic plates. Spectroscopic
observations were carried out with the 2dF instrument, which is a
multi-object spectrograph at the Anglo-Australian Telescope. The 2QZ
covers a total area of 721.6 deg$^2$ arranged in two $75^{\circ}\times
5^{\circ}$ strips across the South Galactic Cap (SGP strip), centered
on $\delta = -30^{\circ}$, and North Galactic Cap (NGP strip, or
equatorial strip), centered at $\delta=0^{\circ}$. The NGP overlaps
the SDSS footprint, corresponding to roughly half of the 2QZ area. By
combining the SDSS quasar catalog with 2QZ quasars in the NGP we
arrive at a combined sample of 12,933 quasars with $z>2.2$, of which
11,742 are from the SDSS and 1,191 from the 2QZ.

The 2QZ spectroscopic survey is also biased against close quasar
pairs: their fiber collision limit is $30\arcsec$. The fiber collision
limits of both the SDSS and 2QZ can be partly circumvented by searching for
SDSS-2QZ projected quasar pairs in the region where the two surveys
overlap.  

\subsection{Keck, Gemini, and MMT Spectroscopic Observations}

Another approach to overcome the fiber collision limits is to use the
SDSS five band photometry to search for candidate companion quasars
around known, spectroscopically confirmed quasars.  \citet{binary}
used the 3.5m telescope at Apache Point Observatory (APO) to
spectroscopically confirm a large sample of photometrically selected
close quasar pair candidates.  This survey discovered both physically
associated, binary quasars, as well as projected quasar pairs, and
produced the largest sample of close pairs in existence.

We have obtained high signal-to-noise ratio, moderate resolution
spectra of a subset of the \citet{binary} quasar pairs from Keck,
Gemini, and the MMT. Thus far, 88 quasars with $z > 1.8$ have been
observed, which is the operational lower limit for detecting
Ly$\alpha$ set by the atmospheric cutoff.  We primarily targeted the
closest quasar pairs with small separations below the fiber-collision
limit ($\Delta \theta<55\arcsec$).  In some cases other nearby quasars
or quasar candidates were also observed at wider separations from a
known close pair. This was most often the case with the Keck
observations, where a multi-slit configuration was used, such that
other nearby known quasars or quasar candidates could be
simultaneously observed on a single mask.  Because some of the 88
quasars we observed are in triples or quadruples, the total number of
pairs is greater than 44.  About half of our pairs targeted consisted
of projected pairs of quasars ($\Delta v > 2500\kms$) at different
redshifts; the rest were physically associated binary quasars.

This spectroscopy program has several science goals: to measure small
scale transverse Ly$\alpha$ forest correlations, to constrain the dark
energy density of the Universe with the Alcock-Paczy{\'n}ski test
\citep{AP79,Pat99,HSB99}, and to characterize the transverse proximity
effect.  None of these projected pairs were specifically targeted
based on the presence or absence of an LLS. Thus these projected
sightlines constitute an unbiased sample for searching for optically
thick absorbers near foreground quasars. 

For the Keck observations, we used the Low Resolution Imaging
Spectrograph \citep[LRIS;][]{LRIS}, in multi-slit mode with custom
designed slitmasks, which allowed placement of slits on other known
quasars or quasar candidates in the field.  LRIS is a double
spectrograph with two arms giving simultaneous coverage of the near-UV
and red.  We used the D460 dichroic with the $1200$ lines mm$^{-1}$
grism blazed at $3400$~\AA\ on the blue side, resulting in wavelength
coverage of $\approx 3300-4200$~\AA. The dispersion of this grism is
$0.50$~\AA\ per pixel, giving a resolution of FWHM$\simeq 125\kms$.
On the red side, we used the 300 lines mm$^{-1}$ grating blazed at
$5000$~\AA, which covered the wavelength range $4700-10,000$~\AA,
resulting in $2.4$~\AA\ per pixel dispersion or a FWHM$\simeq
500\kms$.  All the LLSs discovered in the Keck LRIS data were found in
the blue side spectra, owing to the low redshift ($z\sim 2$) of our
Keck targets. We used the longer wavelength coverage on the red side
to aid with the identification of new quasars and to determine
accurate systemic redshifts (see \S~\ref{sec:systemic}).  The Keck
observations took place during two runs on UT 2004 November 7-8 and UT
2005 March 8-9.

The Gemini data were taken with the Gemini Multi-Object Spectrograph
\citep[GMOS;][]{GMOS} on the Gemini North facility. We used the
B$1200\_G5301$ grating which has 1200 lines mm$^{-1}$ and is blazed at
5300~\AA. The detector was binned in the spectral direction resulting
in a pixel size of 0.47~\AA~with the $1\arcsec$ slit, corresponding to
a FWHM$\simeq 125\kms$.  The slit was rotated so that both quasars in
a pair could be observed simultaneously.  The wavelength center
depended on the redshift of the quasar pair being observed. We
typically observed $z \sim 2.3$ quasars with the grating center at
4500~\AA~, giving coverage from $3750-5225$~\AA~, and higher redshift
$z\sim 3$ pairs centered at 4500~\AA~, covering $4000-5250$~\AA~. The
Gemini CCD has two gaps in the spectral direction, corresponding to
9~\AA~ at our resolution. The wavelength center was thus dithered by
15-50\AA\ so as to obtain full wavelength coverage in the gaps. The
Gemini observations were conducted over three classical runs during UT
2004 April 21-23, UV 2004 November 16-18, and UT 2005 March 13-16.

At the MMT on UT 2003 December 28-29, we used the blue channel
spectrograph with the 832 lines mm$^{-1}$ grating at a second order
blaze wavelength of 3900~\AA. The CuSO4 red blocking filter was used
to block contamination from first order red light. The resolution was
0.36~\AA\ per pixel, or FWHM=75~$\kms$. The grating tilt again depended
on the quasar pair redshift, but a typical wavelength center was
3600~\AA, giving coverage from 3100-4000~\AA.  

Exposure times ranged from $1800-7200$s, for the Keck, Gemini and MMT
observations, depending on the magnitudes of the targets. The SNR in
\lya\ forest region varies considerably,  but it is always SNR$>5$ per 
pixel for the data we consider here. 

One of the projected pairs in our sample, SDSSJ0239-0106, was observed
with LRIS-B at low resolution because because a damped Ly$\alpha$
system was detected in the background quasars' SDSS spectrum, 
coincident with the foreground quasar redshift, but the SNR of the
SDSS spectrum was very low. For this observation (spectrum shown in
Figure~\ref{fig:qsos}), the $300$ lines mm$^{-1}$ grism was used with
the D680 dichroic. The spectral coverage was from $3500-6800$~\AA\
giving 1.43~\AA\ per pixel or a FWHM$\simeq 500\kms$. The date of this
observation was UT 2005 December 1.

Finally, we observed the quasar pair SDSSJ1427-0121 with the DEep
Imaging Multi-Object Spectrograph \citep[DEIMOS;][]{DEIMOS} on the
Keck II telescope on UT 2005 May 5.  We observed the quasar for two
exposures of 300s with the long-slit mask (0.75$''$ slit), using the
600 lines mm$^{-1}$ grating centered at 7500\AA\ and the gg495
blocking filter.  The primary reason for this observation was to acquire
coverage of the \ion{Mg}{2} emission line of the foreground quasar so
as to estimate an accurate systemic redshift (see
\S~\ref{sec:systemic}).

\section{Sample Selection}

\label{sec:sample}
\input{tab1.tex}

A large number of projected quasar pairs must be searched to find optically 
thick absorbers near quasars. For example, the number of absorbers
per unit redshift at $z\sim 2.5$ with column densities $\log \mnhi >
19$ is $dN\slash dz \simeq 0.5$ \citep{Peroux05,Omeara06}. In
practice, we search within a velocity window $\Delta v =
1500\kms$, because of our uncertainty of the foreground quasar's
systemic redshift (see \S~\ref{sec:systemic}). The random probability
of finding an absorber in the background quasar spectrum within 
$\Delta v$ of a foreground quasar is $\sim 2\%$. Thus of order fifty
projected quasar pair sightlines are required to find a single
quasar-absorber pair, with absorbers of lower (higher) column density
being more (less) abundant.  As we will see, LLSs are indeed
\emph{clustered} around quasars, so that the probability of finding an
absorber will increase substantially for scales smaller than the
correlation length \citep{LLS2}.

Our goal is to construct a \emph{statistical} sample of absorbers near
quasars which is nearly complete above some column density
threshold. One approach would be to start with complete samples of
absorption line systems and quasars, and simply search for
quasar-absorber pairs. Indeed, \cite{phw05} recently published a large
complete sample of DLAs found by searching the SDSS spectroscopic
quasar catalog. However, by restricting the quasar sample to be
complete in a flux or volume limited sense, we would substantially reduce
the number of usable quasars, and furthermore, this would not exploit
the large number of close projected sightlines provided by our
Keck/Gemini/MMT  observations. Instead, our approach is to
search all projected pair sightlines for which an absorber could be
detected in the background quasar spectrum, at the redshift of the
foreground quasar. This is of course a question about the signal-to
-noise ratio and resolution of the \emph{background} quasar spectrum.

\cite{phw05} demonstrated that the spectral resolution (FWHM $\simeq
150\kms$), signal-to-noise ratio (SNR), and wavelength coverage of the
SDSS DR3 spectra are well suited to detecting DLAs at $z > 2.2$.  They
limited their search to objects with SNR$>4$ per pixel corresponding
to roughly $r\lesssim 19.5$, for which they were complete at $> 95\%$ for
column densities $\log \mnhi > 20.2$. Here we must be more aggressive
because of the limited number of projected sightlines and the paucity
of known LLSs near quasars.  Consequently, our sample may not yet achieve
such a high completeness and it is more susceptible to false-positive
detections. 

We begin by finding all unique projected quasar pair sightlines which
have a \emph{comoving} transverse separation of $R <
5~\hMpc$\footnote{A comoving distance limit is imposed rather than a
  proper one because the clustering analysis in \citep{LLS2} is
  carried out in comoving units.} at the redshift of the foreground
quasar. The list of potential quasar pair members includes all of the
quasars in the SDSS+2QZ sample, all 88 of the quasars for which we
have Keck/Gemini/MMT spectra, and all the quasars confirmed from APO
follow up spectra published in \cite{binary}. Any known quasar can
serve as a foreground quasar, provided we are confident of its
redshift.  Only objects which satisfied the SNR criterion described
below could serve as background quasars. The 2QZ spectra and the APO
spectra do not have sufficient resolution or SNR to find high column
density absorbers, so these quasars could only serve as foreground
quasars.  Furthermore, all of our Keck/Gemini/MMT spectra easily
satisfy our SNR criteria, so in practice, we only apply a SNR
statistic to the SDSS spectra.

\subsection{SNR Statistic}

We define a signal-to-noise statistic SNR$_{\rm bg}$ in the background
quasar spectrum which is an average of the median signal-to-noise
ratio blueward and redward of the Ly$\alpha$ transition at the
foreground quasar redshift. 

For the blue side, we begin at the wavelength $\lambda_{\rm blue}= (1
+ z_{\rm fg})(1215.67 - 20)~{\rm \AA}$, and take the median SNR of the
150 pixels blueward of this wavelength. The 20~\AA~offset
(4936~$\kms$) is applied so that the SNR is not biased by the
presence of a potential absorber. If there are not 150 available
pixels blueward of $\lambda_{\rm blue}$ because of the blue cutoff of
the spectrum, we take the median of the $n_{\rm blue} > 50$ pixels
which remain. If less than 50 pixels are available, we set
SNR$_{\rm blue}=0$ and $n_{\rm blue}=0$.

Similarly, on the red side we begin at $\lambda_{\rm red}= (1 + z_{\rm
  fg})(1215.67 +20)~{\rm \AA}$, and take the median SNR of the 150
pixels redward of this wavelength. If there are not 150 pixels redward
of $\lambda_{\rm red}$ which also have $\lambda < (1 + z_{\rm
  bg})~1190$~\AA, we compute the median SNR$_{\rm red}$ of the $n_{\rm
  red}$ pixels available. Wavelengths larger than $(1 + z_{\rm
  bg})1190$~\AA are avoided because the SNR rises at the Ly$\alpha$
emission line in the background quasar spectrum. If $n_{\rm red} <
150$, we then also compute the median SNR$_{\rm 1275}$ of the $n_{\rm
  1275}= 150-n_{\rm red}$ remaining pixels redward of the wavelength
$\lambda_{\rm 1275}=(1+z_{\rm bg})1275$~\AA, which is free of emission
lines and a good place to estimate the red continuum SNR. Our SNR statistic
is defined to be the average 
\be 
{\rm SNR}_{\rm bg}\equiv \frac{n_{\rm
    blue}~{\rm SNR}_{\rm blue} + n_{\rm red}~{\rm SNR}_{\rm red} +
  n_{\rm 1275}~{\rm SNR}_{\rm 1275}}{n_{\rm blue} + n_{\rm red} +
  n_{\rm 1275}}.  
\ee

We require that the foreground quasar's Ly$\alpha$ must be redward of
the background quasar's Lyman limit, $(1 + z_{\rm fg})1215.67 >
(1+z_{\rm bg})912$, to avoid searching in the highly absorbed low SNR
region blueward of $912~$\AA.  A minimum velocity difference between
the two quasars of $\Delta v > 2500\kms$ is chosen to exclude binary
quasars.  These pairs with small velocity separation are excluded to
avoid confusion about which object is in the background and to avoid
distinguishing absorption intrinsic to the background quasar from 
absorption associated with the foreground quasar.  Because the
small angular separation projected pairs are particularly rare, we set a more
liberal minimum SNR of SNR$_{\rm bg} > 1.5$ for projected pairs which
have (comoving) transverse separation $R < 1~\hMpc$. For wider
separation pairs $1~\hMpc < R < 5~\hMpc$ (comoving), we require
SNR$_{\rm bg} > 2$.

\subsection{Visual Inspection}

All projected quasar pairs satisfying the aforementioned criteria were
visually inspected and we searched for significant Ly$\alpha$
absorption within a velocity window of $|\Delta v| = 1500\kms$ about
the foreground quasar redshift. This velocity range because it
brackets the uncertainties of the foreground quasar systemic redshift
(see \S~\ref{sec:systemic}). Strong broad absorption line (BAL)
quasars with large \ion{C}{4} equivalent widths (EWs) were excluded
from the analyses. Mild BALs were excluded if the BAL absorption
clearly coincided with the velocity window about the foreground quasar
redshift which was being searched.

Systems with significant Ly$\alpha$ absorption were flagged for
\ion{H}{1} absorption profile fitting. In the SDSS spectra, all
systems which had an absorber with rest equivalent width $W_{\lambda}
> 2$~\AA\ were flagged to be fit. We adopted a lower threshold of
$W_{\lambda} > 1.5$~\AA\ for the Keck/Gemini/MMT spectra, which have
higher SNRs and slightly better resolution. These equivalent width
thresholds correspond to column densities of roughly $\log \mnhi
\gtrsim 19$ and $\log \mnhi \gtrsim 18.5$, respectively.

The \ion{H}{1} search was complemented by a search for metal lines at
the foreground quasar redshift, in the clean continuum region redward
of the Ly$\alpha$ forest of the background quasar.  The narrow metal
lines provide a redshift for the absorption line system and, if
present, they can help distinguish optically thick absorbers from
blended Ly$\alpha$ forest lines.  We focused on the strongest low-ion
transitions commonly observed in DLAs \citep[e.g.][]{Proch03}:
\ion{Si}{2}~$\lambda 1260, 1304, 1526$, \ion{O}{1}~$\lambda 1302$,
\ion{C}{2}~$\lambda 1334$, \ion{Al}{2}~$\lambda 1670$,
\ion{Fe}{2}~$\lambda 1608,2382,2600$, \ion{Mg}{2}~$\lambda 2796,2803$;
and the strong high-ionization transitions commonly seen in LLSs:
\ion{C}{4}~$\lambda 1548,1550$ and \ion{Si}{4}~$\lambda
1393,1402$. Any systems with secure metal-line absorption were also
flagged to be fit.

The Lyman limit at $912$~\AA\ is redshifted into the SDSS spectral
coverage for $z > 3.2$. Although we did not apply any specific SNR
criteria on the spectra at these bluer wavelengths, special attention
was paid to projected pairs for which the Lyman limit was
detectable. Systems which showed Lyman limit absorption at the
redshift of the foreground quasar were also flagged, regardless of the
equivalent width of their Ly$\alpha$ absorption or the strength or
presence of metal lines.

\begin{figure}
  \centerline{\epsfig{file=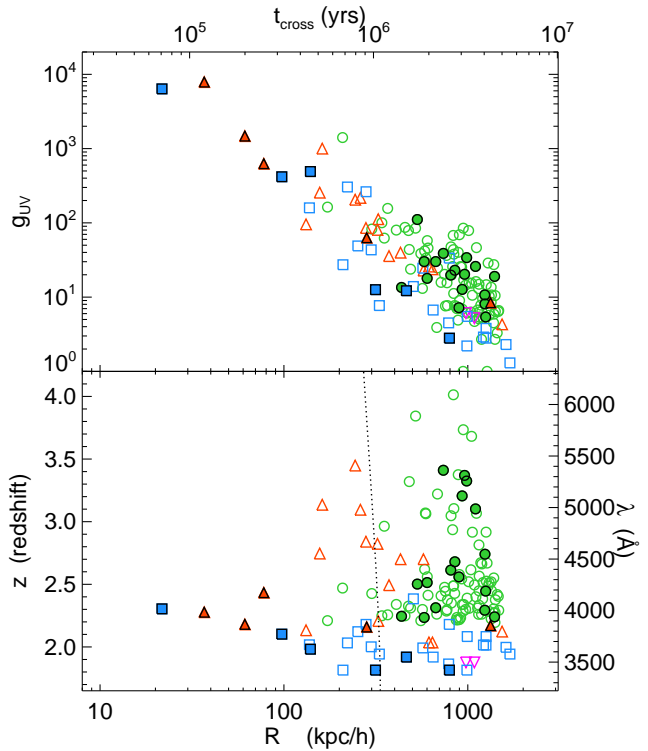,bb= 100 80 650
      700,width=0.50\textwidth}}
  \caption{Distribution of foreground quasar redshifts, transverse
    separations, and ionizing fluxes probed by the background quasar
    sightlines.  The upper plot shows ionizing
        flux versus proper separations, which explains the general
        $R^{-2}$ trend.  The lower plot shows foreground quasar
        redshift versus proper separations and the y-axis on the right
        indicates the wavelength of the Ly$\alpha~\lambda1215.67$~\AA
        ~transition at this redshift. The (blue) squares have a Keck (LRIS-B)
        spectrum of the background quasar, (red) triangles have Gemini
        (GMOS) background spectra, (magenta) upside down triangles
        have MMT (Blue Channel) background spectra, and (green)
        circles have SDSS background spectra. Filled symbols outlined
        in black have an optically thick absorber at the foreground
        quasar redshift (see Table~\ref{table:thick}) and open symbols have 
        no absorber.  The region to the left of the dotted line is excluded 
        by the SDSS fiber
        collision limit of $\theta=55\arcsec$, which explains the
        paucity of SDSS background spectra there. The follow-up 
        Keck/Gemini/MMT spectra probe angular separations 
        an order of magnitude smaller than the fiber collision
        limit, allowing us to probe the foreground quasar environment down 
        to $20~\kpc/h$ where the ionizing flux is $\sim 10,000$ times the UV 
        background.\label{fig:scatter}}
\end{figure}

\subsection{Voigt Profile Fitting}

For all of the systems which were flagged by the initial visual
inspection, we estimated the \ion{H}{1} column density by fitting the
\lya\ profiles using standard practice.  Namely, we over-plotted a
Voigt profile on the \lya\ transition, and centered the profile
according to the redshift of metal-lines, if present.  Otherwise, the
redshift of the absorber was allowed to be a free parameter in the
fit. The fits are done `by-eye', which is to say we do not minimize a
$\chi^2$ because the error in the fit is dominated by systematic
uncertainty related to the quasar continuum placement and
line-blending.  Conservative error estimates are adopted to account for
this uncertainty.  In all cases, we assume a Doppler parameter $b$,
which is typical of the high $z$ \lya\ forest
\citep[e.g.][]{Kirkman97}.  In general, the fits are insensitive to
the Doppler parameter parameter because most of the leverage in the
fit comes from the damping wings of the line-profile; we assume $b= 30
\mkms$.  See \cite{phw05} for more discussion on Voigt profile fits to 
\lya\ absorption profiles.

The completeness and false positive rate of our survey are sources of
concern.  Line-blending, in particular, can significantly depress the
continuum near the \lya\ profile and mimic a damping wing, biasing
the column density high.  To investigate these issues we consider
three objects in our sample for which we have both SDSS spectra and
independent echelle observations. We obtained archived spectra of the
background quasars SDSSJ~0303$-$0023, and SDSSJ~1204$+$0221, observed
with the High Resolution Echelle Spectrometer \citep[HIRES;][]{HIRES}
(FWHM $\approx 10\kms$) on the Keck-I telescope, and
SDSSJ~1429$-$0145, observed with the Magellan Inamori Kyocera Echelle
spectrograph \citep[MIKE;][]{MIKE} (FWHM $\approx 10\kms$).  For
SDSSJ0303$-$0023, the column density of $\log \mnhi = 18.95 \pm 0.2$
measured form the SDSS spectrum is in good agreement $\log \mnhi =
18.9 \pm 0.2$ from the echelle spectrum. Likewise, for
SDSSJ~1204$+$0221 we measured $\log \mnhi = 19.7 \pm 0.25$ from the
SDSS data and $\log \mnhi = 19.7 \pm 0.2$ from HIRES.  However, for
SDSSJ1429$-$0145 we measure a total column of $\log \mnhi = 19.55 \pm
0.3$ from the SDSS data; whereas the echelle data gives $\log \mnhi =
18.8 \pm 0.2$, a value $2.5\sigma$ lower than the SDSS value.  The
source of the error is line blending, but we note that this absorber
was located blueward of the quasars Ly$\beta$\ emission line, in a
`crowded' part of the spectrum because of the presence of both the
Ly$\alpha$\ and Ly$\beta$ forests.

Based on visually inspecting 149 background quasar spectra and and the
comparison with the echelle data for three systems, we suspect that
our survey is $\sim 90\%$ complete for $\log \mnhi > 19.3$ for all the
Keck/Gemini/MMT spectra and the SDSS spectra with SNR$_{\rm bg} > 3$
(75\%), or 123 of the 149 spectra we searched.  For SDSS spectra with
lower SNR$_{\rm bg} > 2$, our completeness limit is probably closer to
$\log \mnhi > 20$.  A more careful examination of the completeness and
false positive rate of Lyman limit systems identified in spectra of
the resolution and SNR used here is definitely warranted. Statistical
studies based on our sample which attempt to quantify the abundance of
absorbers near foreground quasars \citep{LLS2}, will suffer from
a `Malmquist' type bias because line-blending biases lower column
densities upward, and the line density of absorbers $dN\slash dz$, is
a steep function of column density limit.  Finally, our completeness
is likely to be higher and the false positive rate lower at $z\sim 2$,
as compared higher redshifts ($z\sim 3$), because the mean flux
decrement of the Ly$\alpha$ forest absorption decreases with
decreasing redshift, and thus line blending causes less confusion.

\subsection{Quasar-Absorber Sample}

We present relevant quantities for the quasar-absorber pairs
discovered in our survey in Table~\ref{table:thick}.  All systems with
$\log \mnhi > 17.2$ are included; however those systems with column
densities $\log \mnhi < 18.3$ are included as LLSs based on the
identification of definite Lyman limit absorption (i.e. $z_{\rm fg} >
3.2$). In the next section (\S~\ref{sec:systemic}), we describe in
detail how we estimated the foreground quasar redshifts and redshift
errors which are listed in Table~\ref{table:thick}. The emission line
which was used to compute this redshift is also noted in the Table.

The quantity $g_{\rm UV}= 1 + F_{\rm QSO}\slash F_{\rm UVB}$ in
Table~\ref{table:thick}, is the maximum enhancement of the quasars
ionizing photon flux over that of the extragalactic ionizing
background\footnote{We compare to the UV background computed by
  F. Haardt \& P. Madau (2006, in preparation)}, at the location of
the background quasar sightline, assuming that the quasar emission is
isotropic.  It is the maximal enhancement because we assumed the
distance is given by the transverse component alone. A discussion of
how $g_{\rm UV}$ was computed is provided in Appendix A.

The distribution of foreground quasar redshifts, transverse separations, and
ionizing fluxes probed by all of our projected pair sightlines
is illustrated by the scatter plot in Figure~\ref{fig:scatter}.  
The filled symbols outlined in black indicate the
sightlines which have an optically thick absorption line system (see
Table~\ref{table:thick}) and open symbols are sightlines without an
absorber. Note that the closest pairs are predominantly at redshift $z
\sim 2$, because this is where quasar pair selection is most efficient
\citep[see][]{qsoselect,binary}.

A list of tentative absorbers near quasars, for which we could not be
sure that $\log \mnhi > 17.2$, is published in Table~\ref{table:poss}
of Appendix B. Higher SNR and higher resolution spectra are
required to make definitive conclusions about these systems; they are
a valuable set of targets for future research. Coordinates
and SDSS five band photometry of the quasar pairs in
Tables~\ref{table:thick} and \ref{table:poss} are provided in
Tables~\ref{table:thick_coords} and \ref{table:poss_coords},
respectively, of Appendix C.  

To summarize, we searched 149 projected quasar pair sightlines with
transverse proper distances $22~\hkpc < R < 1.7~\hMpc$ away from
foreground quasars in the redshift range $1.8 < z < 4.0$. Keck spectra
accounted for 25 of the background quasar spectra, 19 came from
Gemini, 2 from the MMT and the remaining 103 were from the SDSS.  Of
these sightlines, 25 had angular separations below the SDSS fiber
collision limit ($\Delta \theta < 55\arcsec$) with the Keck/Gemini/MMT
accounting for all but three of these. We discovered 27 LLSs within
$\Delta v=1500\kms$ of the foreground quasar redshifts, of which 17
are super-LLSs with $\log \mnhi > 19$.

\section{Estimating Systemic Redshifts}
\label{sec:systemic}

The primary rest-frame ultraviolet quasar emission lines which are
redshifted into the optical for $z\gtrsim 2$ quasars are:
Ly$\alpha$~$\lambda 1216$, \ion{N}{5}~$\lambda 1240$,
\ion{C}{4}~$\lambda 1549$, \ion{C}{3}]~$\lambda 1908$, and
\ion{Mg}{2}~$\lambda 2798$.  The redshifts determined from these lines
can differ by up to $\sim 3000\kms$ from systemic, due to
outflowing/inflowing material in the broad line regions of quasars
\citep{Gaskell82,TF92,vanden01,Richards02}.  A redshift determined
from the narrow ($\sigma \lesssim 200\kms$) forbidden emission lines
[\ion{O}{2}]~$\lambda 3727$ or [\ion{O}{3}]~$\lambda 5007$, are the
best predictors of systemic redshift; but at $z \gtrsim 2$,
measurements of these lines would require spectra covering the near
infrared.

For example, redshifts are most uncertain when estimated from the
\ion{C}{4} emission line: \citet{Richards02} found a median blueshift
of 824$\kms$ from \ion{Mg}{2} with a dispersion about the median of
511$\kms$, but with a tail extending to blueshifts as large as
3000~$\kms$. Furthermore, \ion{Mg}{2} has a median shift of 97$\kms$
and a dispersion of 269$\kms$ about [\ion{O}{3}], which is used to
define the systemic frame \citep{Richards02}.  The implied average
redshift uncertainty between \ion{C}{4} and [\ion{O}{3}] (systemic) is
thus $\gtrsim 600\kms$. In addition, the \ion{C}{4} blueshift is
luminosity dependent \citet{Gaskell82;Richards02}, such that simply
adding a median offset to a \ion{C}{4} line center can bias redshifts
and result in larger errors. The quasar redshifts computed by the SDSS
spectroscopic pipeline are the result of a maximum likelihood
procedure which involves fitting of multiple emission lines
simultaneously \citep[see e.g. ][]{Stoughton02}, a procedure which 
will not result in robust systemic redshift estimates.

In light of these issues, we recompute the systemic redshifts of all
the foreground quasars published in Table~\ref{table:thick}. Note that
the spectra of foreground quasars used to compute these redshifts come
from a variety of instruments (see Table~\ref{table:thick}) and they
have varying SNRs and spectral coverage. Some are from the SDSS, for
others we have high SNR Keck LRIS-R spectra covering
$\lambda=5000-9500$~\AA, and for others we have only Gemini GMOS
spectra with $\sim 1500$~\AA~of coverage centered on the Ly$\alpha$
forest and typically extending to $\sim 1350$~\AA\ in the quasar 
rest-frame.

For our spectra with full wavelength coverage (SDSS and LRIS-R), we
begin by fitting the sum of a Gaussian plus a linear continuum (both
in $\log \lambda$) to the primary emission lines \ion{C}{4},
\ion{C}{3}], and \ion{Mg}{2} if present, where the line-widths and
centers are free parameters.  We also include components for the
weaker lines \ion{He}{2}~$\lambda 1640$ and \ion{Al}{3}$\lambda 1857$,
since they can be significantly blended with \ion{C}{4} and
\ion{C}{3}], respectively, thus contributing to the background and
influencing the placement of the continuum.  Guided by these fits, we
calculate the mode of each line using the relation mode = 3 $\times$
median - 2$\times$ mean, applied to the upper $\approx$ 60\% of the
emission line. This is a more robust estimator than the centroid or
median for slightly skewed profiles in noisy data. Specifically, we
compute the mode of all spectral pixels within $\pm 1.5\sigma$, of the
Gaussian line center which have flux 
\be f_{\rm \lambda} > \frac{0.6
  A_i}{\sqrt{2\pi\sigma_i}} + C_{\rm \lambda} + \sum_{j \ne i}
\frac{A_j}{\sqrt{2\pi\sigma_j}}e^{-\left(\log_{10}\lambda-\log_{10}
    \lambda_j\right)^2\slash 2\sigma_j^2} 
\ee 
where $A_{i}$, $\log_{10}\lambda_i$, and $\sigma_i$, are the amplitude, central
wavelength, and dispersion, of the best fit Gaussian to the $i$th
emission line and $C_{\rm \lambda}$ is the linear continuum. The sum
represents the effective background due to other nearby lines.

Given these line centers, a redshift is computed for each line.
Redshifts are computed using the average velocity shift from systemic,
defined with respect to the [\ion{O}{3}] emission lines, where we use
the shifts measured by \citet{vanden01} and \citet{Richards02}.  A
median SNR $> 5$ over the line is required for it to be used as a
redshift estimator.  At high SNR, the intrinsic velocity shifts and
line asymmetries will dominate over errors in line centering.  In this
case, our error estimates are motivated by the dispersion measurements
of \cite{Richards02}. For lower SNR, line centering errors can become
significant, however we do not explicitly estimate this error
contribution. Instead, we simply flag low SNR redshift determinations
and dilate their errors by `hand'. Redshifts and errors are assigned
to the 27 foreground quasars in Table~\ref{table:thick} according to
the following procedure:

\begin{itemize}
  
\item[--] Use \ion{Mg}{2} if it is present ($z \lesssim 2.2$). The
  redshift error is $\pm 300\kms$ (6 systems).
  
\item[--] If \ion{Mg}{2} is not present or has low SNR, \ion{C}{3}] is
  used. The error is $\pm 500\kms$\footnote{The dispersion of
    \ion{C}{3}] about systemic has not been quantified, but it is
    likely comparable to that of \ion{Mg}{2} (G.  Richards, private
    communication 2005), motivating our choice of $\pm 500\kms$.} (13
  systems)
  
\item[--] If neither \ion{Mg}{2} nor \ion{C}{3}] can be used, the
  redshift is computed from \ion{C}{4}The error is $\pm
  1000\kms$\footnote{A conservative uncertainty of $\pm
    1000\kms$ is assumed for \ion{C}{4} to account for redshift errors
    due to asymmetries, self-absorption, BAL features, and large
  intrinsic velocity shifts.} (3 systems).

\item[--] If neither \ion{Mg}{2}, \ion{C}{3}], or \ion{C}{4} have SNR $<$
  5, all are fit simultaneously for the redshift, using Gaussians plus
  continua (similar to SDSS pipeline procedure). The error is $\pm 1500\kms$
  (1 system).
  
\item[--] If \ion{C}{4} is unavailable because of limited spectral
  coverage (i.e. Gemini GMOS spectra), the redshift is computed from
  cross correlation with a composite quasar template \citep{vanden01}.
  The redshift error is $\pm 1500\kms$ (4 systems).
\end{itemize}

%JXP should this be in an appendix
% The redshift section?

\section{Notes on Individual Absorption Systems}
\label{sec:notes}

The Keck, Gemini, and SDSS spectra of both the foreground and
background quasars from Table~\ref{table:thick} are shown in
Figure~\ref{fig:qsos}.  The right panels show show closeups of the
\ion{H}{1} absorption in the background quasar spectra and our best
fit \lya\ profiles. Below we describe how each system was fit and
highlight interesting details.

\noindent {\rm \bf SDSSJ~0036$+$0839} The SDSS spectrum of the
background quasar reveals strong absorption features of
\ion{C}{2}~$\lambda 1334$, \ion{Si}{4}~$\lambda 1393,1403$,
\ion{C}{4}~$\lambda 1548,1550$, \ion{Si}{2}~$\lambda 1526$ and
possibly \ion{O}{1}~$\lambda 1302$ and \ion{Si}{2}~$\lambda 1304$. A
strong feature is coincident with \ion{Si}{2}~$\lambda 1260$, but this
could be a Ly$\alpha$ forest absorption line. Centered on the redshift
of the metal lines at $z=2.5647\pm 0.00005$, our Voigt profile fit to
the \ion{H}{1} Ly$\alpha$ gives $\log \mnhi = 18.95 \pm 0.35$. A
redshift of $z_{\rm fg}=2.569$ is measured from the foreground quasars
\ion{C}{3}] emission line with an error of $\pm 500\kms$. The
velocity difference of $|\Delta v| = 360\kms$ from the absorber is
within this error estimate.

\noindent {\rm \bf SDSSJ~0127$+$1507} While most of the systems in
Table~\ref{table:thick} contain only two quasars, there are five
quasars within $5'$ of each other in the region surrounding
SDSSJ~0127+1507 at redshifts $2.60$, $2.38$, $2.18$,$2.08$, and $1.81$
(denoted ABCDE). Along these sightlines,
we identify two absorption systems in SDSSJ~0127+1507A ($z_{\rm
  bg}=2.60$; $\Delta\theta = 131.0\arcsec$) and SDSSJ~0127+1507B
($z_{\rm bg}=2.38$; $\Delta\theta = 51.9\arcsec$) with \lya\ profiles
indicating an optically thick absorber near the redshift of the
foreground quasar SDSSJ~0127+1507E ($z_{\rm fg}=1.818$).  No
significant metal-line absorption is observed in either background
quasar spectrum.  Voigt profile fits to the two profiles give $z=1.819
\pm 0.001$ and $\log \mnhi = 18.6 \pm 0.4$ toward SDSSJ0127+1508A and
$z=1.1818 \pm 0.001$ with $\log \mnhi = 18.9 \pm 0.3$ toward
SDSSJ0127+1507B. We further identify additional strong \lya\
absorption features in the spectrum of the background quasar
J0127+1507C $(z_{\rm bg}= 2.18; \Delta\theta=35\arcsec)$, but we
cannot precisely determine their \nhi\ values, and hence whether an
optically thick absorber is present (see Table~\ref{table:poss}). The
fourth background quasar SDSSJ~0127+1507D ($(z_{\rm bg}= 2.18;
\Delta\theta=163.1\arcsec)$ shows no signs of significant \lya\
absorption at the redshift $z_{\rm fg}=1.818$.  Coordinates and
photometry for all five members of this projected group of quasars are
given in Table~\ref{table:0127} of Appendix B.

\noindent {\rm \bf SDSSJ~0225$-$0739} We detect strong
\ion{Si}{2}~$\lambda 1526$ and \ion{Al}{2}~$\lambda 1670$ in the
background quasar spectrum, which identify the redshift of the
absorber at $z=2.4476 \pm 0.0007$, coincident with the foreground
quasar at $z_{\rm fg}=2.440$.  Our fit to the \lya\ profile is complicated by
\ion{O}{6} emission from the quasar but yields the value $\log \mnhi =
19.55 \pm 0.2$~dex. The velocity difference of $|\Delta v| =
690\kms$ from the absorber is in excess of our estimated foreground
quasar redshift error of $\pm 500\kms$. 

\noindent {\rm \bf SDSSJ~0239$-$0106} The SDSS spectrum of the
background quasar reveals strong absorption features of
\ion{Si}{2}~$\lambda 1526$, \ion{Fe}{2}~$\lambda 1608$ and
\ion{Al}{2}~$\lambda 1670$ at a redshift $z=2.3025 \pm 0.0007$. Our
estimate of the foreground quasar redshift is $z_{\rm fg}=2.308$ is
particularly uncertain ($\pm 1500\kms$) because of large BAL
features.  There is an apparently strong \lya\
profile at this redshift in the SDSS spectrum but the SNR$_{\rm
  fg}=1.8$ in this region is too low for a meaningful
fit.  As described in \S~\ref{sec:observe}, we acquired a
low-dispersion LRIS-B spectrum of this pair to improve the fit, and it
is these spectra which are shown in Figure~\ref{fig:qsos}.  The
metal-lines are detected in the LRIS-B spectrum of the background
quasar, and a fit to the \lya\ profile gives $\log \mnhi = 20.45 \pm
0.2$~dex.  This absorber, therefore, satisfies the damped \lya\
threshold ($\log \mnhi > 20.3$).  It was not included in the
\cite{phw05} compilation, however, because of the low SNR of the SDSS
spectrum in this region.

\noindent {\rm \bf SDSSJ~0256$+$0039} The SDSS spectrum of the
background quasar shows a weak \ion{C}{4} absorption system at
$z=3.387\pm 0.001$ and a Lyman limit system with consistent redshift.
There may also be weak absorption from \ion{Si}{2}~$\lambda 1260$ and
\ion{Si}{4}~$\lambda 1393,1402$.  An examination of the \lya\ profile
at this redshift shows a significant feature, centered at
$z_{Ly\alpha} = 3.387$.  A fit indicates an \ion{H}{1} column density
of $\log \mnhi = 19.25\pm 0.25$ is allowed by the data, but this value
is very uncertain because of the low SNR and lack of evidence for
significant damping wings. The presence of an LLS sets a lower limit
of to the \ion{H}{1} column density $\log \mnhi > 17.2$, unless the
limit corresponds to another absorber nearby in redshift. The redshift
of the foreground quasar $z_{\rm fg} =3.387$ is consistent with the
redshift of the absorber.

\noindent {\rm \bf SDSSJ~0303$-$0023} There is an absorption system
showing very strong \ion{Si}{4} and \ion{C}{4} lines at $z=2.7243 \pm
0.0007$ in the spectrum of the background quasar, which is $500\kms$
away from the foreground quasar at $z=2.718$ ($\pm 1000\kms$).  
Weak absorption consistent with \ion{Al}{2}~$\lambda 1670$ is observed
at this redshift, but no significant detection of \ion{Fe}{2}~$\lambda
1608$ ($W_\lambda < 100$m\AA).  A strong \lya\ profile is
consistent with the \ion{C}{4} lines, although with nominal line
centroid blueward of the metal-line absorption ($z_{Ly\alpha} =
2.723$).  If we constrain the fit by the \ion{C}{4} redshift, the red
wing of \lya\ provides the only constraint on the fit and we find
$\log \mnhi = 18.95 \pm 0.20$.  If we allow the redshift to be a free
parameter, the best fit solution is $\approx 0.1$~dex larger.  As
mentioned in the previous section, we have independent echelle data
for this background quasar and a fit to the Ly$\alpha$ profile gave
$\log \mnhi = 18.9 \pm 0.20$, consistent with our measurement from the
lower resolution SDSS spectrum.

\noindent {\rm \bf SDSSJ0338-0005} This damped \lya\ system shows a
series of metal-line transitions including the \ion{Mg}{2} doublet.
The redshift of the gas is $z=2.2290 \pm 0.0007$ and the foreground
quasar is at $z=2.239$, but with large error, $\pm 1500\kms$.  The
absorber is a member of the SDSS-DR3 DLA catalog and \cite{phw05}
report an \ion{H}{1} column density of $\log \mnhi = 20.9 \pm
0.15$~dex.

\noindent {\rm \bf SDSSJ0800+3542} This absorber is notable for
showing strong \ion{N}{5}~$\lambda 1238,1242$ absorption, which is
rarely ever observed in LLSs. There is also significant absorption due
to \ion{Si}{2}~$\lambda 1260$ and \ion{C}{2}~$\lambda 1334$, as well
as the \ion{Fe}{2}~$\lambda 1608$ and \ion{Al}{2}~$\lambda 1670$
transitions.  The metal line profiles are centered at $z=1.9828 \pm
0.0007$ and are accompanied by a \lya\ profile with $\log \mnhi = 19.0
\pm 0.15$, according to our fit. The absorber redshift is in good
agreement with that of the foreground quasar, $z_{\rm fg}=1.983$.

\noindent {\rm \bf SDSSJ0814+3250} Similar to SDSSJ0800+3524, this
absorption system shows remarkably strong \ion{N}{5} absorption in the
spectrum of the background quasar, as well as several other metal
transitions (\ion{Si}{2}, \ion{C}{2}, \ion{Si}{4}).  The redshift of
these metal-lines is $z=2.1792 \pm 0.0007$.  At the corresponding
position of \ion{H}{1} \lya\ there is significant absorption, although
oddly, the largest optical depth occurs 1.2\AA\ away from the line centroid
predicted by the metal lines.  We have fit the \lya\ profile with two
\ion{H}{1} components, one centered at $z=2.1792$ and an additional
component at $z=2.1778$.  Our best solution has $\log \mnhi = 18.50
\pm 0.2$ for each component.  There is significant degeneracy between
the two components, especially if one does not demand a component at
$z=2.1792$.  However, the total column density is well constrained
and has a value of $\log \mnhi = 18.80 \pm 0.2$~dex. The foreground
quasar is at $z_{\rm fg}=2.182 \pm 1500\kms$, which is $|\Delta v|=280\kms$
away from the absorber (at $z=2.1792$), but well within the
uncertainty of the foreground quasar redshift.

\noindent {\rm \bf SDSSJ0833+0813} This system shows likely
\ion{Si}{2}~$\lambda 1526$ and possible \ion{C}{4} absorption at
$z=2.505 \pm 0.001$ and a strong corresponding \lya\ profile.  If we
adopt the metal line redshift, our best fit is a multi-component model
with $\log \mnhi = 19.45 \pm 0.3$.  But, there is a small chance that
this profile is the blend of several $\log \mnhi \approx 18$ absorbers
and that the total $\log \mnhi < 19$. The metal line redshifts are
$980\kms$ from the foreground quasar at $z_{\rm fg}=2.516$, but this velocity
difference is comparable to its redshift error $\pm 1000\kms$.

\noindent {\rm \bf SDSSJ0852+2637} The foreground quasar redshift
($z=3.203$) is sufficiently high that the SDSS spectrum covers the
Lyman limit region.  We do detect a break due to the Lyman limit at
$912$~\AA~ and its corresponding \lya\ profile at $z=3.211 \pm 0.001$,
in the background quasar spectrum.  There are no detected metal-line
transitions, which is not uncommon for Lyman limit systems at the
resolution of the SDSS spectra.  We estimate the column density to be
$\log \mnhi = 19.25 \pm 0.4$. There is a $|\Delta v|=550\kms$ between
the absorber and the foreground quasar redshift ($z_{\rm fg}=3.203$),
but this is within our estimated $\pm 1500\kms$ margin of error on the
latter.

\noindent {\rm \bf SDSSJ0902+2841} Similar to SDSSJ0852+2637, the
background quasar spectrum covers the Lyman limit at the redshift of
the foreground quasar. The SDSS spectrum of the background quasar
reveals a Lyman limit system at $z=3.342 \pm 0.005$, corresponding to
an offset by $|\Delta v|=1200\kms$ from the foreground quasar redshift
$z_{\rm fg}=3.325$. This offset is more than twice the estimated
redshift error ($\pm 500\kms$) from the \ion{C}{3}] emission line of
the foreground quasar. No significant metal-line absorption is
identified.  There is a \lya\ absorption profile with centroid
consistent with the Lyman limit redshift. There appears to be residual
flux in the line profile, but the data has insufficient SNR in this
region to say definitively.  While a column density $\log \mnhi > 19$
is permitted by the data, we suspect the actual value is less than
this.  Thus we set a conservative lower limit of $\log \mnhi > 17.2$ based
on the presence of a limit.

\noindent {\rm \bf SDSSJ1134+3409} This quasar exhibits a very strong
\ion{C}{4} system at $z=2.2879 \pm 0.005$ and possible detections of
\ion{Si}{2}~$\lambda 1526$ and \ion{Al}{2}~$\lambda 1670$ in the
background quasar spectrum.  There is a corresponding \lya\ profile
with rest equivalent width $W_{\lambda} = 4.9 \pm 0.2$\AA.  Our fit to this
profile gives $\log \mnhi = 19.5 \pm 0.3$ if we tie the redshift to
the \ion{C}{4} absorption.  We derive a $+0.3$~dex larger value if we
allow the redshift to vary as a free parameter to $z=2.2865$. Either
redshift is in good agreement ($|\Delta v| < 400\kms|$) with the
foreground quasar at $z_{\rm fg}=2.291$.

\noindent {\rm \bf SDSSJ1152+4517} The background quasar spectrum has
a strong \lya\ absorption feature at $z=2.3158 \pm 0.0010$ and a
\ion{C}{4} absorption system at $z=2.31323 \pm 0.0007$, significantly
offset from the \lya\ feature.  Our profile fit to \lya\ gives $\log
\mnhi = 19.1 \pm 0.3$ assuming the redshift $z=2.3158$.  
The absorber is offset by $|\Delta v|= 370\kms$ from the 
foreground quasar at $z_{\rm fg}=2.312\pm 500\kms$, but this is 
within the redshift error. 

\noindent {\rm \bf SDSSJ1204+0221} For this quasar, we have both GMOS
and SDSS spectra.  Figure~\ref{fig:qsos} shows the GMOS spectra
(left) of both pairs, but we used the SDSS data to fit the \lya\
profile (right), because this spectrum gave coverage of more of the metal
lines. A series of strong metal transitions were identified at
$z=2.4402 \pm 0.0003$ and our fit to the \lya\ profile gives $\log
\mnhi = 19.7 \pm 0.25$. A fit to the column density using an echelle
spectrum (HIRES) of this object gives the same value.  The absorber
redshift is offset by $|\Delta v|=370\kms$ from that of the
foreground quasar $z_{\rm fg} = 2.436\pm 1500\kms$, but the error
on this determination is large. 

\noindent {\rm \bf SDSSJ1213+1207} A strong \lya\ absorption feature
is identified at $z=3.4110 \pm 0.0007$ in the background quasar
spectrum, which is in good agreement with the foreground quasar
redshift $z_{\rm fg}=3.411$. A likely coincident \ion{C}{4} absorption feature
is also identified.  The SDSS data provide coverage of the Lyman
limit at this redshift, and the break is detected, although the SNR of
the data in this region is poor.  Our Voigt profile fit to the \lya\
feature gives $\log \mnhi = 19.25 \pm 0.3$.

\noindent {\rm \bf SDSSJ1306+6158} 
The LRIS-B spectrum of the background quasar shows a damped \lya\
system at $z=2.1084 \pm 0.0007$, consistent ($|\Delta v|=200\kms$)
with the redshift of the foreground quasar, $z_{\rm fg}=2.111 \pm 300\kms$. 
A series of metal-line transitions outside the \lya\ forest are
identified, which constrain the redshift well.  Our single component
fit to the \lya\ profile gives $\log \mnhi=20.30 \pm 0.15$.

\noindent {\rm \bf SDSSJ1312+0002} 
The low SNR spectrum of the background quasar shows the following
metal-line transitions at $z=2.6688 \pm 0.0007$: \ion{O}{1}~$\lambda
1302$, \ion{Si}{2}~$\lambda 1304$, \ion{C}{2}~$\lambda 1334$,
\ion{Si}{4}~$\lambda 1393,1402$, \ion{Si}{2}~$\lambda 1526$,
\ion{C}{4}~$\lambda 1548,1550$, \ion{Al}{2}~$\lambda 1670$, and likely
\ion{Fe}{2}~$\lambda 2382$. A strong \lya\ absorption profile is
consistent with this redshift.  Although the SNR is low ($\approx 2$
per pixel), the \ion{H}{1} column density is consistent with the DLA
threshold $\log \mnhi = 20.3 \pm 0.3$.  This profile was fit with
a similar value by \cite{phw05} but not included in their sample
because of the low SNR. The absorber redshift is offset by $|\Delta
v|=200\kms$ from that of the foreground quasar $z_{\rm fg}= 2.671$, but within 
its error ($\pm 500\kms$). 

\noindent {\rm \bf SDSSJ1426+5002} We identify an absorption system in
this SDSS background quasar spectrum with very strong metal-lines
including detections of the weak \ion{Si}{2}~$\lambda 1808$ and
\ion{Zn}{2}~$\lambda 2026$ transitions at $z=2.2247 \pm 0.0007$.
\cite{phw05} reported an \nhi\ column density for this absorber below
the DLA threshold ($\log \mnhi > 20.3$) and we similarly find $\log
\mnhi = 20.00 \pm 0.15$ here.  This low column density is rather
remarkable relative to the the strong metal-line absorption. The
redshift of the absorber is offset by $|\Delta v|=1330\kms$ from the
foreground quasar redshift $z_{\rm fg}=2.239$, which is more than twice our
estimated redshift error of $\pm 500$ from the \ion{C}{3}] emission
line of the foreground quasar.

\noindent {\rm \bf SDSSJ1427-0121} Our high SNR GMOS spectrum of the
background quasar reveals several metal-line absorption features
redward of \lya. Near the redshift of the foreground quasar, we
identify \ion{Si}{2}~$\lambda 1260$, \ion{C}{2}~$\lambda 1334$, and
\ion{Si}{4} absorption.  These lines show multiple components spanning
over $1000\kms$.  We identify similar structure in the \lya\ profile,
although, the bluest component of the metal lines does not correspond
to significant \ion{H}{1} absorption, similar to the case for
SDSSJ~0814$+$3250.  We have fit the \lya\ complex with 3 components at
$z=2.2762, 2.2788$, and $2.2828$ with $\log \mnhi = 14, 18.85$, and
$18.80$, respectively.  The low column density $\log \mnhi = 14$
component results from the lack of significant optical depth in the bluest
component.  Because the features are well separated, there is minimal
degeneracy in the fit, although the uncertainty is still $\approx
0.25$~dex for each component. The two features at $z=2.2788$ and
$z=2.2828$ differ in redshift by $366\kms$, greater than the $300\kms$
maximum value suggested by \cite{Omeara06} in combining multiple
component absorption systems to one.  Thus, we take the component at
$z=2.2788$ with $\log \mnhi = 18.85$ as the quasar companion, since it
is closest to the foreground quasar systemic redshift $z_{\rm
  fg}=2.278$, differing by only $|\Delta v|=50\kms$.

\noindent {\rm \bf SDSSJ1429-0145} The SDSS spectrum of the background
quasar shows an absorption system with large EW metal-line features of
\ion{Si}{2}~$\lambda 1526$, \ion{C}{4}~$\lambda 1548,1550$,
\ion{Fe}{2}~$\lambda 1608$ and \ion{Al}{2}~$\lambda 1670$ at a
redshift $z=2.6235 \pm 0.0007$, in relatively good agreement with the
foreground quasar redshift at $z_{\rm fg}=2.628$.  We identify a strong \lya\
profile with rest equivalent width $W_{\lambda} = 3.8$\AA; however,
the centroid of this profile lies blueward of the peak in the optical
depth of the metal-line transitions. There is also 
evidence for weak absorption in the metal-line profiles redward of the 
dominant absorption.  Fitting the profile with two components, we derive 
$\log \mnhi = 19.3 \pm 0.3$ at $z=2.6235$ and $\log \mnhi = 19.2 \pm 0.3$ 
at $z=2.6267$.  Similar to the absorbers associated with SDSSJ0814+3250, 
the two values for the two components are degenerate, but their total 
\ion{H}{1} column density, $\mnhi = 19.55 \pm 0.3$ is well constrained.  
As mentioned in \S~\ref{sec:sample}, we have also examined echelle spectra 
of this background quasar, and a significantly lower column density was 
measured, $\log \mnhi = 18.8 \pm 0.2$, with the difference being caused by 
unresolved line-blending in the SDSS spectrum. It is this lower value
which is listed in Table~\ref{table:thick}.

\noindent {\rm \bf SDSSJ1430-0120} The background quasar exhibits a
pair of damped \lya\ systems from the \cite{phw05} compilation at
$z=3.0238$ and $z=3.115$ with $\log \mnhi = 21.10 \pm 0.2$ and $\log
\mnhi = 20.50 \pm 0.2$, respectively. The DLA at $z=3.115$ is $|\Delta
v|=960\kms$ away from the foreground quasar $z_{\rm fg}=3.102$;
however the foreground quasar redshift has a large associated error
$\pm 1500\kms$. Also, there is no apparent metal-line absorption for
the DLA coincident with the quasar,  and the redshift is not tightly
constrained by the \lya\ profile fit: we estimate an uncertainty of
$\delta z = \pm 0.002$ ($\pm 150\kms$).

\noindent {\rm \bf SDSSJ1545+5112} We identify a strong \lya\ profile
at $z=2.243 \pm 0.001$ in the background quasar spectrum of this pair,
in good agreement with the foreground quasar redshift $z_{\rm
  fg}=2.240$.  Weak absorption features consistent with
\ion{O}{1}~$\lambda 1302$, \ion{C}{2}~$\lambda 1334$ and possibly
\ion{C}{4} are present; but we note the absence of any obvious
\ion{Si}{2}~$\lambda 1260$, calling these identifications into
question, because the oscillator strengths of these transitions are
comparable.  A Voigt profile fit to the \lya\ gives $\log \mnhi =
19.45 \pm 0.3$.

\noindent {\rm \bf SDSSJ1621+3508} Our LRIS-R spectrum of this bright
background quasar exhibits a \ion{Mg}{2} system at $z=1.9309 \pm
0.0007$, in good agreement with the foreground quasar redshift
$z_{\rm fg}=1.931$. It also shows metal absorption from \ion{C}{2}, 
\ion{C}{4}, and \ion{Si}{2}. A corresponding strong \ion{H}{1} \lya\ 
absorption profile is identified in our LRIS-B spectra and the best fit column 
density is $\log \mnhi = 18.7 \pm 0.2$.

\noindent {\rm \bf SDSSJ1635+3013} The SDSS spectrum of this
background quasar shows metal-line absorption from
\ion{Si}{2}~$\lambda 1526$ and \ion{Al}{2}~$\lambda 1670$ at a
$z=2.5025 \pm 0.001$. The foreground quasar at $z_{\rm fg}=2.493\pm
500\kms$ is offset from the absorber by $|\Delta v|=820\kms$ which is
significant compared to the foreground quasar redshift error. The
\lya\ profile at the redshift of the metals lines appears
significantly blended with other coincident \lya\ absorption
transitions.  The \ion{H}{1} value is thus poorly constrained. We
obtain a fit of $\log \mnhi = 19.35 \pm 0.3$ but report a conservative
lower limit of $\log \mnhi > 19$.

\noindent {\rm \bf SDSSJ2347+1501} This triple system consists of a
background quasar at SDSSJ2347$+$1501A at $z_{\rm bg}=2.29$, and two
foreground quasars: SDSSJ2347$+$1501B ($z_{\rm fg}=2.157$; $\Delta
\theta = 47.3$) and SDSSJ2347$+$1501C ($z_{\rm fg}=2.171$; $\Delta
\theta = 223.0$).  We identify \ion{S}{4} absorption at $z=2.176 \pm
0.001$, near the redshifts of the foreground quasars.  There is a
corresponding strong \lya\ absorption feature, but it appears to be a
blend of two components.  Our best fit profile, which is not very well
constrained, gives $\log \mnhi = 18.55$ at $z=2.174$ and $\log \mnhi =
18.1$ at $z=2.177$.  These values are highly correlated but the
absence of significant damping wings limits the total \ion{H}{1}
column density to be $\log \mnhi< 19$, and we place a lower limit on
the total column density of $\log \mnhi< 18.3$. The absorber is offset
by $|\Delta v|= 1770\kms$ from SDSSJ2347$+$1501B at $z_{\rm fg}=2.157\pm
1000\kms$, and $|\Delta v|= 380\kms$ from SDSSJ2347$+$1501C at
$z_{\rm fg}=2.171\pm 500\kms$. The large offset from SDSSJ2347$+$1501B exceeds
the limit of our search ($\pm 1500\kms$). If part of this offset is
due to Hubble flow, the quasar would be separated from the absorber by
$\gtrsim > 5 \hMpc$ in the radial direction, much larger than the
$R=282~\hkpc$ transverse separation.

\section{Cosmological Applications of Optically Thick Absorbers Near Quasars}
\label{sec:apps}

The cosmological applications of optically thick absorption line
systems near quasars are many. They include measurement of the
clustering of absorbers around quasars, determining the spatial
distribution of neutral gas in LLSs and DLAs, fluorescent Ly$\alpha$
emission from LLSs, constraints on the emission geometry and lifetimes
of quasars, and studying the enrichment and velocity fields in the
environs of luminous quasars. We discuss each of these in turn.

\subsection{Clustering of Absorbers around Quasars}

A conspicuous feature of Figure~\ref{fig:scatter}
is the high covering factor of absorbers for small separations.  Six 
out of eight projected sightlines with $R < 150~\hkpc$ have a
LLS coincident with the foreground quasar, of which four are
super-LLSs ($\log \mnhi > 19$).  We previously estimated that the
probability of a random quasar-super-LLS coincidence is $\sim 2\%$
(see beginning of \S~\ref{sec:sample}); thus  the average number expected in
eight sightlines is $\sim 0.16$. This factor of $\sim 25$
increase in the number of small scale quasar-absorber coincidences
provides significant evidence that these absorbers are strongly
clustered around quasars. It is particularly interesting that in the
projected pairs which show absorption, an LLS is seen in the
transverse direction only -- none of the foreground quasars
show an \emph{intrinsic} or \emph{proximate} optically thick 
absorber along the line of sight. 

Although they are rare, proximate absorbers with $z_{\rm abs}\sim
z_{\rm em}$ are occasionally observed in single lines of sight. In a
recent study, \citet{REB05}, found that the abundance of DLAs is
enhanced by a factor of $\sim 2$ near quasars, which they attributed
to the clustering of the DLA counterpart galaxies around quasars.  The
high covering factor for $R < 150~\hkpc$ in
Figure~\ref{fig:scatter} suggests a significantly larger enhancement in
the transverse direction.  We measure the \emph{transverse}
quasar-absorber correlation function and compare it to the abundance
of proximate absorbers in the second paper of this series
\citep{LLS2}.

An asymmetry between the strength of clustering in the radial and
transverse direction could be the hallmark of anisotropy or
obscuration of quasar emission.  If the emission were highly
anisotropic, optically thick absorbers along the line-of-sight might be
evaporated by the ionizing flux; whereas, transverse absorbers could
lie in shadowed regions, and hence survive.  This
speculation gains some credibility in light of the recent null
detections of the transverse proximity effect in the Ly$\alpha$
forests of projected quasar pairs \citep[][but see Jakobsen \etal
2003]{Crotts89,DB91,FS95,LS01,Schirber04,Croft04}.  Although these
studies are each based only on a handful of projected pairs, they all
come to similar conclusions: the amount of (optically thin) Ly$\alpha$
forest absorption, in the background quasar sightline near the redshift
of the foreground quasar, is \emph{larger} than average rather than smaller --
the opposite of what is expected from the transverse proximity
effect. Our Keck, Gemini, MMT, and SDSS spectra will be used to study
the transverse proximity effect in an upcoming paper
\citep{transverse}.  

Finally, if obscuration of the quasars emission is indeed the
explanation for the anisotropic clustering pattern suggested by
Figure~\ref{fig:scatter}, we would naively expect the transverse
covering factor to be approximately equal to the average fraction of
the solid angle obscured. Studies of Type II quasars and the X-ray
background suggest that quasars with luminosities comparable to our
foreground quasar sample ($M_{\rm B} < -23$) have $\sim 30\%$ of the
solid angle obscured \citep{Ueda03,Barger05,TU05}, although these
estimates are highly uncertain. This seems to be at odds with the high
covering factor (6/8) for having an absorber with $\log \mnhi > 17.2$) that
we observe on the smallest scales ($R <  150~\hkpc$), although
our statistics are clearly very poor.

\subsection{Fluorescent Ly$\alpha$ Emission: Shedding Light On Lyman Limit Systems}

Are there LLSs or DLAs that can self-shield the intense ionizing flux
of a nearby quasar? How do we know if these systems are being
illuminated? -- By observing the fluorescent glow of their
recombination radiation. Optically thick \ion{H}{1} clouds in
ionization equilibrium with a photoionizing background, re-emit $\sim
60\%$ of the ionizing radiation they absorb as fluorescent Ly$\alpha$
recombination line photons
\citep{GW96,ZM02,Cantalupo05,Adel05,Juna06}.  All attempts to detect
this fluorescent radiation from LLSs illuminated by the ambient UV
background have been unsuccessful \citep{Bunker98,Mar99,Becker05},
despite $\gtrsim 60$ hour integrations on 10m telescopes
\citep{Becker05}. However, if a nearby quasar illuminates an absorber,
as could be the case for the quasar-absorber pairs in
Table~\ref{table:thick} the fluorescent surface brightness is enhanced
by a factor of $g_{\rm UV}$, making it much easier to detect.  Indeed,
\citet{Adel05} reported a possible detection of fluorescence from a
quasar-DLA pair at $z=2.84$, which was serendipitously discovered in
their LBG survey\citep{Steidel99,Adel03}. The ionizing flux is
enhanced by a factor of $g_{\rm UV}\simeq 4850$, owing to their
extremely luminous foreground quasar $r\simeq 16$ \citep[but also
see][who failed to detect fluorescence around a bright $B=17.6$
quasar]{FB04}.

In Table~\ref{table:thick} we publish seven new quasar-absorber pairs
with enhancements $g_{\rm UV} > 100$, with largest being $\sim 7900$
for SDSSJ1427$-$0121. This corresponds to fluorescent surface
brightnesses of $\mu_{{\rm Ly}\alpha}=19.5-24.3~{\rm
  mag~arcsec}^{-2}$, if the foreground quasars emit isotropically.
This potential 5-10 magnitude increase in the expected surface
brightness would constitute a breakthrough for the study of
fluorescence from optically thick absorbers.  It is intriguing that
two of the three systems which have enhancements of $g_{\rm UV}>1000$
(SDSSJ0814$+$3250 and SDSSJ1427$-$0121) for which we have high SNR
ratio moderate resolution spectra, both have \ion{H}{1} \lya\
absorption misaligned with the regions of highest metal line optical
depth. These peculiar absorption profiles are suggestive of emission
in the LLS trough (see Figure~\ref{fig:qsos} and
\S~\ref{sec:notes}). This suggestion is particularly intriguing when
one considers that \emph{proximate} DLAs observed along the line of
sight to single quasars, seem to preferentially exhibit \lya\ emission
superimposed on the \lya\ absorption trough \citep{Moller98,Ellison02}.
We present a detailed discussion of fluorescent emission from our
quasar-LLS pairs in the third paper of this series \citep{LLS3}. 

Finally, because the fluorescent emission from a LLS comes from the a
thin ionized `skin' where $\tau_{\rm LL}\sim 1$ \citep{GW96}, the size
of this region can be measured if fluorescent emission is
detected. Indeed, \citet{Adel05} constrained the emitting region from
their fluorescing DLA to be $\gtrsim 1.5\arcsec$, or a proper size of
$r_{\rm DLA} = 8.5~\hkpc$.  A statistical study of fluorescent
emission from LLSs illuminated by quasars, can measure the distribution
of sizes for optically thick absorbers, as well as the relationship
between size and observed \ion{H}{1} column. The fluorescent flux also
measures the ionizing flux at the self-shielding boundary, which, when
combined with the size, constrains the density and distribution of gas
in LLSs and DLAs, providing physical insights into the morphologies of
high redshift galaxies which are difficult to derive from the
resonance line observations of absorption line spectroscopy.  

\subsection{Constraining Quasar Lifetimes}

In addition to serving as a valuable laboratory for studying
fluorescent emission and the size and nature of LLSs, the detection of
fluorescence from an LLS illuminated by a quasar can place interesting
constraints on the lifetimes of \emph{individual} quasars.  A LLS near
a quasar acts like a mirror, `reflecting' $\sim 60\%$ of the ionizing
photons absorbed from the quasar towards us. However, this reflection
can only be detected provided that ionizing flux of the quasar has not
changed significantly for a time longer than the 
light crossing time to the LLS. The light crossing time corresponding to
$60\arcsec$ ($1.2~\hMpc$) from a $z=2.5$ quasar is $1.1\times 10^6$~yr.

Currently, the lower limit on the intermittency of quasar emission
comes from observations of the proximity effect \citep{BDO88,Scott00},
in the Ly$\alpha$ forests near quasars.  The presence of a proximity
effect implies that the IGM has had time to reach ionization
equilibrium with the quasars increased ionizing flux, giving the lower
limit $t_{\rm burst} \gtrsim 10^4$~yr \citep{Martini04}.  However,
owing to the large amplitude of intrinsic Ly$\alpha$ forest
fluctuations, many quasars must be averaged over to detect the
proximity effect, hence the intermittency of individual quasars cannot
be constrained with this technique \citep{Scott00}.  A positive
detection of fluorescence allows one to place much stronger
constraints on quasar intermittency and lifetimes, with the added
advantage that the lifetimes of individual quasars can be constrained.

\subsection{Probing the AGN Environment with High-Resolution Spectroscopy}

If the background quasar is sufficiently bright ($R \lesssim 19$), 
then one can obtain a high-resolution spectrum on a 10m class telescope.
These data provide more accurate measurements of \nhi\ for absorbers
with $\log \mnhi < 20$.  More importantly, these data yield accurate
ionic column density measurements critical to the analysis 
of physical conditions (e.g.\ ionization state, gas density, metallicity).
It is noteworthy that several of the absorbers exhibit the \ion{N}{5}
doublet in the low-dispersion data.  This ion is rarely observed in
quasar absorption line systems and suggests an environment with
a significantly enhanced UV radiation field.
The spectra also reveal the velocity field of gas in the vicinity
of bright AGN where feedback processes (e.g.\ jets, merger activity)
may be important.  In the case of SDSS1204 we measure a velocity 
width $\Delta v \approx 600 \kms$ for the Si$^+$, C$^+$, and C$^{+3}$ ions.
We will present and analyze these observations  in Paper~IV of the series
\citep{LLS4}.

\section{Summary}
\label{sec:conc}

We searched 149 moderate resolution background quasar spectra, from
Gemini, Keck, MMT, and the SDSS for optically thick absorbers in the
vicinity of $1.8 < z < 4.0$ foreground quasars.  We presented 27 new
quasar-LLS pairs with column densities $17.2 < \log \mnhi < 20.9$ and
transverse distances of $22~\hkpc < R < 1.7~\hMpc$ from the quasars,
corresponding to factors of $g_{\rm UV}=5-8000$ larger ionizing fluxes
than the ambient UV background, if the quasars emit isotropically.
The observed probability of intercepting an absorber is very high for
small separations: six out of eight projected sightlines with
transverse separations $R < 150~\hkpc$ have an absorber
coincident with the foreground quasar, of which four have $\mnhi >
10^{19}{\rm cm}^{-2}$. The covering factor of $\mnhi > 10^{19}{\rm
  cm}^{-2}$ absorbers is thus $\sim 50\%$ (4/8) on these small scales,
whereas $\lesssim 2\%$ would have been expected at random.

Techniques for estimating quasar systemic redshifts from noisy rest
frame UV spectra were presented and used to compute the systemic
redshifts of the foreground quasars in our sample. An important area
for future work is to obtain near-IR ($H$-band) spectra of the $z
\gtrsim 2$ foreground quasars to determine more accurate
redshifts from the narrow ($\sigma \lesssim 200\kms$) forbidden
emission lines [\ion{O}{2}]~$\lambda 3727$ or [\ion{O}{3}]~$\lambda
5007$, which are the best predictors of systemic redshift.

Much larger samples of optically thick absorbers near quasars are
easily within reach, and can be compiled in a modest amount of
observing on a 6-10m class telescopes. \citet{Richards04} constructed
a photometric catalog of $z < 3$ quasars from the SDSS imaging data
which has high completeness and efficiency, and photometric redshifts
accurate to $|\Delta z| \lesssim 0.3$. Indeed, many of the projected
quasar pairs shown in Figure~\ref{fig:scatter} were selected from this
catalog \citep[see][]{binary}.  The number density of $z_{\rm phot} >
1.8$ quasars in the photometric catalog is $10.5$~{\rm deg}$^2$, so
that the total number of projected sightlines expected with $\Delta
\theta < 25\arcsec$ ($R \lesssim 150~\hkpc$) for the current $\sim
8000~{\rm deg}^2$ of SDSS imaging is 140, of which $\sim 70$ would
have super-LLSs if we extrapolate from the $\sim50\%$ covering factor
of such systems observed in this study.  At moderate resolution,
sufficient SNR can be obtained in a 30 minute exposure at Keck, and
about 2.5 times longer at the MMT. Thus, a 7 night program on a 10m
class telescope, or 18 nights at 6m class,  targeting the closest
projected pairs would result in a factor of about 20 more quasar-absorber
pairs, with $R <150~\hkpc$ and $\log \mnhi \gtrsim 19$, than the
four published here.  If a multi-slit setup were used, wider
separation projected pairs come for `free', since these quasars can be
targeted and observed simultaneously on the same masks, as was done
for our Keck observations.  Finally, we note that a similar survey of
projected quasar pairs using the techniques presented here can be used
to search for \ion{Mg}{2}, \ion{C}{4} or other absorption lines sytem near
foreground quasars \citep{Bowen06,Prochter06}. 
 
A sample of $\sim 100$ LLSs near quasars would be of tremendous
cosmological interest, providing new opportunities to characterize the
environments, emission geometry, and radiative histories of quasars,
as well as new laboratories for studying fluorescent emission from optically 
thick absorbers and the physical nature of LLSs and DLAs. 

\begin{figure*}
  \centerline{\epsfig{file=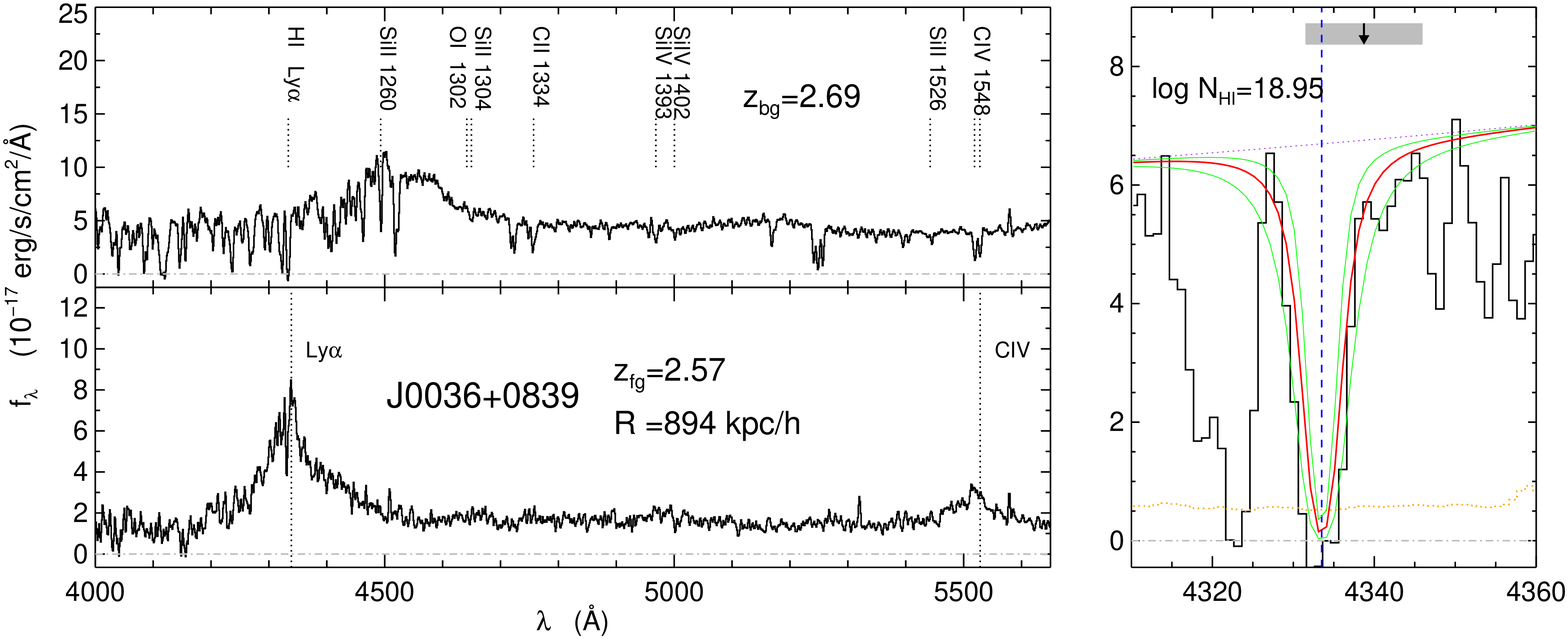,bb=0 0 1225 500,width=\textwidth}}
  \centerline{\epsfig{file=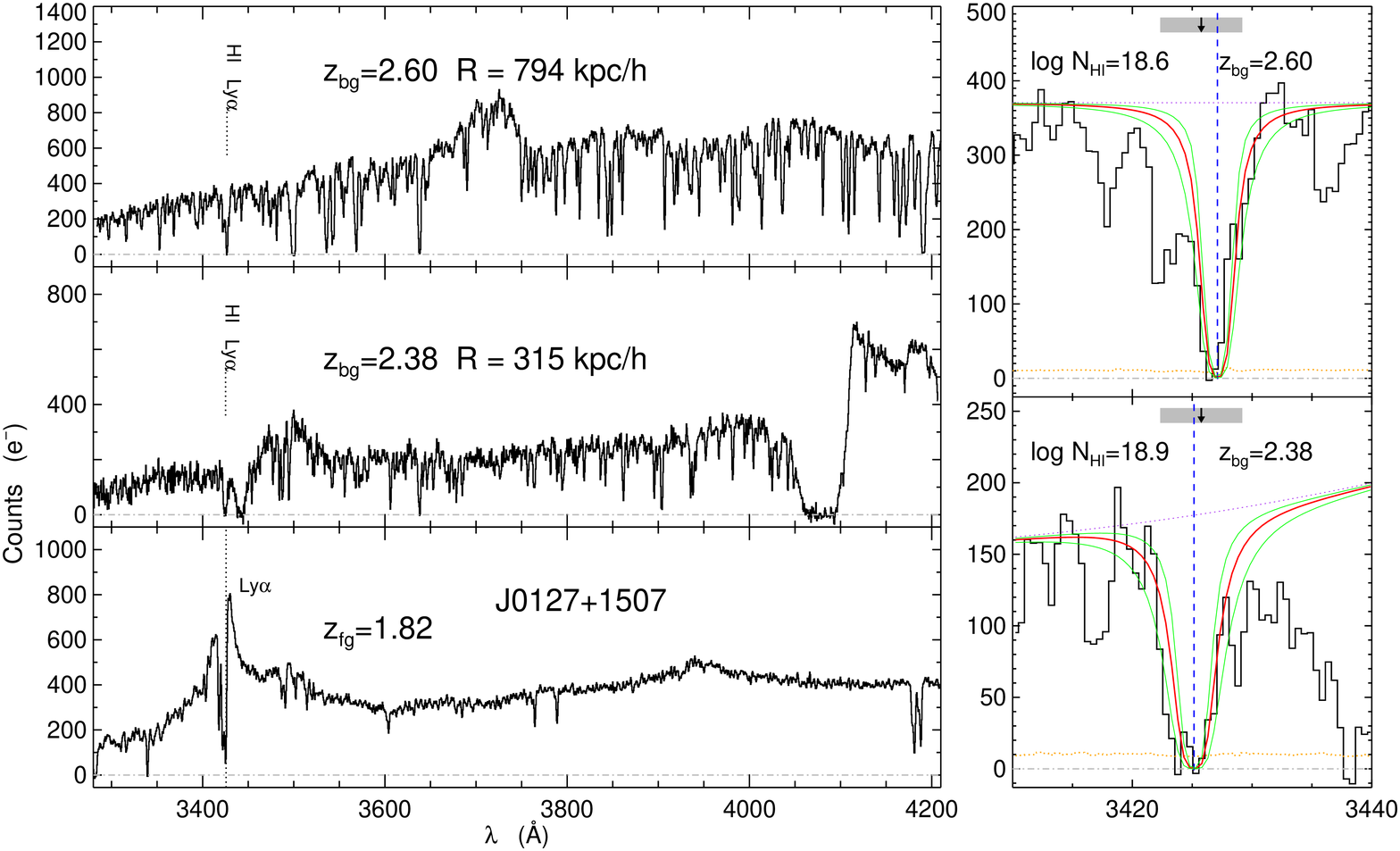,bb=0 0 1225 800,width=\textwidth}}
  \caption{Spectra of projected quasar pairs with optically thick
    absorbers coincident with the foreground quasar redshift. The left
    panels show the background (upper) and foreground (lower)
    quasars. Some of the metal line transitions discussed in
    \S~\ref{sec:notes} and indicated by labels and dotted lines on the
    background quasar spectrum. Quasar emission lines are indicated on
    the foreground quasar spectrum. The panels on the right show a
    closeup of the \ion{H}{1} Ly$\alpha$ absorption in the background
    quasar spectrum. The (red) central curves indicate the best Voigt
    profile fit, (green) upper and lower curves show the fits
    corresponding to our $\pm 1\sigma$ column density errors, and
    (blue) dashed lines indicate the redshifts of absorption
    components. The upper (purple) dotted line indicates the continuum
    level used for the fit and the lower (orange) dotted line
    indicates the noise level. The arrow indicates the foreground
    quasar redshift and the (gray) rectangle indicates our estimate of
    the redshift error.  Three quasars are shown on the left for the
    triple systems SDSSJ0127$+$1507 and SDSSJ2347$+$1501. Cross hatched
    regions in the spectra of SDSSJ1204$+$0221 and SDSSJ2347$+$1501
    indicate gaps in the spectral coverage of the Gemini GMOS
    spectrograph. \label{fig:qsos}}
\end{figure*}
\addtocounter{figure}{-1}
\begin{figure*}
  \centerline{\epsfig{file=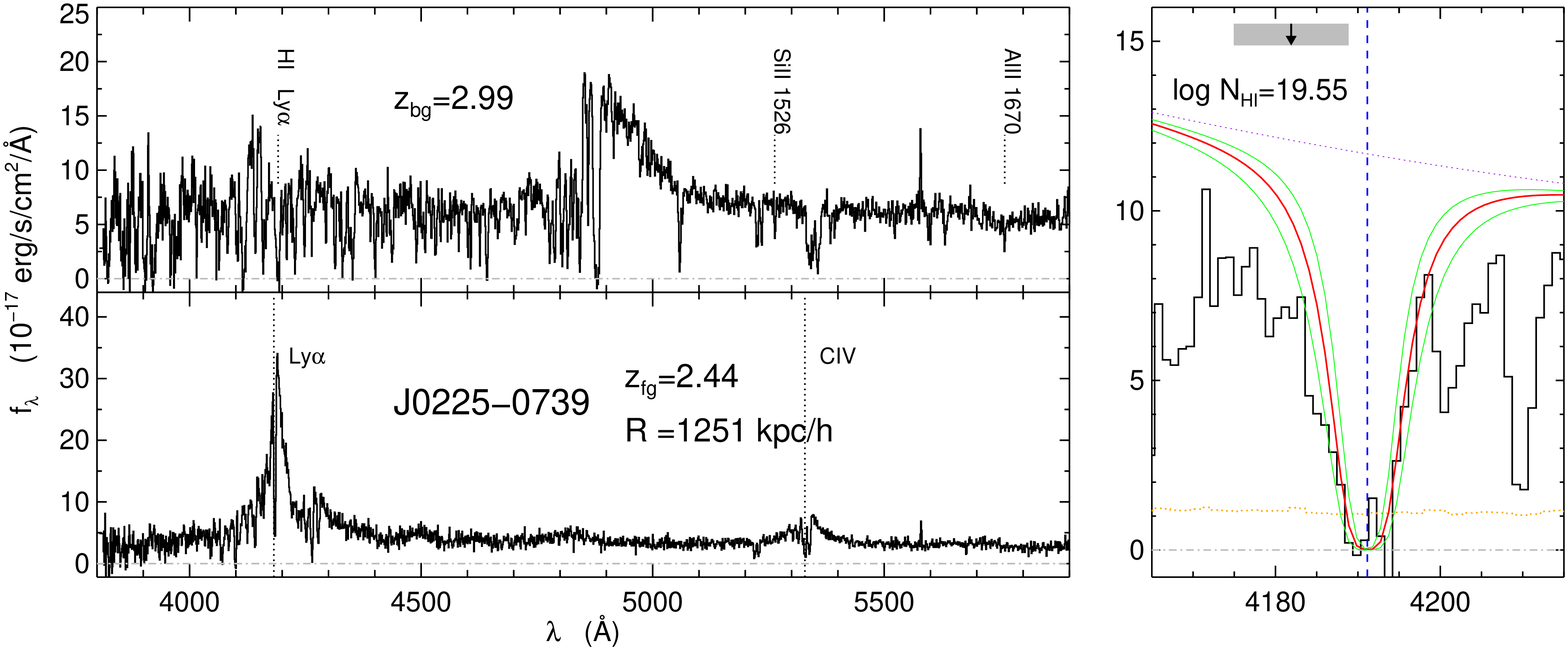,bb=0 0 1225 500,width=\textwidth}}
  \centerline{\epsfig{file=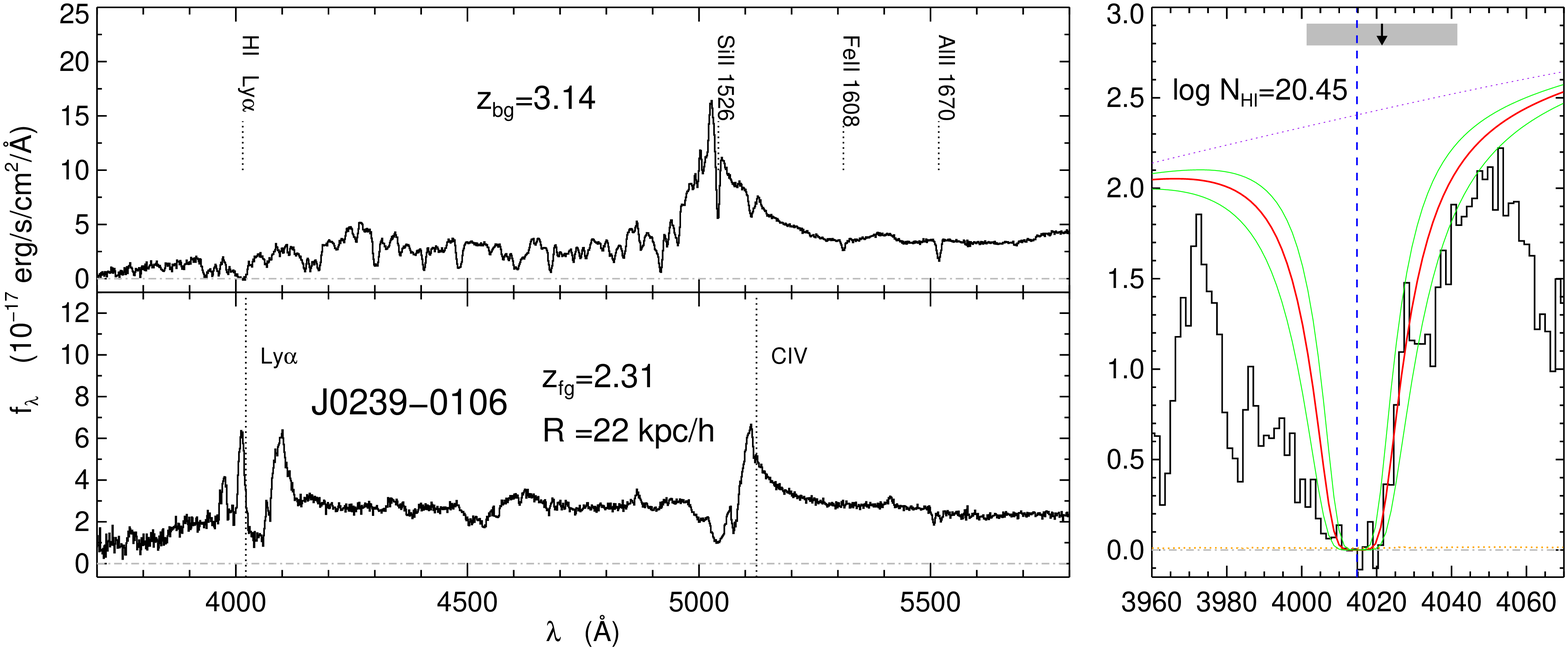,bb=0 0 1225 500,width=\textwidth}}
  \centerline{\epsfig{file=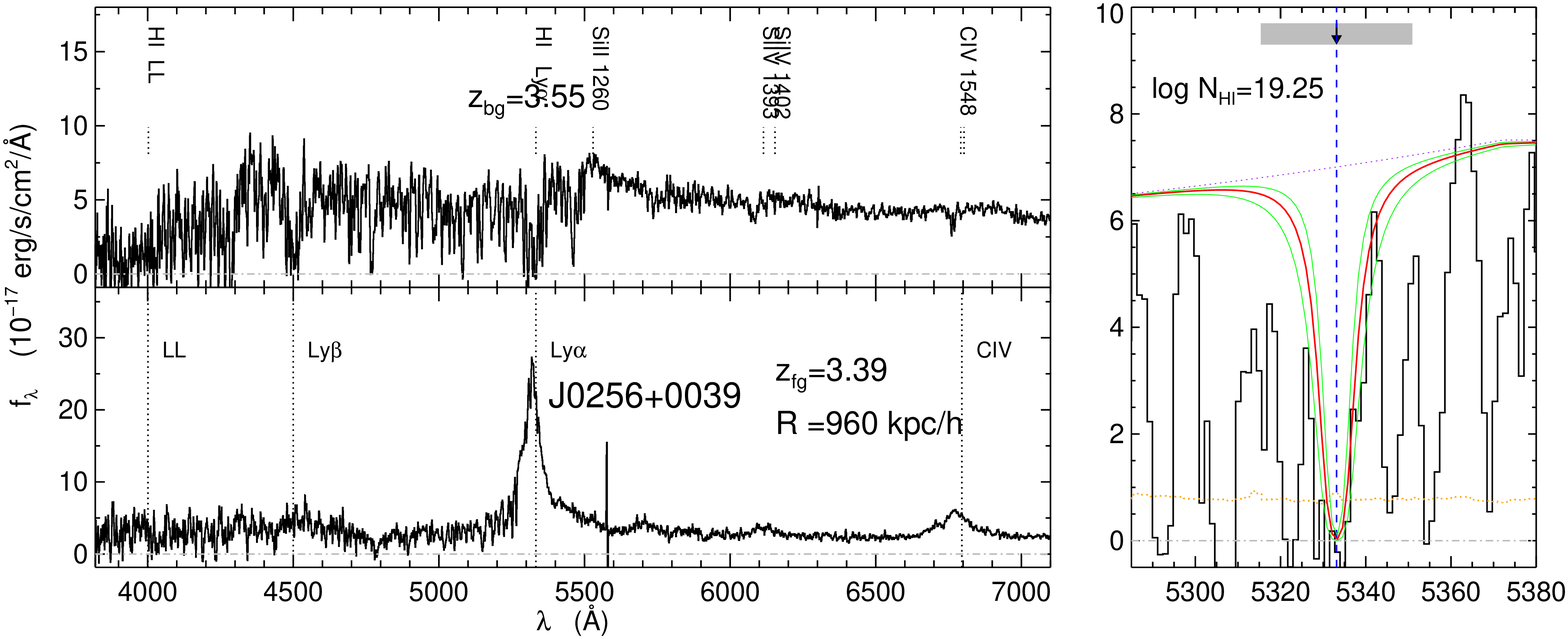,bb=0 0 1225 500,width=\textwidth}}
  \caption{ continued.}
\end{figure*}
\addtocounter{figure}{-1}
\begin{figure*}
  \centerline{\epsfig{file=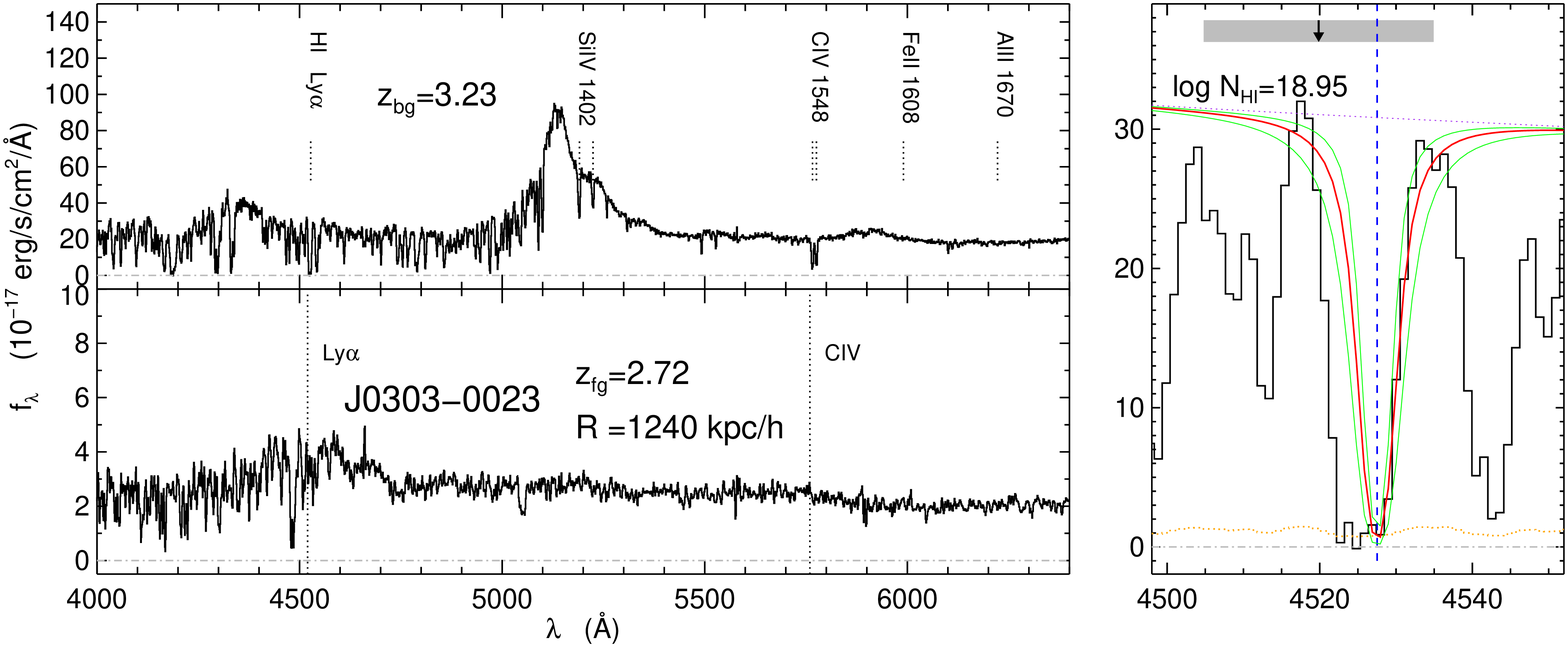,bb=0 0 1225 500,width=\textwidth}}
  \centerline{\epsfig{file=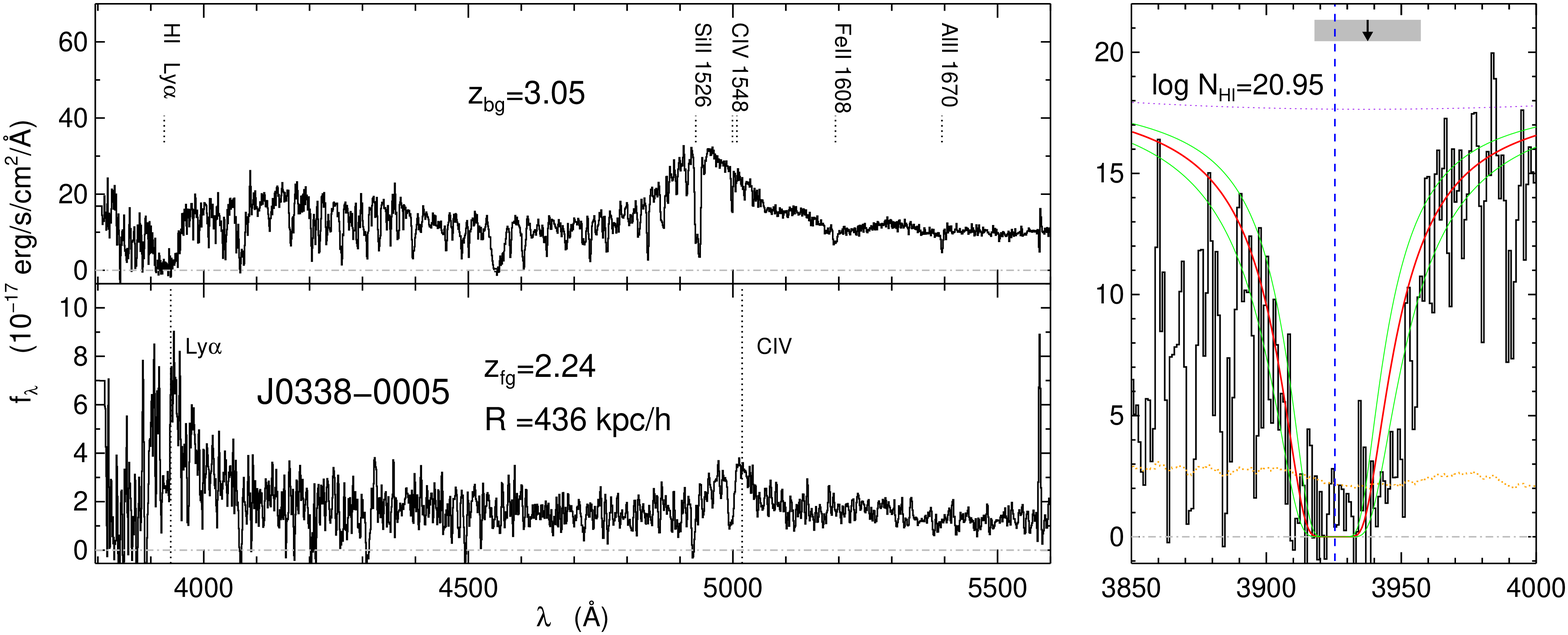,bb=0 0 1225 500,width=\textwidth}}
  \centerline{\epsfig{file=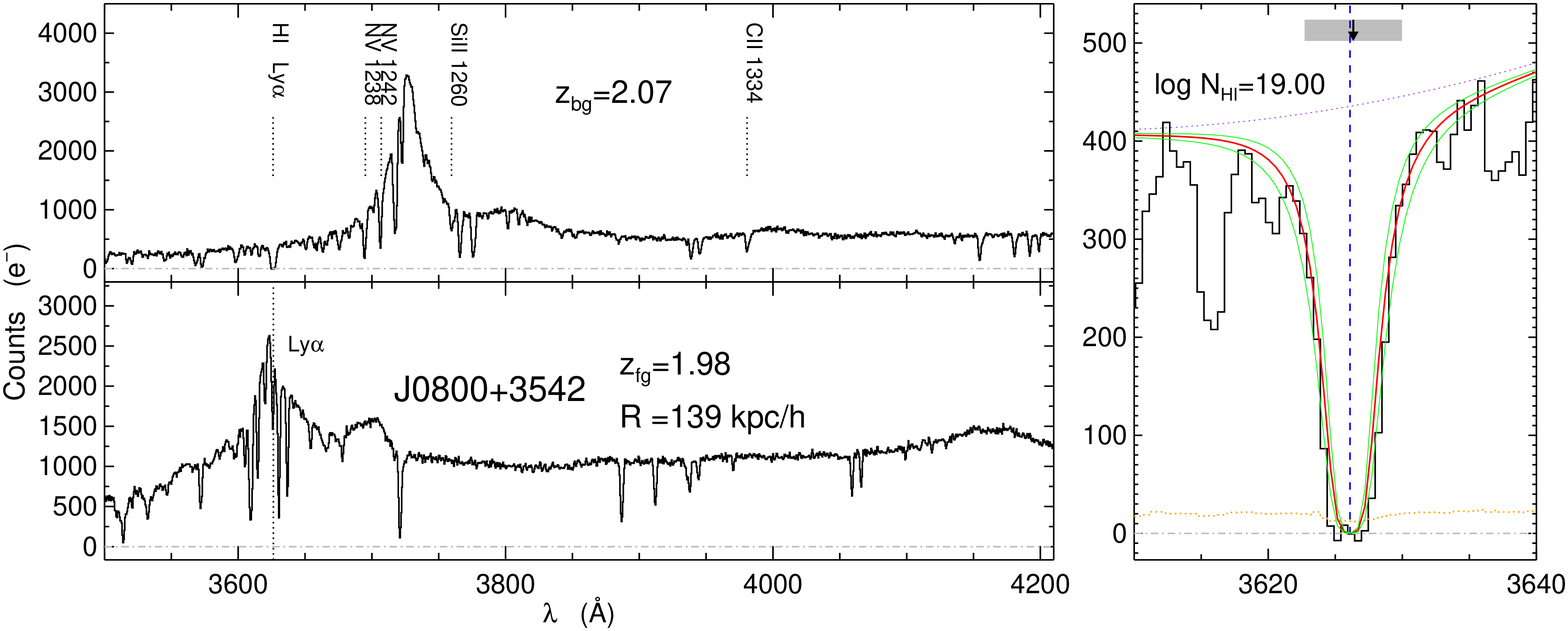,bb=0 0 1225 500,width=\textwidth}}
  \caption{ continued. }
\end{figure*}
\addtocounter{figure}{-1}
\begin{figure*}
  \centerline{\epsfig{file=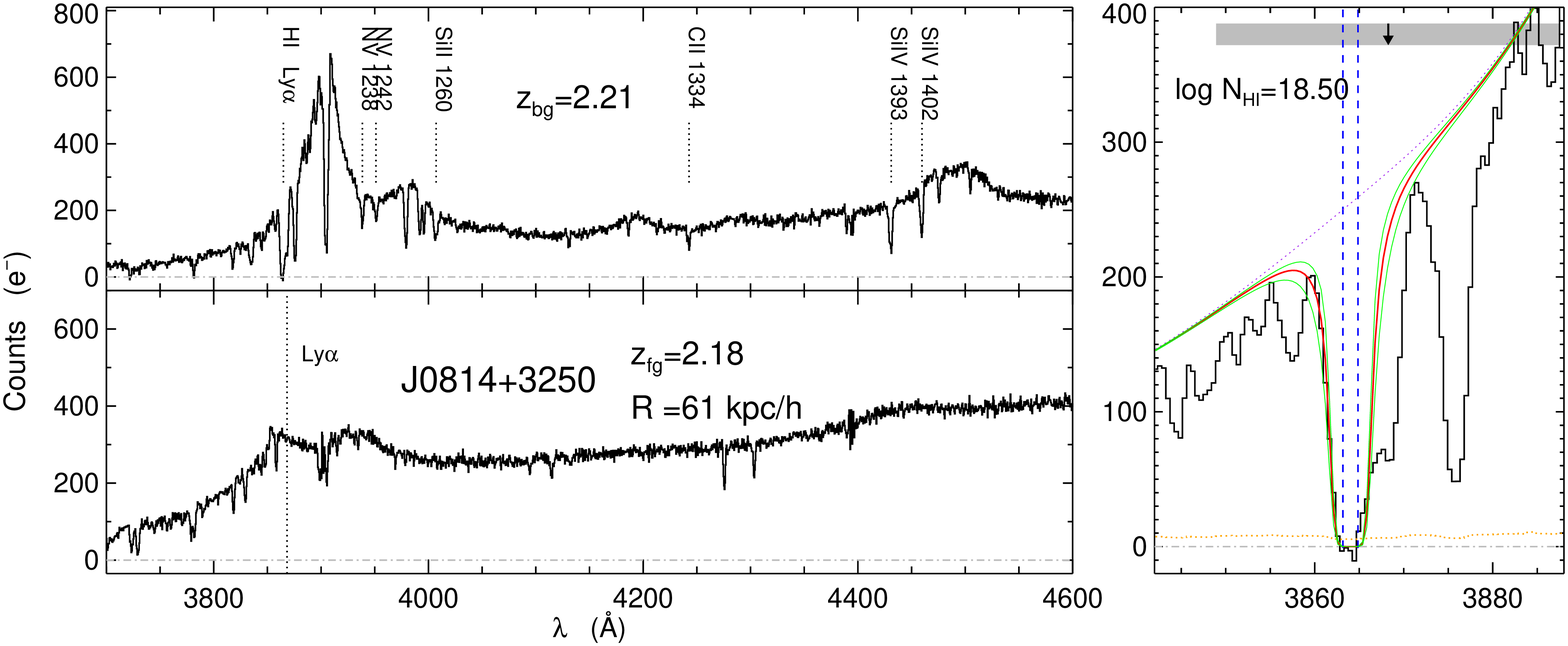,bb=0 0 1225 500,width=\textwidth}}
  \centerline{\epsfig{file=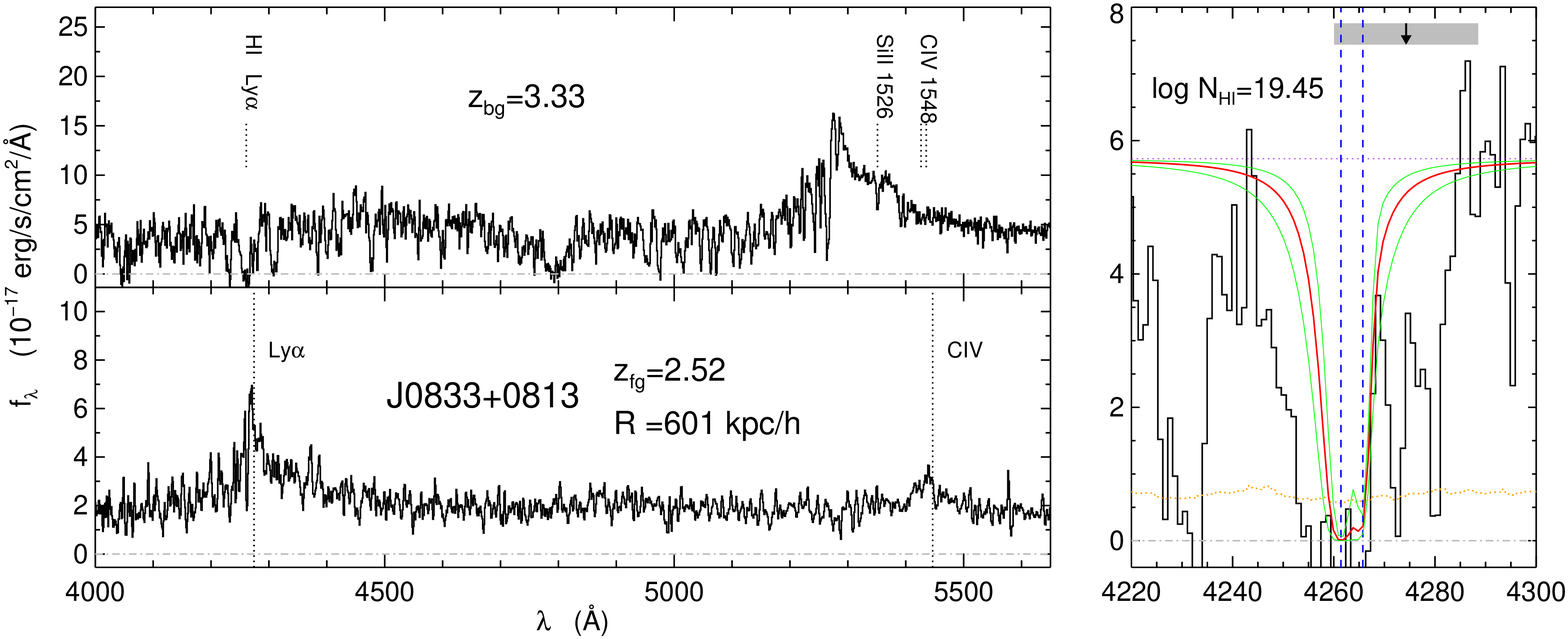,bb=0 0 1225 500,width=\textwidth}}
  \centerline{\epsfig{file=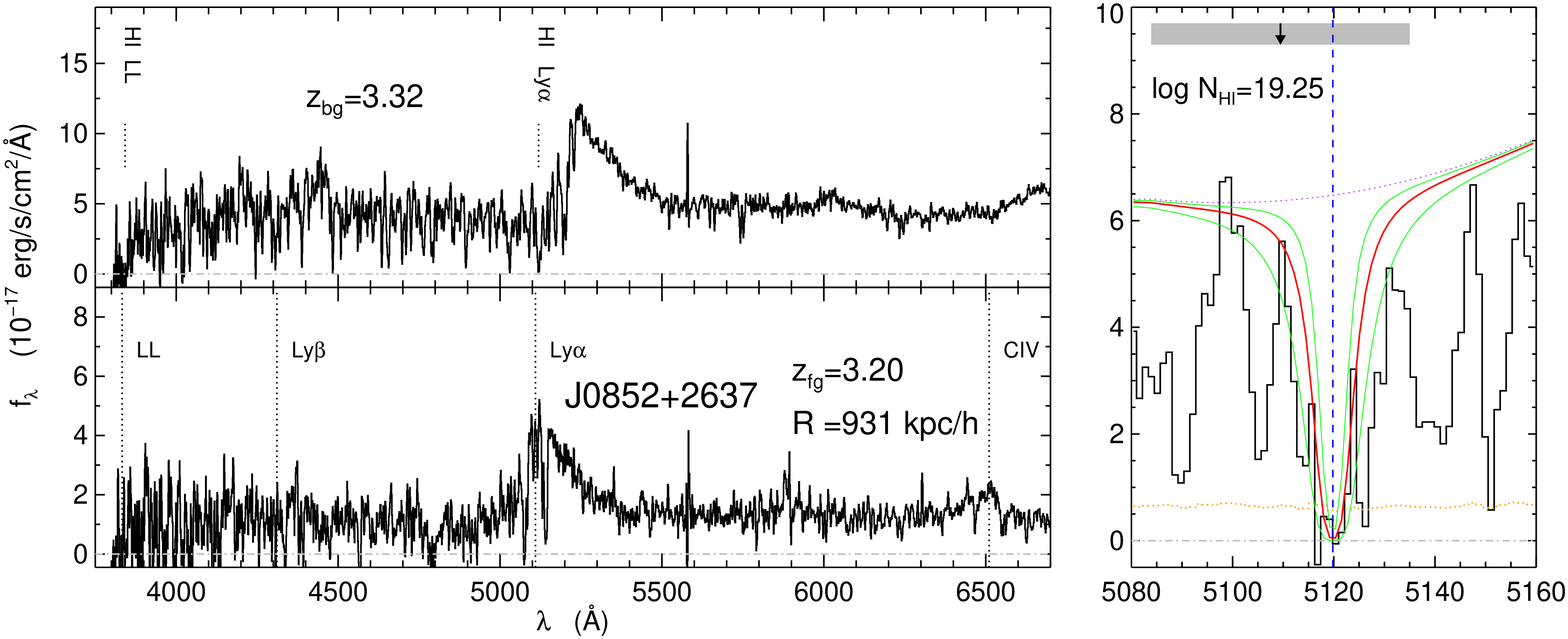,bb=0 0 1225 500,width=\textwidth}}
  \caption{ continued. }
\end{figure*}
\addtocounter{figure}{-1}
\begin{figure*}
  \centerline{\epsfig{file=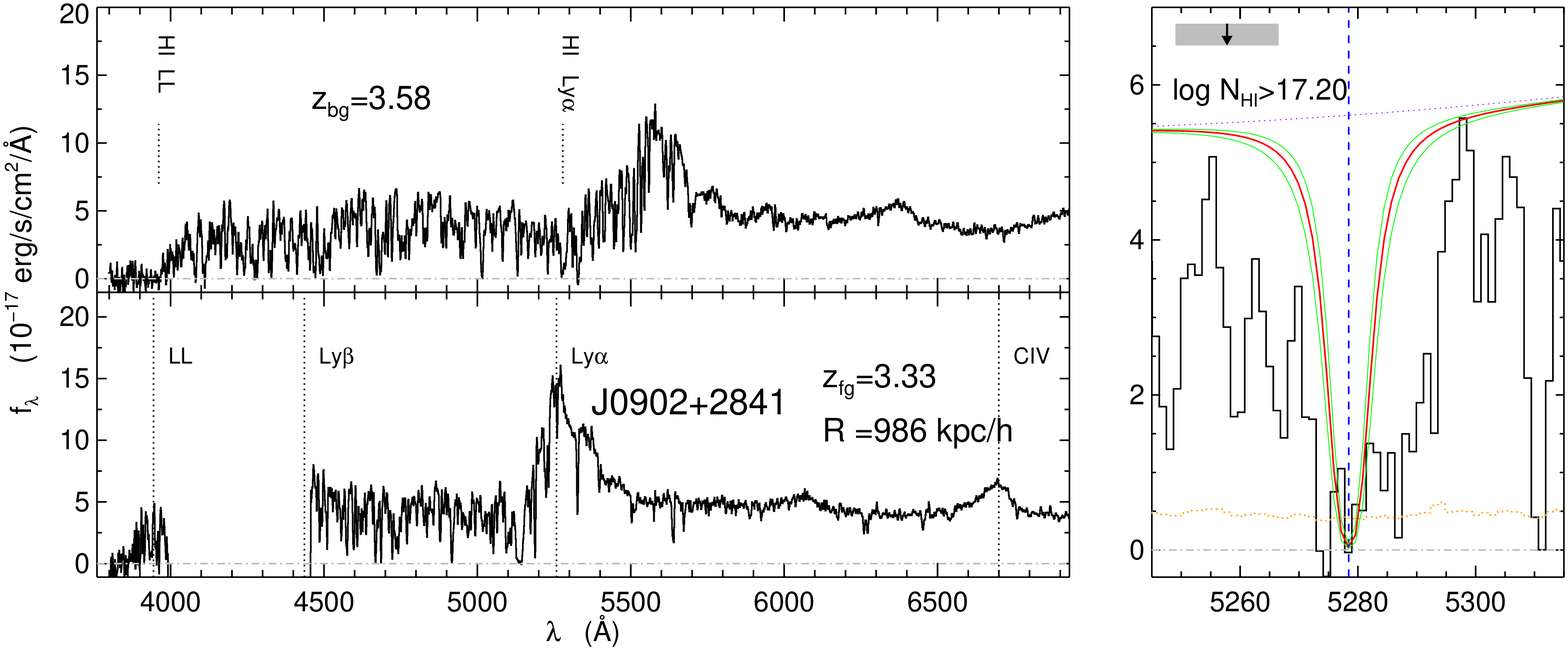,bb=0 0 1225 500,width=\textwidth}}
  \centerline{\epsfig{file=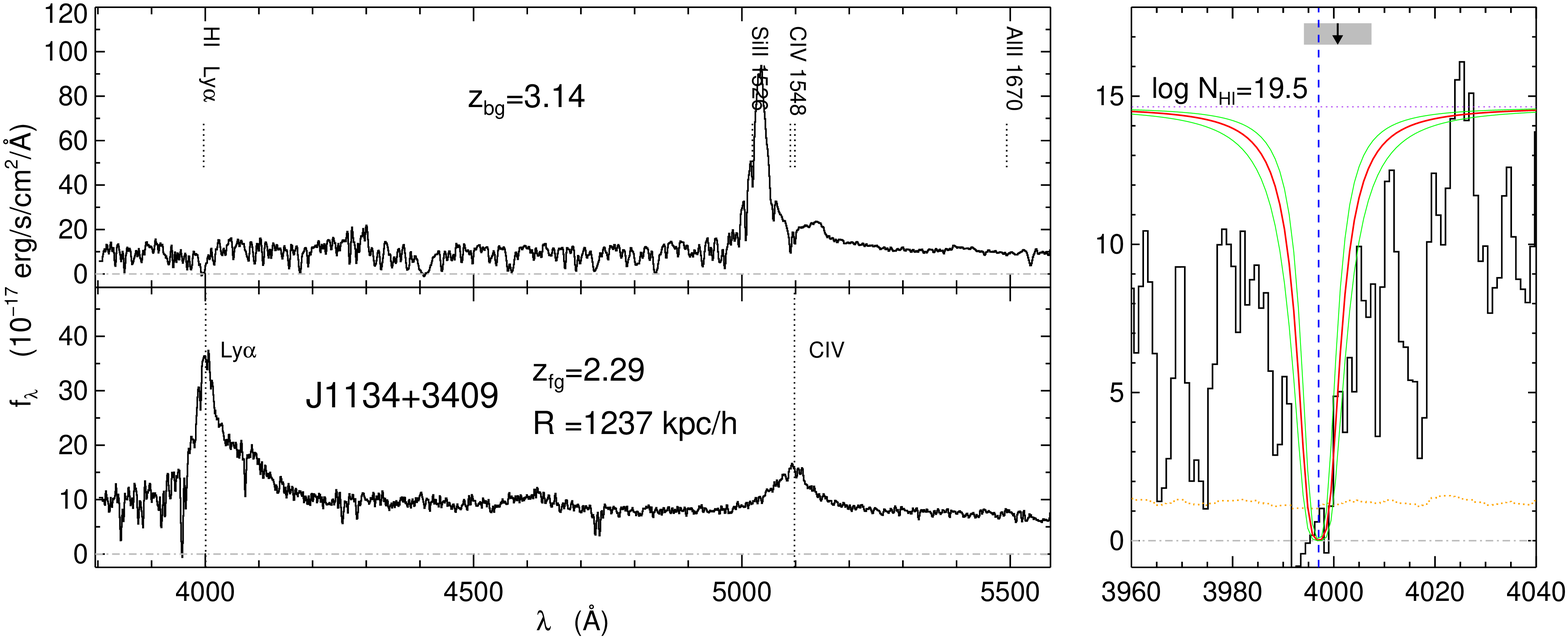,bb=0 0 1225 500,width=\textwidth}}
  \centerline{\epsfig{file=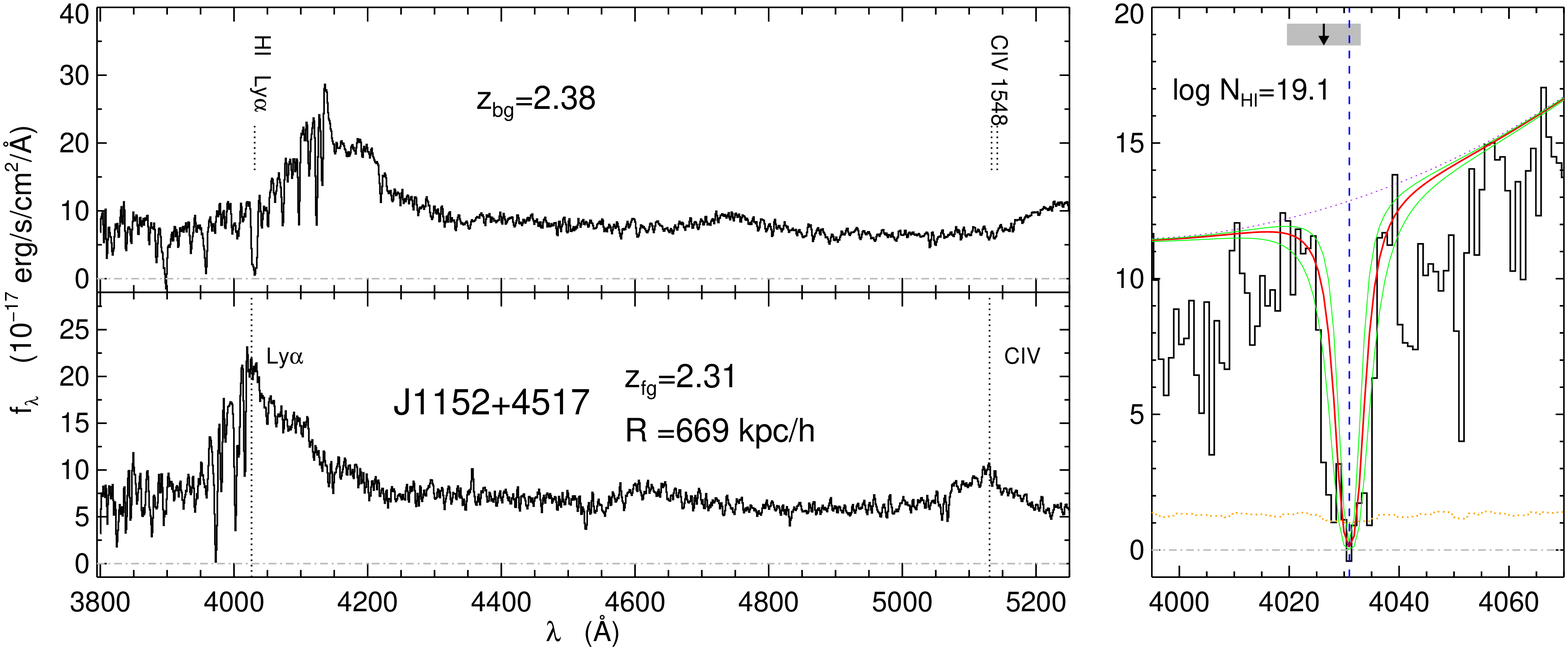,bb=0 0 1225 500,width=\textwidth}}
  \caption{ continued.}
\end{figure*}
\addtocounter{figure}{-1}
\begin{figure*}
  \centerline{\epsfig{file=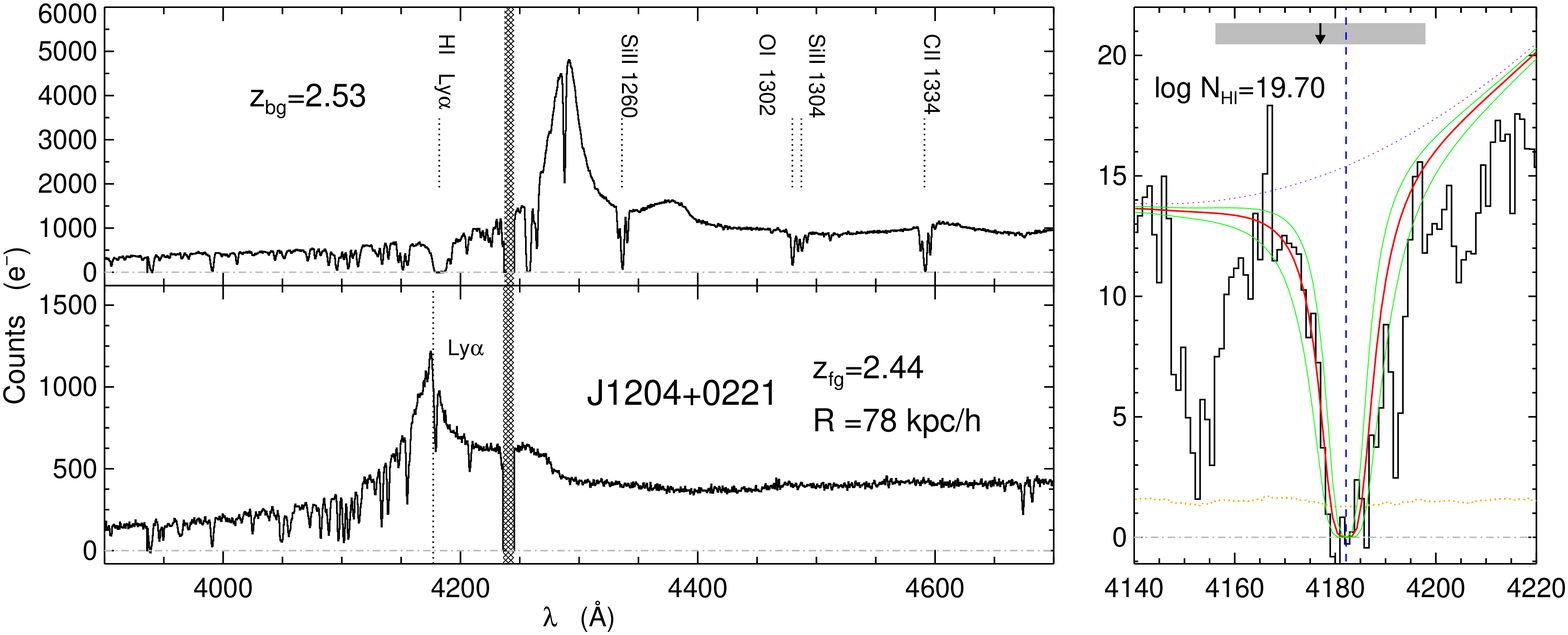,bb=0 0 1225 500,width=\textwidth}}
  \centerline{\epsfig{file=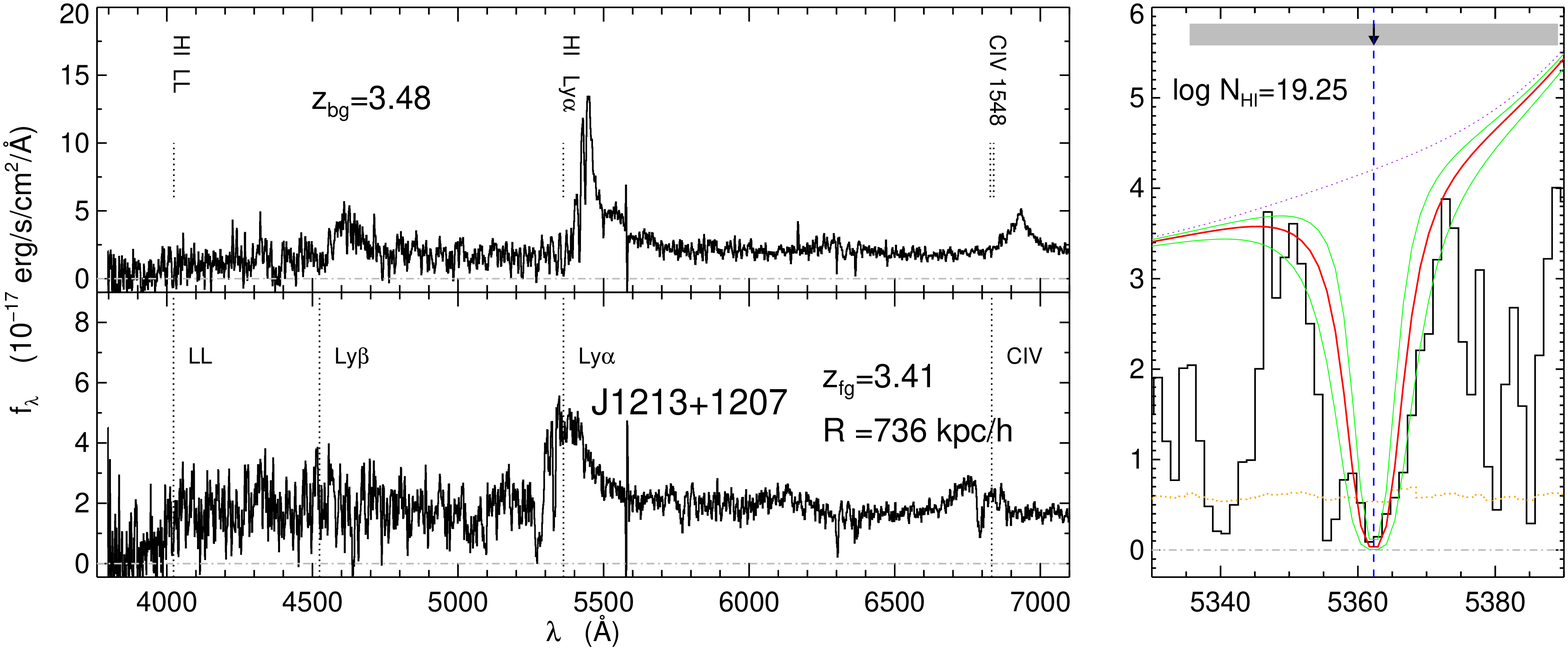,bb=0 0 1225 500,width=\textwidth}}
  \centerline{\epsfig{file=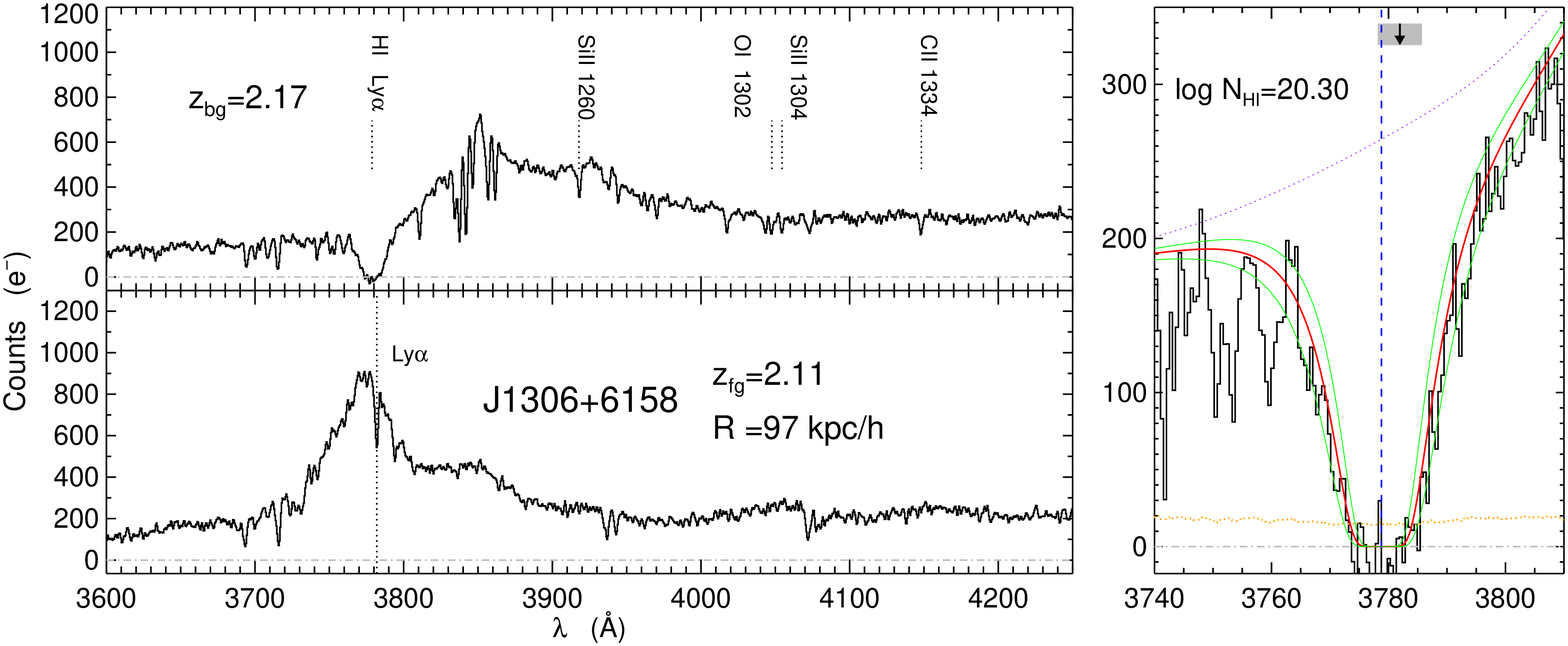,bb=0 0 1225 500,width=\textwidth}}
  \caption{ continued. }
\end{figure*} 
\addtocounter{figure}{-1}
\begin{figure*}
  \centerline{\epsfig{file=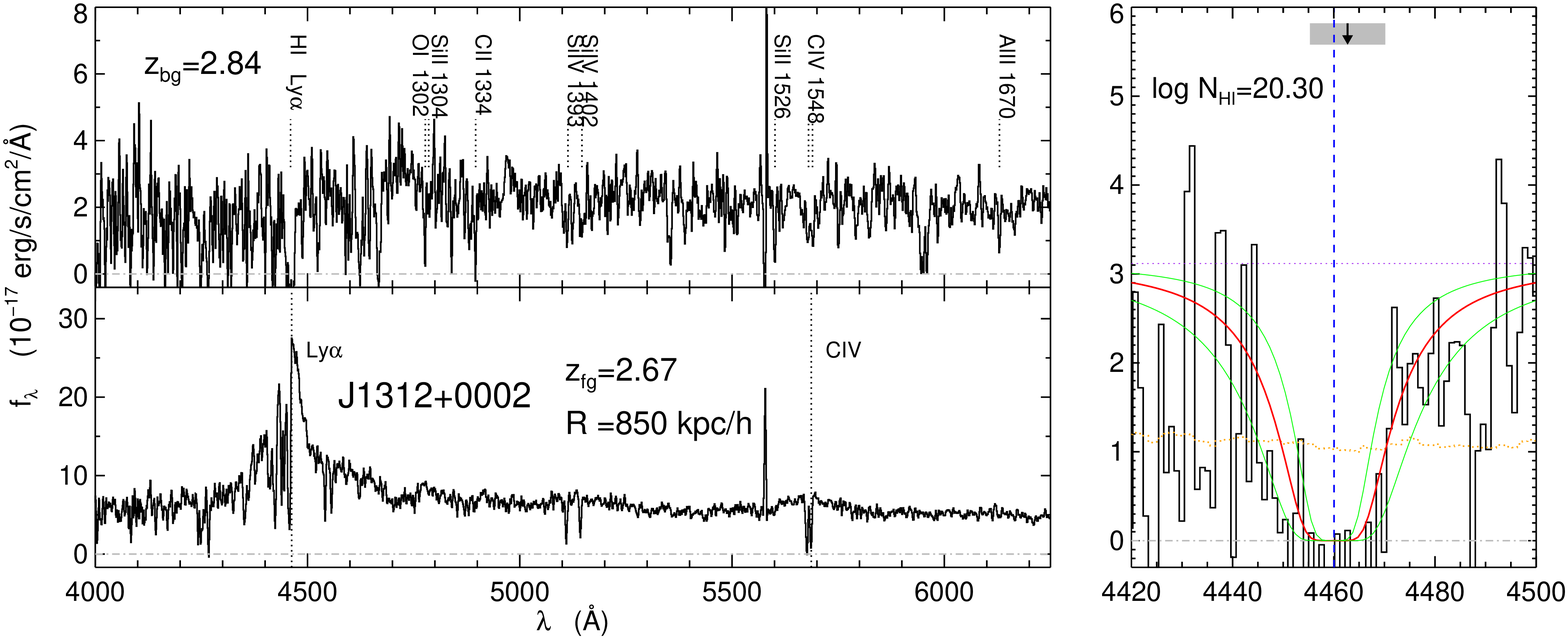,bb=0 0 1225 500,width=\textwidth}}
  \centerline{\epsfig{file=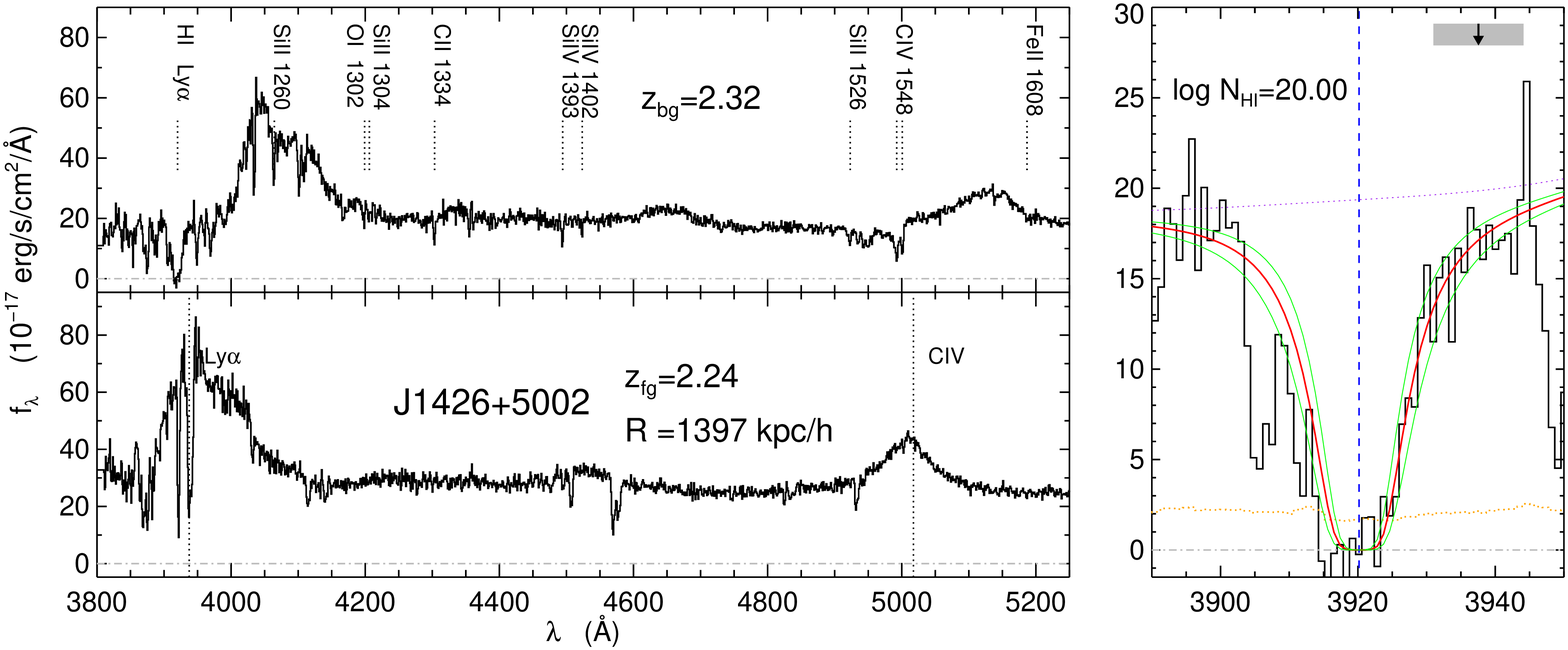,bb=0 0 1225 500,width=\textwidth}}
  \centerline{\epsfig{file=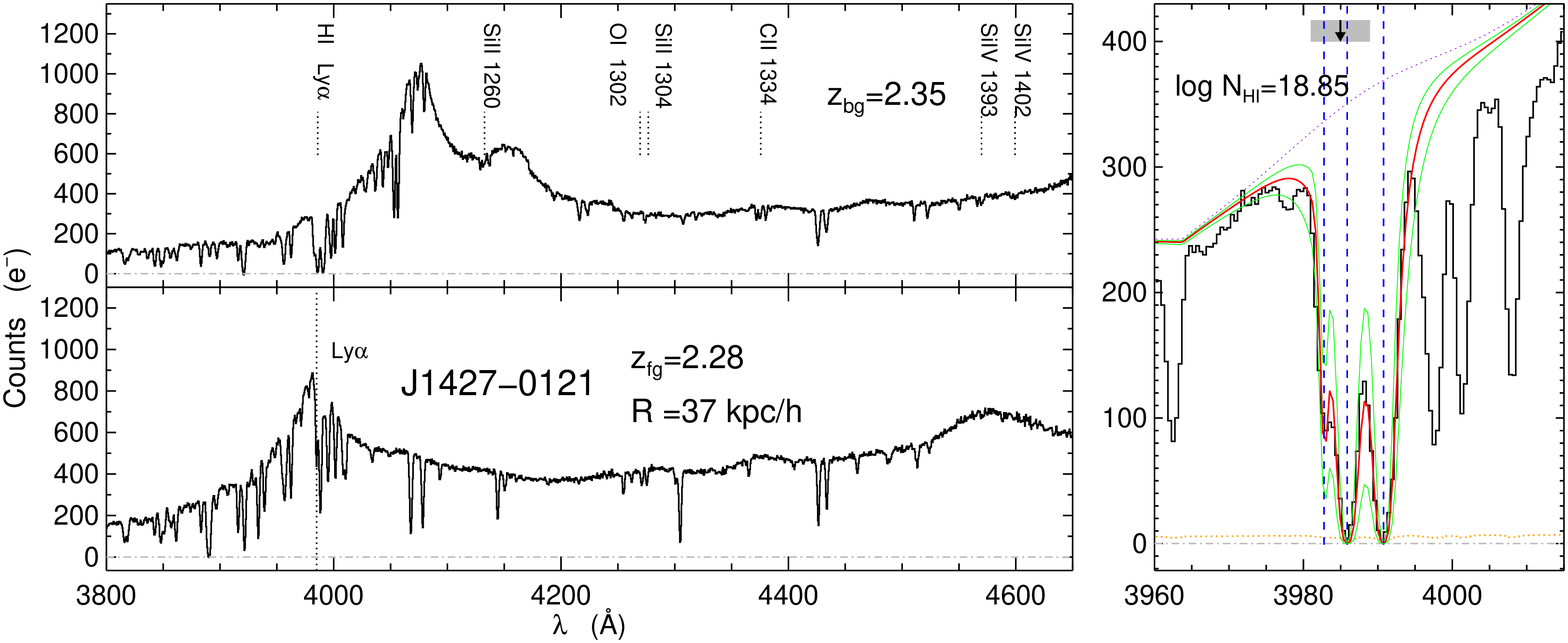,bb=0 0 1225 500,width=\textwidth}}
  \caption{ continued.}
\end{figure*}
\addtocounter{figure}{-1}
\begin{figure*}
  \centerline{\epsfig{file=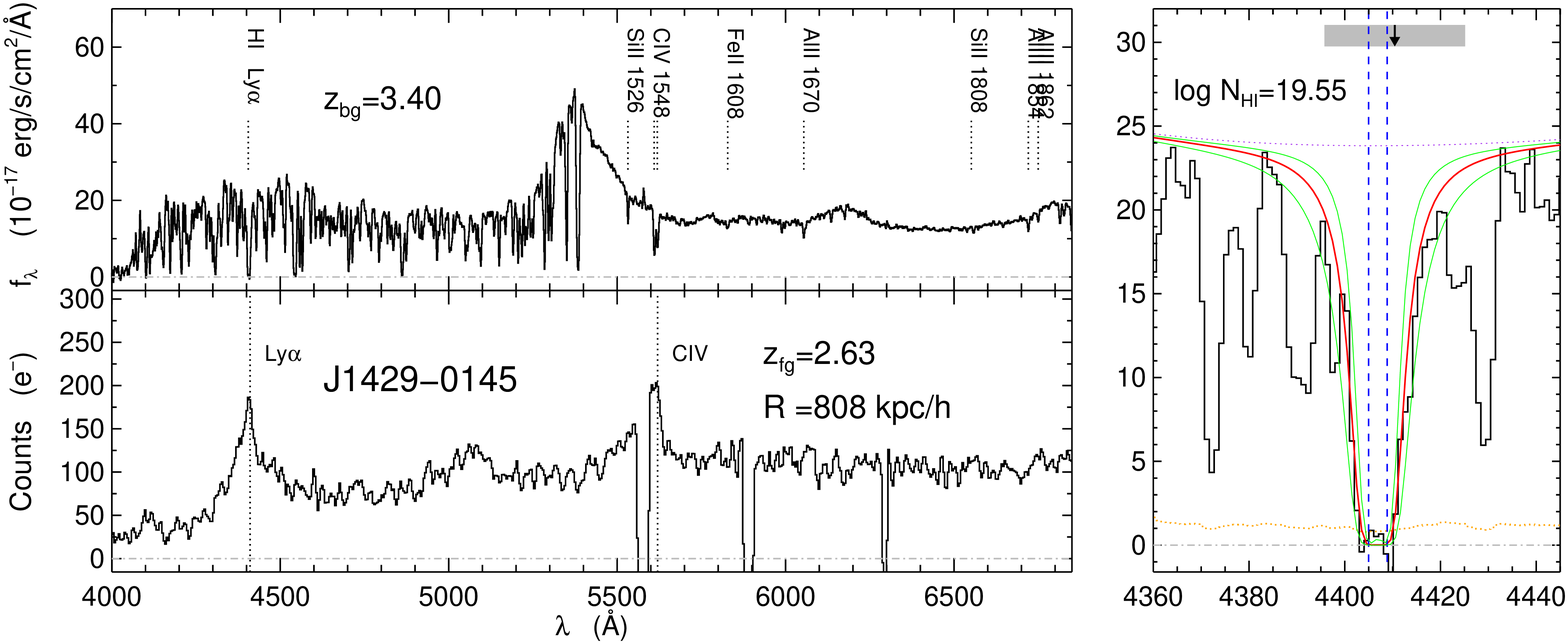,bb=0 0 1225 500,width=\textwidth}}
  \centerline{\epsfig{file=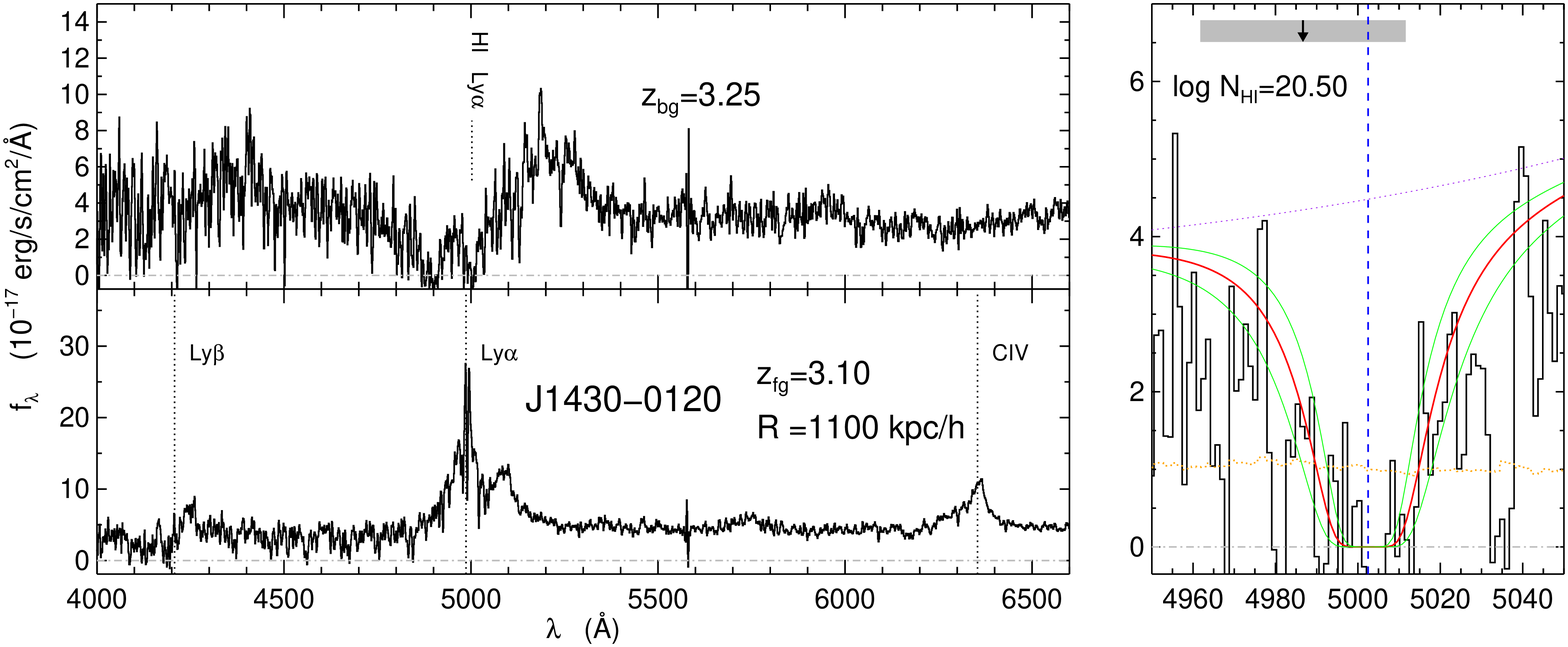,bb=0 0 1225 500,width=\textwidth}}
  \centerline{\epsfig{file=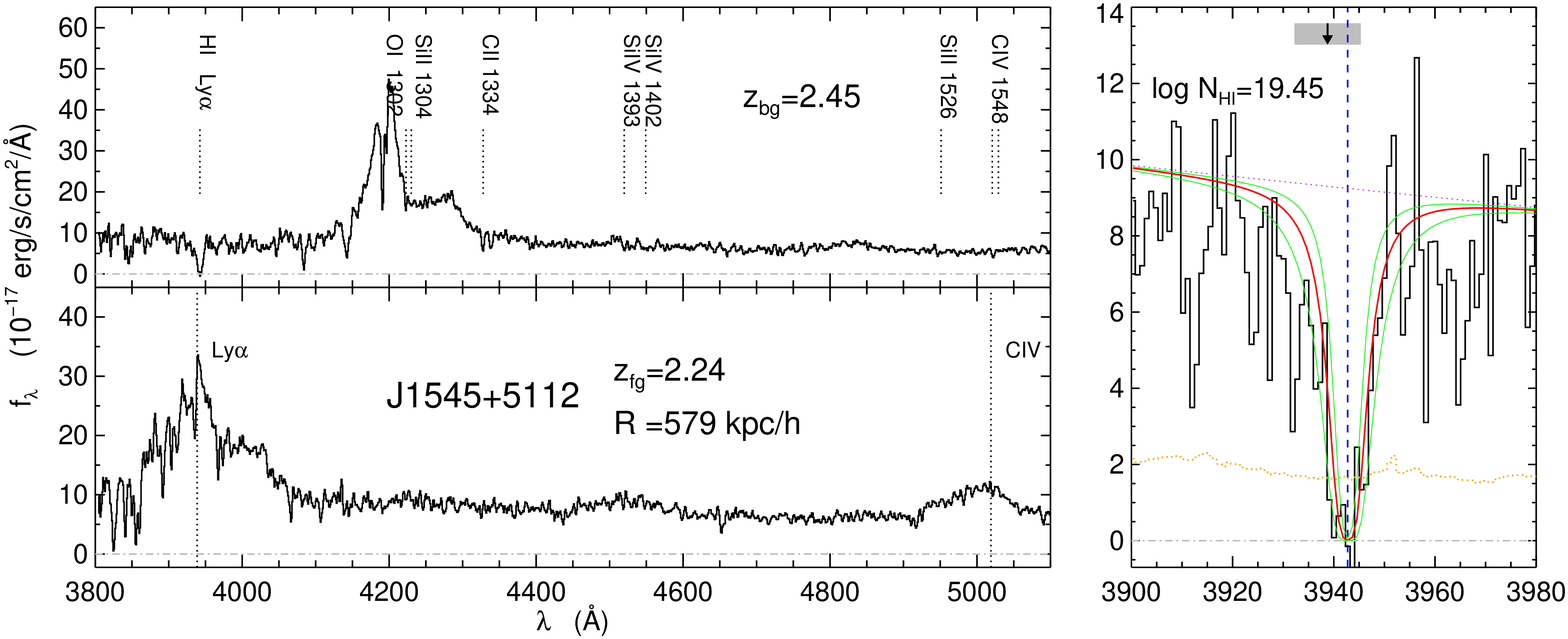,bb=0 0 1225 500,width=\textwidth}}
  \caption{ continued.}
\end{figure*}
\addtocounter{figure}{-1}
\begin{figure*}
  \centerline{\epsfig{file=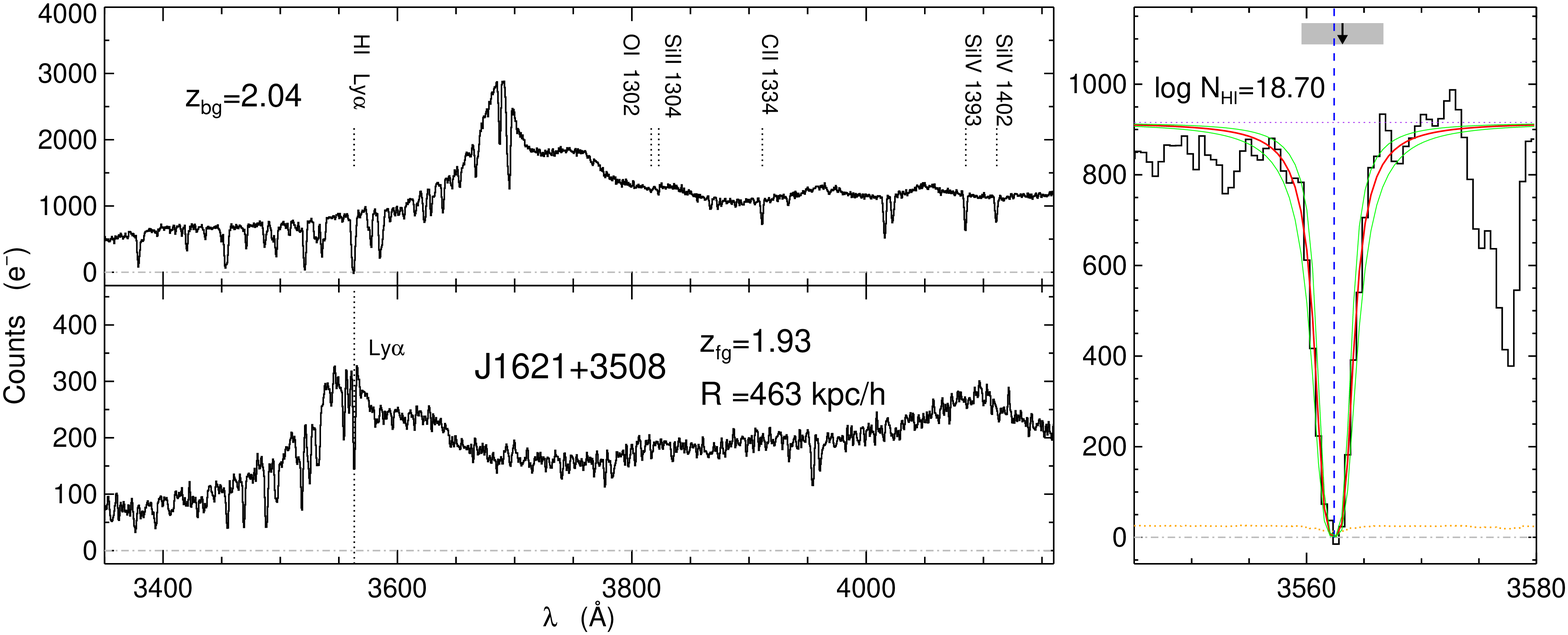,bb=0 0 1225 500,width=\textwidth}}
  \centerline{\epsfig{file=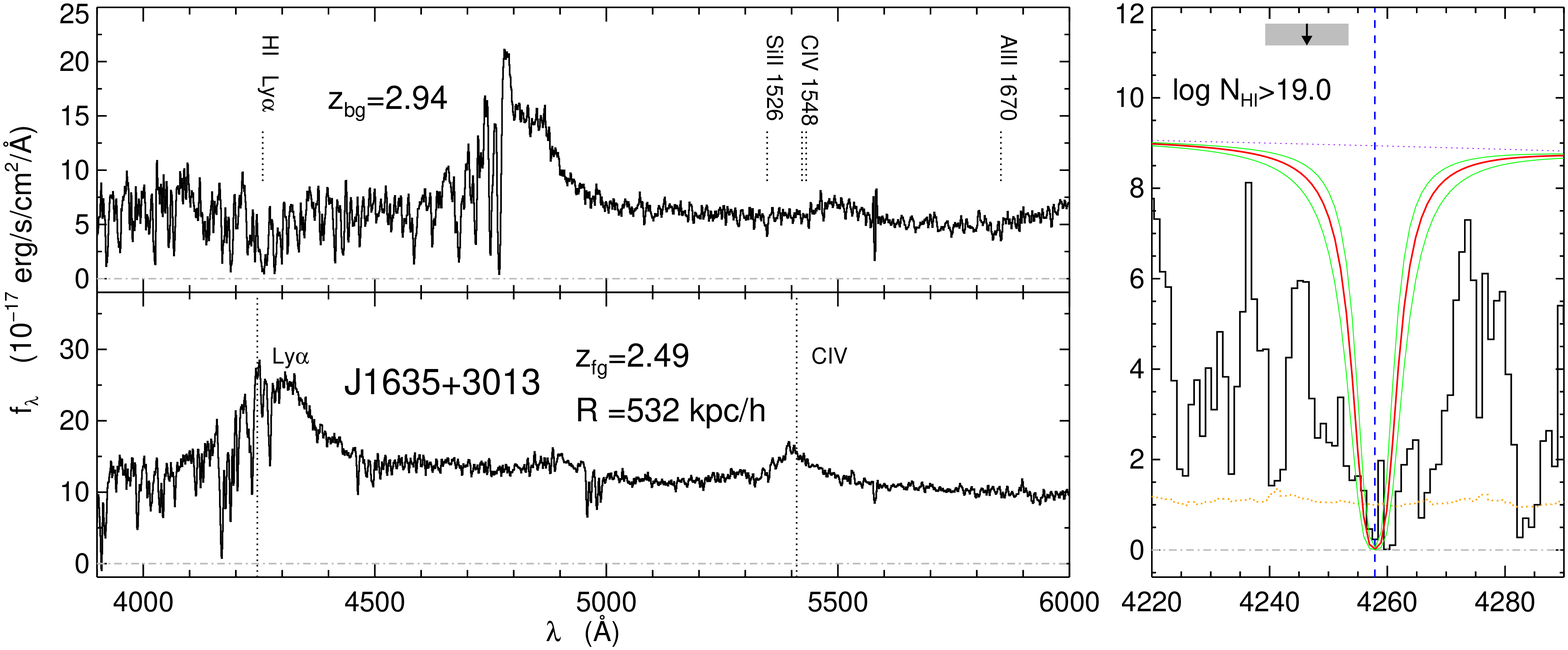,bb=0 0 1225 500,width=\textwidth}}
  \caption{ continued.}
\end{figure*}
\addtocounter{figure}{-1}
\begin{figure*}
  \centerline{\epsfig{file=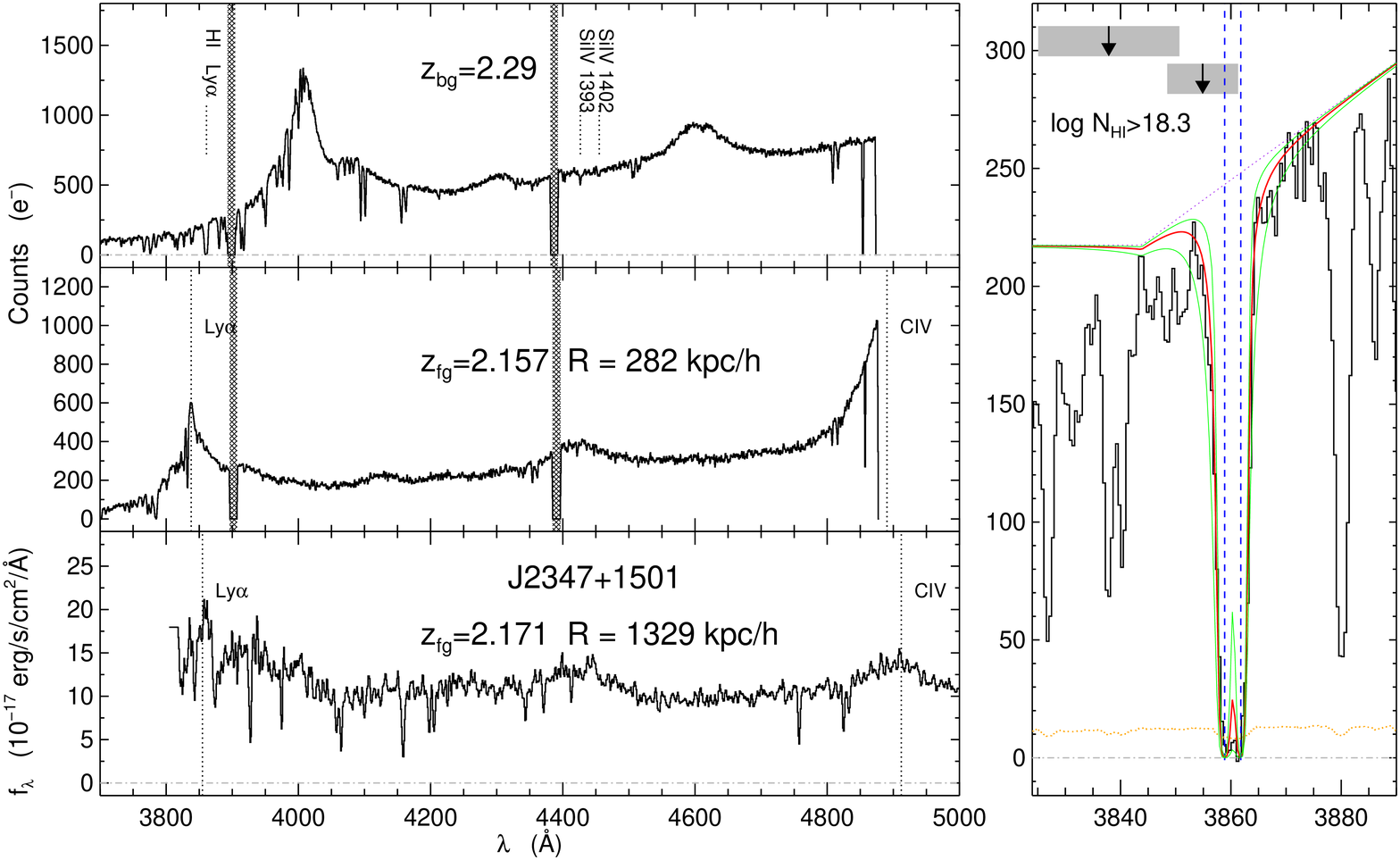,bb=0 0 1225 810,width=\textwidth}}
  \caption{ continued.}
\end{figure*}

\acknowledgements
We are grateful to Kurt Adelberger and John O'Meara for sharing their 
results prior to publication. JFH acknowledges helpful discussions with 
Taotao Fang, Juna  Kollmeier, Piero Madau, Chris McKee, Martin White, and 
Zheng Zheng. 

For part of this study JFH was supported by the Proctor Graduate
fellowship at Princeton University.  JFH is currently supported by
NASA through Hubble Fellowship grant \# 01172.01-A awarded by the
Space Telescope Science Institute, which is operated by the
Association of Universities for Research in Astronomy, Inc., for NASA,
under contract NAS 5-26555. JXP wishes to acknowledge funding through
NSF grant AST-0307408.

Funding for the SDSS and SDSS-II has been provided by the Alfred
P. Sloan Foundation, the Participating Institutions, the National
Science Foundation, the U.S. Department of Energy, the National
Aeronautics and Space Administration, the Japanese Monbukagakusho, the
Max Planck Society, and the Higher Education Funding Council for
England. The SDSS Web Site is http://www.sdss.org/.

The SDSS is managed by the Astrophysical Research Consortium for the
Participating Institutions. The Participating Institutions are the
American Museum of Natural History, Astrophysical Institute Potsdam,
University of Basel, Cambridge University, Case Western Reserve
University, University of Chicago, Drexel University, Fermilab, the
Institute for Advanced Study, the Japan Participation Group, Johns
Hopkins University, the Joint Institute for Nuclear Astrophysics, the
Kavli Institute for Particle Astrophysics and Cosmology, the Korean
Scientist Group, the Chinese Academy of Sciences (LAMOST), Los Alamos
National Laboratory, the Max-Planck-Institute for Astronomy (MPA), the
Max-Planck-Institute for Astrophysics (MPIA), New Mexico State
University, Ohio State University, University of Pittsburgh,
University of Portsmouth, Princeton University, the United States
Naval Observatory, and the University of Washington.  

The authors wish to recognize and acknowledge the very significant
cultural role and reverence that the summit of Mauna Kea has always
had within the indigenous Hawaiian community.  We are most fortunate
to have the opportunity to conduct observations from this mountain.

This paper was partly based on observations obtained at the Gemini
Observatory, which is operated by the Association of Universities for
Research in Astronomy, Inc., under a cooperative agreement with the
NSF on behalf of the Gemini partnership: the National Science
Foundation (United States), the Particle Physics and Astronomy
Research Council (United Kingdom), the National Research Council
(Canada), CONICYT (Chile), the Australian Research Council
(Australia), CNPq (Brazil) and CONICET (Argentina)

Some of the data presented herein were obtained at the W.M. Keck
Observatory, which is operated as a scientific partnership among the
California Institute of Technology, the University of California and
the National Aeronautics and Space Administration. The Observatory was
made possible by the generous financial support of the W.M. Keck
Foundation. 

Some of the Keck data was obtained through the National Science
Foundation's Telescope System Instrumentation Program (TSIP), supported
by AURA through the National Science Foundation under AURA Cooperative
Agreement AST 0132798 as amended.

\begin{appendix}
 
\section{UV Enhancement}

In this appendix we describe how we computed the quantity 
\be g_{\rm UV}\equiv 1 + \frac{F_{\rm QSO}}{F_{\rm UVB}}, 
\ee 
which is the maximum enhancement of the quasars' ionizing photon flux
over that of the extragalactic ionizing background, at the location of
the background quasar sightline, assuming that the quasar emits
isotropically.

The ionizing photon number flux $F_{\rm UVB}$ (photons~s$^{-1}$~cm$^{-2}$) 
due to the UV background is 
\be
F_{\rm UVB}=\pi \int_{\nu_{\rm LL}}^{\infty} \frac{J_{\nu}}{h\nu}  d\nu,
\ee
where $\nu_{\rm LL}=13.6~{\rm eV}\slash h$ is the Lyman limit
frequency, $h$ is Planck's constant, and $J_{\nu}$ is the mean
specific intensity of the UV background
(erg~s$^{-1}$~cm$^{-2}$~ster$^{-1}$~Hz$^{-1}$). For the intensity of
the UV background, we use the version computed by F. Haardt \&
P. Madau (2006, in preparation), which considers the emission from
observed quasars and galaxies after it is filtered through the IGM to
yield the UVB as a function of redshift\footnote{This model uses
  recent results for the quasar luminosity function, assumes that
  $f_{\rm esc}=0.1$ of ionizing photons escape from galaxies, and
  assumes a power law index of $\alpha=1.8$ to the spectral energy
  distribution at wavelengths shortward of 912~\AA\. The spectrum is
  available at http://pitto.mib.infn.it/\~{}haardt/refmodel.html.}.

The ionizing photon number flux from a quasar of specific luminosity
$L_{\rm \nu}$  (erg~s$^{-1}$~cm$^{-2}$~Hz$^{-1}$) at a proper distance 
$r_{\rm Q}$ is 
\be
F_{\rm QSO} = \frac{1}{4\pi r^2_{\rm Q}}\int_{\nu_{\rm LL}}^{\infty} 
\frac{L_{\nu}}{h\nu}  d\nu . 
\ee
We assume that the quasars' spectral energy distribution obeys
the power law form $L_{\nu} = L_{\nu_{\rm LL}} (\nu\slash
\nu_{\rm LL})^{-\alpha_{\rm Q}}$, blueward of  $\nu_{\rm LL}$. 
\citet{Telfer02} measured an average power law 
slope of $\alpha_{\rm Q}= 1.57$ blueward of wavelengths $\lesssim 1200$~\AA,
with the transition to this slope occurring in the range $\lambda =
1200-1300$~\AA.

The luminosity density at the Lyman limit for a quasar at redshift $z$ is, 
\be
L_{\nu_{\rm LL}} = \frac{4\pi d_{L}^2(z)}{\left(1+z\right)}
f_{\nu_0}, 
\ee
where $d_{\rm L}$ is the luminosity distance and $f_{\nu_0}$ is the 
specific flux of the quasar as observed on Earth at the redshifted 
frequency $\nu_0 = \nu_{\rm LL}\slash (1+z)$. We define the apparent
magnitude at the Lyman limit, measured by an observer on Earth 
at the redshifted frequency $\nu_0$ according to 
\be
f_{\nu_0}= 10^{-0.4\left(48.60 + m_{912}\right)}~{\rm erg}~{\rm s}^{-1}~{\rm cm}^{-2}~{\rm Hz}^{-1}.
\ee

For the majority of the quasars in Table~\ref{table:thick} the Lyman
limit is below the atmospheric cutoff.  Furthermore, even for higher
redshifts, determining $m_{912}$ from our spectroscopic observations
or the SDSS broad band magnitudes is complicated by the absorption
from the Lyman-$\alpha$ forest and LLSs.  Our goal is then to
compute $m_{\rm 912}$ given the SDSS apparent magnitude of a quasar,
measured in the rest frame near-UV. To this end, we tie the
\citet{Telfer02} power law spectrum to the composite quasar spectrum
of \citet{vanden01} at the wavelength $1285$\AA, which is a clean
region free of emission lines and in the wavelength range
($\lambda=1200-1300$~\AA) where \citep{Telfer02} observed the
transition to a spectral index of $\alpha_{\rm Q}$. We then
compute the `k-correction' between $m_{\rm 912}$ and the bluest SDSS
filter, which is also redward of the Ly$\alpha$ line in the observed
frame, by convolving the \citet{vanden01} template with the SDSS
filter curve.  We work with a filter redward of Ly$\alpha$ to
avoid the Ly$\alpha$ forest of the quasar, which is not correctly
included in the quasar template.

\section{Tentative Optically Thick Absorbers Near Quasars}

Relevant quantities for a sample of tentative quasar-absorber pairs, 
for which we could not be sure that $\log \mnhi > 17.2$ are given
in Table~\ref{table:poss}.  Higher SNR and higher resolution spectra are
required to make definitive conclusions about these systems, making
them an interesting set of targets for future research. Coordinates
and SDSS five band photometry of these objects are given in 
Table~\ref{table:poss_coords} of Appendix C.  

\input{tab2.tex}

\section{Coordinates and Photometry}
Coordinates and SDSS five band photometry of the 
quasars in Tables~\ref{table:thick} and \ref{table:poss} are listed 
below. We also list coordinates and photometry for all five members of the
projected group of quasars associated with SDSSJ~0127$+$1507 in Table~\ref{table:0127}. 

\input{tab3.tex}
\input{tab4.tex}
\input{tab5.tex}

\end{appendix}

\end{document}

%% file: defs.tex
\newcommand{\beq}{\begin{equation}}
\newcommand{\eeq}{\end{equation}}
\def\be{\begin{equation}}
\def\ee{\end{equation}}
\def\bea{\begin{eqnarray}}
\def\eea{\end{eqnarray}}



\def \logTd6 {\hbox{log$( T/6 \kev)$} }

\def\myputfigure#1#2#3#4#5%
{\vskip#5pt\makebox[0pt]{\hskip#2in
\includegraphics[width=#3\textwidth]{#1}}\vskip#4pt\hfill}

\def \arcsec    {^{\prime\prime}}

\def \kpc       {{\rm\ kpc}}

\def \kms            {~{\rm km~s}^{-1}}

\def \etal      {et al.}

\def \kev       {{\rm\ keV}}

\def \hMpc      {h^{-1}{\rm\ Mpc}}
\def \hkpc      {h^{-1}{\rm\ kpc}}
\def \mkms      {\rm\ km\ s^{-1}}

\newcommand{\mnhi}{N_{\rm HI}}
\newcommand{\nhi}{$N_{\rm HI}$}
\newcommand{\lya}{Ly$\alpha$}

%% file: tab1.tex
\begin{deluxetable*}{lcccccccccccc}
\tablecolumns{13}
\tablewidth{0pc}
\tablecaption{Optically Thick Absorbers Near Quasars \label{table:thick}}
\tablehead{\hskip -0.025in Name & z$_{\rm bg}$ & z$_{\rm fg}$ & $\Delta \theta$ & $R$ & z$_{\rm abs}$ & $\left|\Delta v\right|$ & $\Delta v_{\rm fg}$ & $\log N_{\rm HI}$ & $g_{\rm UV}$ & Redshift & {\footnotesize Fg} & {\footnotesize Bg}\\
  &      &       &  {\footnotesize ($^{\prime\prime}$)}    &  {\footnotesize ($\hkpc$)}   &         &   {\footnotesize (km s$^{-1}$)}   &  {\footnotesize (km s$^{-1}$)} &   {\footnotesize (cm$^{-2}$)}             &                    &         &  {\footnotesize Inst.} & \footnotesize {Inst.} }
\startdata
\hskip -0.05in SDSSJ0036+0839       & 2.69 & 2.569 &  154.5      &  \phn894  &  2.5647 & \phn 360    &\phn500& $18.95\pm 0.35$  & \phn\phn\phn7 & \ion{C}{3}]  & SDSS    & SDSS\\ 
\hskip -0.05in SDSSJ0127$+$1507$^1$ & 2.60 & 1.818 &   131.0     &  \phn794  &  1.8188 & \phn\phn30  &\phn300&  $18.6 \pm 0.3$  & \phn\phn\phn3 & \ion{Mg}{2}  & LRIS-R  & LRIS-B\\
                     & 2.38 & 1.818 &   51.9      &  \phn315  &  1.8175 &  \phn100   &\phn300&  $18.9 \pm 0.3$   & \phn\phn 13    & \ion{Mg}{2}  & LRIS-R  & LRIS-B\\
\hskip -0.05in SDSSJ0225$-$0739     & 2.99 & 2.440 &   214.0     &  1251     &  2.4476 &  \phn690    &\phn500&  $19.55\pm 0.2$  & \phn\phn\phn5 & \ion{C}{3}]  & SDSS    & SDSS\\ 
\smallskip
\hskip -0.05in SDSSJ0239$-$0106$^2$ & 3.14 & 2.308 & \phn\phn3.7 & \phn\phn22 &  2.3025 &  \phn540    & 1500 &  $20.45\pm 0.2$  &  6369          & \ion{C}{4}   & SDSS    & LRIS-B\\ 
\hskip -0.05in SDSSJ0256$+$0039  & 3.55 & 3.387 &   179.0     &  \phn960      &  3.387  &   20   & 1000 &  $19.25\pm 0.25$      &   \phn\phn20    & \ion{C}{4}   & SDSS    & SDSS\\ 
\hskip -0.05in SDSSJ0303$-$0023  & 3.23 & 2.718 &   217.6     &   1240      &  2.7243 &  \phn500    & 1000 &  $18.95   \pm 0.2$ & \phn\phn\phn8  & \ion{C}{3}]  & SDSS    & SDSS\\ 
\hskip -0.05in SDSSJ0338$-$0005  & 3.05 & 2.239 & \phn73.5    &  \phn436      &  2.2290 &  \phn960    & 1500 &  $20.9 \pm 0.2$  &  \phn\phn 13    & \ion{C}{4}-\ion{C}{3}] & SDSS & SDSS\\ 
\hskip -0.05in SDSSJ0800$+$3542  & 2.07 & 1.983 & \phn23.1    &  \phn139   &  1.9828 & \phn\phn40  &\phn300&  $19.0 \pm 0.15$   & \phn 488      & \ion{Mg}{2}  & LRIS-R  & LRIS-B\\
\smallskip
\hskip -0.05in SDSSJ0814$+$3250  & 2.21 & 2.182 & \phn10.3    &  \phn61   &  2.1792 &  \phn280    &  1500 &  $18.8 \pm 0.2$     &  1473         &  Template    & GMOS    & GMOS\\ 
\hskip -0.05in SDSSJ0833$+$0813  & 3.33 & 2.516 &   103.4     &  \phn601     &2.505\phn&  \phn 980   &  1000 &  $19.45 \pm 0.3$ & \phn\phn 18   & \ion{C}{3}]  & SDSS    & SDSS\\ 
\hskip -0.05in SDSSJ0852$+$2637  & 3.32 & 3.203 &   170.9     &  \phn931     &3.211\phn&  \phn550    &  1500 &  $19.25 \pm 0.4$ & \phn\phn 13   & \ion{C}{4}  & SDSS    & SDSS\\ 
\hskip -0.05in SDSSJ0902$+$2841  & 3.58 & 3.325 &   183.0     &  \phn986     &3.342\phn&    1200     &   500 &     $>17.2$      & \phn\phn 34   & \ion{C}{3}]  & SDSS    & SDSS\\ 
\hskip -0.05in SDSSJ1134$+$3409  & 3.14 & 2.291 &   209.2     &   1237       &2.2879 &  \phn320    &   500 &  $19.5 \pm 0.3$    & \phn\phn 11   & \ion{C}{3}]  & SDSS    & SDSS\\ 
\smallskip
\hskip -0.05in SDSSJ1152$+$4517  & 2.38 & 2.312 &   113.4     &  \phn669     &2.3158 &  \phn370    &   500 &  $19.1 \pm 0.3$    & \phn\phn 30   & \ion{C}{3}]  & SDSS    & SDSS\\
\hskip -0.05in SDSSJ1204$+$0221  & 2.53 & 2.436 & \phn13.3    &  \phn\phn78 &2.4402 &  \phn370    &  1500 &  $19.7\pm 0.15$     & \phn 625      &  Template    & GMOS    & GMOS\\ 
\hskip -0.05in SDSSJ1213$+$1207  & 3.48 & 3.411 &   137.8     &   736       &3.4105 &  \phn\phn30 &  1500 &  $19.25\pm 0.3$     &  \phn\phn39   &  Template    & SDSS    & SDSS\\ 
\hskip -0.05in SDSSJ1306$+$6158  & 2.17 & 2.111 & \phn16.3    &  \phn\phn97 &2.1084 &  \phn200    &   300 &  $20.3\pm 0.15$     & \phn 420      & \ion{Mg}{2}  & LRIS-R  & LRIS-B\\
\hskip -0.05in SDSSJ1312$+$0002  & 2.84 & 2.671 &   148.5     &   850     &  2.6688 &  \phn200    &   500 &  $20.3\pm 0.3$      & \phn\phn23    & \ion{C}{3}]  & SDSS    & SDSS\\
\smallskip
\hskip -0.05in SDSSJ1426$+$5002  & 2.32 & 2.239 &   235.6     &   1397     &  2.2247 &   1330      &   500 &  $20.0\pm 0.15$      & \phn\phn19    & \ion{C}{3}]  & SDSS    & SDSS\\ 
\hskip -0.05in SDSSJ1427$-$0121  & 2.35 & 2.278 & \phn\phn6.2 &   \phn\phn37  &  2.2788 &  \phn50    &   300 &  $18.85\pm 0.25$         & 7871         & \ion{Mg}{2}  & DEIMOS  & GMOS\\
\hskip -0.05in SDSSJ1429$-$0145  & 3.40 & 2.628 &   140.2     &   \phn808     &  2.6235 &  \phn400    &  1000 &  $18.8\pm 0.2$   & \phn\phn 20  & \ion{C}{3}]  &  2QZ    &  SDSS\\ 
% Should we change to the value to the ESI fit?
\hskip -0.05in SDSSJ1430$-$0120  & 3.25 & 3.102 &   200.0     &   1100     &3.115\phn&  \phn960    &  1500 &  $20.5\pm 0.2$   &  \phn\phn 26  &  Template    &  SDSS   &  SDSS\\ 
\hskip -0.05in SDSSJ1545$+$5112  & 2.45 & 2.240 & \phn97.6    &   579     &2.243\phn&  \phn320    &   500 &   $19.45\pm 0.3$             &  \phn\phn30   & \ion{C}{3}]  &  SDSS   &  SDSS\\
\smallskip
\hskip -0.05in SDSSJ1621$+$3508  & 2.04 & 1.931 & \phn76.7    &   463     &  1.9309 &  \phn10     &   300 &  $18.7\pm 0.2$   &  \phn\phn 12  & \ion{Mg}{2}  & LRIS-R  & LRIS-B\\
\hskip -0.05in SDSSJ1635$+$3013  & 2.94 & 2.493 & \phn91.4    &   532     &  2.5025 &  \phn820    &   500 &     $>19$              &  \phn111       & \ion{C}{3}]  &  SDSS   &  SDSS\\
\hskip -0.05in SDSSJ2347$+$1501$^3$ & 2.29 & 2.157 & \phn47.3 &  282   &  2.176  &    1770   &  1000 &     $>18.3$                &   \phn\phn63    & \ion{C}{3}]  &  APO    & GMOS\\
& 2.29 & 2.171 &   223.0  & 1329   &  2.176  &  \phn380    &   500 &     $>17.2$  & \phn\phn\phn 8 & \ion{Mg}{2}  &  SDSS   & GMOS\\
\enddata
\tablecomments{\footnotesize
  Optically thick absorption line systems near foreground quasars.\\
  $^{1}$ In the systems SDSSJ0127$+$1507 there are two distinct background quasars at $z=2.38$ and $z=2.60$, which 
  show absorption in the vicinity of the same foreground quasar at $z=1.818$.\\ 
  $^{2}$ The foreground quasar for this system has large BAL troughs in the Ly$\alpha$ and \ion{C}{4} emission lines. 
  The redshift was computed by comparing the peak of \ion{C}{4}, determined by eye, to the shifted wavelength $\lambda=1545.3$~\AA.  We apply a conservative
  redshift uncertainty of $\pm 1500~\kms$. \\
  $^{3}$ Voigt profile fits to the \lya\ absorption in the SDSS spectrum of the background quasar gave $\log \mnhi = 19.55\pm 0.3$. An archive echelle spectrum 
  of this quasar gives the smaller value which is listed in the table $\log \mnhi = 18.8\pm 0.2$.\\
  $^{4}$ In the systems SDSSJ2347$+$1501, there is a single background quasar at $z=2.29$ and two foreground quasars 
  at $z=2.157$ and $z=2.167$, although the velocity separation is larger than our nominal $1500~\kms$ cutoff for the former.\\
}
\end{deluxetable*}

%% file: tab2.tex
\begin{deluxetable*}{lcccccccccccc}
\tablecolumns{13}
\tablewidth{0pc}
\tablecaption{Tenative Optically Thick Absorbers Near Quasars\label{table:poss}}
\tablehead{Name & z$_{\rm bg}$ & z$_{\rm fg}$ & $\Delta \theta$ & $R$ & z$_{\rm abs}$ & $\left|\Delta v\right|$ & $\Delta v_{\rm fg}$ & $g_{\rm UV}$ & Redshift & Fg & Bg\\
                           &      &       &   ($^{\prime\prime}$)    &  ($\hkpc$)   &         &   ($\kms$)   &  ($\kms$)  &         &            &  Inst. & Inst. }
\startdata
SDSSJ0127$+$1507 &   2.38 &  2.173 &   46.9  &   280   &  2.177  & \phn 370  &  300  &  261      &  \ion{Mg}{2}  &  SDSS  &  LRIS-B\\
SDSSJ0225$+$0048 &   2.82 &  2.692 &   100.3 &   574   &  2.710  &    1430   &  500  &  \phn 23  &  \ion{C}{3}]  &  SDSS  &  GMOS\\
SDSSJ0340$+$0018 &   3.34 &  2.661 &   207.8 &  1192   &  2.6788 &    1420   &  500  & \phn\phn7 &  \ion{C}{3}]  &  SDSS  &  SDSS\\ 
SDSSJ1132$+$5338 &   3.34 &  2.919 &  188.9  &  1060   &  2.914  & \phn 350  &  500  & \phn 10   &  \ion{C}{3}]  &  SDSS  &  SDSS\\
SDSSJ1314$+$6213 &   3.14 &  2.335 &   69.3  &   409   &  2.328  &   670     &  500  & \phn 79   &  \ion{C}{3}]  &  SDSS  &  SDSS\\ 
SDSSJ1356$+$6133 &   2.17 &  2.013 &   22.9  &   138   &  2.026  &  1290     &  1500 &   159     & \ion{C}{3}]  &  SDSS  &  LRIS-B\\
SDSSJ1719$+$2919 &   3.29 &  3.072 &  106.4  &   588   &  3.069  & \phn 220  &  1000 & \phn 30   &  \ion{C}{4}  &  SDSS  &  SDSS\\ 
\enddata
%\tablecomments{\footnotesize Tentative opticllay thick absorption line systems near foreground quasars.}
\end{deluxetable*}

%% file: tab3.tex
\begin{deluxetable*}{lcccccccc}
\tablecolumns{9}
\tablewidth{0pc}
\tablecaption{Coordinates and Photometry for Projected Quasar Pairs in Table \ref{table:thick}
  \label{table:thick_coords}}
\tablehead{Name & RA (2000) & Dec (2000) & z & $u$ &$g$ & $r$ & $i$ & $z$ }
\startdata
SDSSJ0036$+$0839A &   00:36:43.45 &   $+$08:39:44.4 &   2.69 &  20.13  &   19.54  &   19.28  &   19.30  &   19.00\\
\smallskip 
SDSSJ0036$+$0839B &   00:36:53.85 &   $+$08:39:36.2 &   2.57 &  20.90  &   20.37  &   20.34  &   20.30  &   20.33\\
SDSSJ0127$+$1508A &   01:27:44.85 &   $+$15:08:58.0 &   2.60 &  20.95  &   20.56  &   20.50  &   20.38  &   20.33\\ 
SDSSJ0127$+$1507B &   01:27:42.57 &   $+$15:07:38.4 &   2.38 &  21.21  &   20.61  &   20.55  &   20.71  &   20.70\\ 
\smallskip 
SDSSJ0127$+$1506E &   01:27:43.53 &   $+$15:06:48.4 &   1.82 &  21.21  &   21.25  &   21.14  &   20.94  &   20.99\\ 
SDSSJ0225$-$0739A &   02:25:59.78 &   $-$07:39:38.9 &   3.04 &  21.92  &   19.53  &   19.17  &   19.07  &   19.06\\ 
\smallskip 
SDSSJ0225$-$0743B &   02:25:56.50 &   $-$07:43:07.3 &   2.44 &  20.70  &   19.95  &   19.82  &   19.58  &   19.30\\ 
SDSSJ0239$-$0106A &   02:39:46.44 &   $-$01:06:44.1 &   3.12 &  22.20  &   20.28  &   19.99  &   20.00  &   19.93\\ 
\smallskip 
SDSSJ0239$-$0106B &   02:39:46.43 &   $-$01:06:40.4 &   2.31 &  21.26  &   20.61  &   20.56  &   20.54  &   20.19\\ 
SDSSJ0256$+$0039A &   02:56:47.15 &   $+$00:39:01.2 &   3.55 &  18.22  &   17.08  &   16.65  &   30.89  &   16.47\\ 
\smallskip 
SDSSJ0256$+$0036B &   02:56:50.89 &   $+$00:36:11.3 &   3.39 &  22.23  &   20.23  &   19.96  &   19.87  &   20.08\\ 
SDSSJ0303$-$0023A &   03:03:41.01 &   $-$00:23:21.9 &   3.22 &  19.48  &   17.48  &   17.27  &   17.25  &   17.26\\
\smallskip 
SDSSJ0303$-$0020B &   03:03:35.42 &   $-$00:20:01.1 &   2.72 &  20.38  &   19.89  &   19.67  &   19.46  &   19.26\\
SDSSJ0338$-$0005A &   03:38:54.78 &   $-$00:05:21.0 &   3.05 &  19.65  &   18.42  &   18.21  &   18.18  &   18.14\\ 
\smallskip 
SDSSJ0338$-$0006B &   03:38:51.83 &   $-$00:06:19.6 &   2.24 &  21.67  &   20.99  &   20.74  &   20.53  &   20.18\\ 
SDSSJ0800$+$3542A &   08:00:48.74 &   $+$35:42:31.3 &   2.07 &  19.55  &   19.54  &   19.55  &   19.42  &   19.27\\ 
\smallskip 
SDSSJ0800$+$3542B &   08:00:49.90 &   $+$35:42:49.6 &   1.98 &  19.25  &   19.14  &   19.13  &   18.94  &   18.75\\ 
SDSSJ0814$+$3250A &   08:14:19.59 &   $+$32:50:18.7 &   2.21 &  20.79  &   20.33  &   20.33  &   20.17  &   19.70\\ 
\smallskip 
SDSSJ0814$+$3250B &   08:14:20.38 &   $+$32:50:16.1 &   2.18 &  20.28  &   19.93  &   19.80  &   19.73  &   19.46\\ 
SDSSJ0833$+$0813A &   08:33:28.07 &   $+$08:13:17.5 &   3.34 &  21.64  &   19.63  &   19.17  &   19.19  &   19.11\\ 
\smallskip
SDSSJ0833$+$0812B &   08:33:21.61 &   $+$08:12:38.6 &   2.52 &  20.82  &   20.20  &   20.04  &   20.11  &   19.78\\ 
SDSSJ0852$+$2637A &   08:52:37.93 &   $+$26:37:58.5 &   3.32 &  22.94  &   19.64  &   19.18  &   19.07  &   19.12\\ 
\smallskip
SDSSJ0852$+$2635B &   08:52:32.16 &   $+$26:35:26.2 &   3.20 &  24.74  &   20.70  &   20.34  &   20.19  &   20.18\\ 
SDSSJ0902$+$2841A &   09:02:34.54 &   $+$28:41:18.6 &   3.58 &  25.96  &   20.09  &   19.14  &   19.00  &   18.77\\
\smallskip
SDSSJ0902$+$2839B &   09:02:23.33 &   $+$28:39:30.2 &   3.33 &  22.93  &   19.71  &   19.26  &   19.19  &   19.16\\
SDSSJ1134$+$3409A &   11:34:44.22 &   $+$34:09:33.8 &   3.14 &  20.58  &   18.58  &   18.37  &   18.43  &   18.42\\ 
\smallskip
SDSSJ1134$+$3406B &   11:34:53.94 &   $+$34:06:42.8 &   2.29 &  19.21  &   18.82  &   18.82  &   18.81  &   18.62\\ 
SDSSJ1152$+$4517A &   11:52:00.54 &   $+$45:17:41.5 &   2.38 &  19.75  &   19.12  &   19.13  &   19.08  &   18.87\\
\smallskip
SDSSJ1152$+$4518B &   11:52:10.43 &   $+$45:18:25.9 &   2.31 &  19.46  &   18.99  &   18.98  &   18.98  &   18.76\\
\smallskip
%SDSSJ1152$+$4517B &   11:52:10.43 &   $+$45:18:25.9 &   2.28 &  19.44  &   18.95  &   18.94  &   18.95  &   18.73\\
SDSSJ1204$+$0221A &   12:04:16.69 &   $+$02:21:11.0 &   2.53 &  19.68  &   19.06  &   19.02  &   18.96  &   18.67\\ 
\smallskip
SDSSJ1204$+$0221B &   12:04:17.47 &   $+$02:21:04.7 &   2.44 &  20.99  &   20.52  &   20.46  &   20.49  &   20.05\\ 
SDSSJ1213$+$1207A &   12:13:10.72 &   $+$12:07:15.1 &   3.48 &  22.85  &   20.15  &   19.58  &   19.50  &   19.53\\ 
\smallskip
SDSSJ1213$+$1208B &   12:13:03.26 &   $+$12:08:39.0 &   3.41 &  25.74  &   20.55  &   19.86  &   19.76  &   19.69\\ 
SDSSJ1306$+$6158A &   13:06:03.55 &   $+$61:58:35.2 &   2.17 &  20.88  &   20.31  &   20.33  &   20.39  &   20.10\\ 
\smallskip
SDSSJ1306$+$6158B &   13:06:05.19 &   $+$61:58:23.7 &   2.11 &  20.33  &   20.23  &   20.17  &   20.19  &   20.01\\ 
SDSSJ1312$+$0002A &   13:12:13.30 &   $+$00:02:31.3 &   2.88 &  22.32  &   20.71  &   20.03  &   19.45  &   19.02\\ 
\smallskip
SDSSJ1312$+$0000B &   13:12:13.84 &   $+$00:00:03.0 &   2.67 &  20.42  &   19.36  &   19.17  &   18.84  &   18.55\\ 
SDSSJ1426$+$5002A &   14:26:28.02 &   $+$50:02:48.3 &   2.32 &  18.87  &   18.27  &   18.16  &   18.06  &   17.85\\ 
\smallskip
SDSSJ1426$+$5004B &   14:26:05.79 &   $+$50:04:26.6 &   2.24 &  18.40  &   17.88  &   17.81  &   17.73  &   17.52\\ 
\smallskip
SDSSJ1427$-$0121A &   14:27:58.73 &   $-$01:21:36.2 &   2.35 &  20.07  &   19.32  &   19.27  &   19.27  &   18.99\\ 
\smallskip
\hfill 2QZJ1427$-$0121B&   14:27:58.89 &   $-$01:21:30.4 &   2.28 &  19.83  &   19.26  &   19.15  &   19.13  &   19.08\\ 
SDSSJ1429$-$0145A &   14:29:03.04 &   $-$01:45:19.4 &   3.39 &  21.51  &   18.58  &   17.96  &   17.78  &   17.64\\ 
\smallskip
\hfill 2QZJ1429$-$0146B&   14:29:10.81 &   $-$01:46:37.3 &   2.63 &  20.08  &   19.45  &   19.41  &   19.43  &   19.27\\ 
SDSSJ1430$-$0120A &   14:30:06.42 &   $-$01:20:20.0 &   3.26 &  22.95  &   20.47  &   20.05  &   19.94  &   19.82\\ 
\smallskip
SDSSJ1429$-$0117B &   14:29:57.09 &   $-$01:17:57.1 &   3.10 &  21.97  &   19.52  &   19.02  &   18.84  &   18.62\\ 
SDSSJ1545$+$5112A &   15:45:34.60 &   $+$51:12:29.3 &   2.45 &  20.07  &   19.43  &   19.38  &   19.35  &   19.15\\ 
\smallskip
SDSSJ1545$+$5113B &   15:45:44.15 &   $+$51:13:07.5 &   2.24 &  19.65  &   19.18  &   19.19  &   19.17  &   18.91\\ 
SDSSJ1621$+$3508A &   16:21:45.42 &   $+$35:08:07.2 &   2.03 &  18.76  &   18.71  &   18.65  &   18.55  &   18.42\\ 
\smallskip
SDSSJ1621$+$3507B &   16:21:41.02 &   $+$35:07:12.8 &   1.93 &  20.73  &   20.57  &   20.54  &   20.45  &   20.23\\ 
SDSSJ1635$+$3013A &   16:35:00.06 &   $+$30:13:21.7 &   2.94 &  20.60  &   19.05  &   18.85  &   18.82  &   18.93\\ 
\smallskip
SDSSJ1634$+$3014B &   16:34:56.16 &   $+$30:14:37.8 &   2.49 &  19.08  &   18.52  &   18.26  &   18.26  &   18.13\\ 
SDSSJ2347$+$1501A &   23:47:03.24 &   $+$15:01:01.5 &   2.29 &  20.22  &   19.52  &   19.44  &   19.34  &   19.06\\ 
SDSSJ2347$+$1501B&   23:47:04.25 &    $+$15:01:46.4 &   2.16 &  20.05  &   20.04  &   20.02  &   19.80  &   19.44\\
SDSSJ2346$+$1457C&   23:46:57.26 &    $+$14:57:35.9 &   2.18 &  19.50  &   18.99  &   18.74  &   18.62  &   18.48\\ 
\enddata
\tablecomments{\footnotesize Coordinates, redshifts, and extinction corrected SDSS five band PSF
  photometry are given in the columns $u$, $g$, $r$, $i$, and $z$ for the projected quasar pairs in Table~\ref{table:thick}. 
  The background and foreground quasars are labeled `A' and `B', respectively.}
\end{deluxetable*}

%% file: tab4.tex
\begin{deluxetable*}{lcccccccc}
\tablecolumns{9}
\tablewidth{0pc}
\tablecaption{Coordinates and Photometry for Projected Quasar Pairs in Table~\ref{table:poss}\label{table:poss_coords}}
\tablehead{Name & RA (2000) & Dec (2000) & z & $u$ &$g$ & $r$ & $i$ & $z$ }
\startdata
SDSSJ0127$+$1507A &   01:27:42.57 &   $+$15:07:38.4 &   2.38 &  21.21  &   20.61  &   20.55  &   20.71  &   20.70\\ 
\smallskip 
SDSSJ0127$+$1506B &   01:27:41.20 &   $+$15:06:55.9 &   2.17 &  18.73  &   18.51  &   18.36  &   18.36  &   18.17\\ 
SDSSJ0340$+$0018A &   03:40:14.25 &   $+$00:18:47.9 &   3.34 &  22.02  &   20.28  &   19.92  &   19.85  &   19.78\\ 
\smallskip 
SDSSJ0340$+$0019B &   03:40:00.92 &   $+$00:19:44.7 &   2.66 &  20.80  &   20.11  &   19.94  &   19.92  &   19.65\\ 
SDSSJ1132$+$5338A &   11:32:06.14 &   $+$53:38:08.7 &   3.34 &  24.66  &   20.42  &   20.08  &   19.91  &   19.74\\ 
\smallskip 
SDSSJ1131$+$5335B &   11:31:50.70 &   $+$53:35:59.0 &   2.92 &  21.17  &   20.02  &   19.92  &   19.80  &   19.57\\ 
SDSSJ1314$+$6213A &   13:14:29.24 &   $+$62:13:00.1 &   3.13 &  22.01  &   19.94  &   19.65  &   19.50  &   19.51\\ 
\smallskip 
SDSSJ1314$+$6212B &   13:14:19.52 &   $+$62:12:46.6 &   2.34 &  19.90  &   19.21  &   19.00  &   19.00  &   18.70\\ 
SDSSJ1356$+$6133A &   13:56:29.54 &   $+$61:33:10.4 &   2.17 &  20.43  &   20.18  &   20.13  &   19.78  &   19.59\\ 
\smallskip 
SDSSJ1356$+$6133B &   13:56:32.44 &   $+$61:33:00.7 &   2.01 &  20.67  &   20.43  &   20.32  &   20.24  &   20.16\\ 
SDSSJ1719$+$2919A &   17:19:32.94 &   $+$29:19:29.7 &   3.29 &  23.00  &   20.16  &   19.86  &   19.71  &   19.77\\ 
SDSSJ1719$+$2918B &   17:19:37.87 &   $+$29:18:05.0 &   3.07 &  22.72  &   20.56  &   20.20  &   20.19  &   20.30\\ 
\enddata
\tablecomments{\footnotesize Coordinates, redshifts, and extinction corrected SDSS five band PSF
  photometry are given in the columns $u$, $g$, $r$, $i$, and $z$ for the projected quasar pairs in Table~\ref{table:poss}. 
  The background and foreground quasars are labeled `A' and `B', respectively.}
\end{deluxetable*}

%% file: tab5.tex
\begin{deluxetable*}{lcccccccc}
\tablecolumns{9}
\tablewidth{0pc}
\tablecaption{Coordinates and Photometry for the Five Quasars In the Projected Group SDSSJ0127$+$1507\label{table:0127}}
\tablehead{Name & RA (2000) & Dec (2000) & z & $u$ &$g$ & $r$ & $i$ & $z$ }
\startdata
SDSSJ0127$+$1508A &   01:27:44.85 &   $+$15:08:58.0 &   2.60 &  20.95  &   20.56  &   20.50  &   20.38  &   20.33\\ 
SDSSJ0127$+$1507B &   01:27:42.57 &   $+$15:07:38.4 &   2.38 &  21.21  &   20.61  &   20.55  &   20.71  &   20.70\\ 
SDSSJ0127$+$1506C &   01:27:41.20 &   $+$15:06:55.9 &   2.18 &  18.73  &   18.51  &   18.36  &   18.36  &   18.17\\ 
SDSSJ0127$+$1504D &   01:27:40.77 &   $+$15:04:10.3 &   2.08 &  19.93  &   20.07  &   20.03  &   19.86  &   19.57\\ 
SDSSJ0127$+$1506E &   01:27:43.53 &   $+$15:06:48.4 &   1.82 &  21.21  &   21.25  &   21.14  &   20.94  &   20.99\\ 
\enddata
\tablecomments{\footnotesize Coordinates, redshifts, and extinction corrected SDSS five band PSF
  photometry are given in the columns $u$, $g$, $r$, $i$, and $z$ for the projected quasar pairs in Table~\ref{table:thick}. 
  The background and foreground quasars are labeled `A' and `B', respectively.}
\end{deluxetable*}